\documentclass[iop,apj]{emulateapj}
\usepackage{apjfonts}
\usepackage{graphicx} 
\usepackage{color} 
\pdfoutput=1

\newcommand{\D}[2]{\frac{\partial #2}{\partial #1}}

\newcommand{\deriv}[2]{\frac{{\rm d} #2}{{\rm d} #1}}
\newcommand\bb[1]{\mathbit{#1}}

\font\syvec=cmbsy10                        
\def\grad{\hbox{{\syvec\char114}}}       

\newcommand{\mc}[1]{\mathcal{#1}}

\newcommand{\msb}[1]{\mathsf{#1}}

\newcommand{\eb}{\hat{\bb{b}}}
\newcommand{\ex}{\hat{\bb{x}}}

\newcommand{\ez}{\hat{\bb{z}}}
\newcommand{\kb}{k_{\rm B}}

\newcommand{\vthsq}{v^2_{\rm th}}
\newcommand{\mfp}{\lambda_{\rm mfp}}
\newcommand{\cond}{\omega_{\rm cond}}
\newcommand{\visc}{\omega_{\rm visc}}
\newcommand{\dyn}{\omega_{\rm dyn}}
\newcommand{\cool}{\omega_{\rm cool}}

\shortauthors{KUNZ ET AL.}

\begin{document}
\title{Buoyancy Instabilities in a Weakly Collisional Intracluster Medium}
\shorttitle{BUOYANCY INSTABILITIES IN THE ICM}
\author{Matthew W. Kunz\altaffilmark{1,4,5}, Tamara Bogdanovi\'{c}\altaffilmark{2,3,5}, Christopher S. Reynolds\altaffilmark{2,3}, and James M. Stone\altaffilmark{1}}
\affil{\altaffilmark{1}{Department of Astrophysical Sciences, Princeton University, Peyton Hall, 4 Ivy Lane, Princeton, NJ 08544, USA; kunz@astro.princeton.edu, jstone@astro.princeton.edu}}
\affil{\altaffilmark{2}{Department of Astronomy, University of Maryland, College Park, MD 20742, USA; tamarab@astro.umd.edu, chris@astro.umd.edu}}
\affil{\altaffilmark{3}{Joint Space Science Institute (JSI), University of Maryland, College Park, MD 20742, USA}}
\altaffiltext{4}{Previous Address: Rudolf Peierls Centre for Theoretical Physics, University of Oxford, 1 Keble Road, Oxford OX1 3NP, U. K.}
\altaffiltext{5}{Einstein Postdoctoral Fellow}

\begin{abstract}
The intracluster medium (ICM) of galaxy clusters is a weakly collisional, high-beta plasma in which the transport of heat and momentum occurs primarily along magnetic-field lines. Anisotropic heat conduction allows convective instabilities to be driven by temperature gradients of either sign, the magnetothermal instability (MTI) in the outskirts of non-isothermal clusters and the heat-flux buoyancy-driven instability (HBI) in their cooling cores. We employ the Athena magnetohydrodynamic code to investigate the nonlinear evolution of these instabilities, self-consistently including the effects of anisotropic viscosity (i.e. Braginskii pressure anisotropy), anisotropic conduction, and radiative cooling. We highlight the importance of the microscale instabilities (firehose, mirror) that inevitably accompany and regulate the pressure anisotropies generated by the HBI and MTI. We find that, in all but the innermost regions of cool-core clusters, anisotropic viscosity significantly impairs the ability of the HBI to reorient magnetic-field lines orthogonal to the temperature gradient. Thus, while radio-mode feedback appears necessary in the central few tens of kpc, heat conduction may be capable of offsetting radiative losses throughout most of a cool core over a significant fraction of the Hubble time. Magnetically-aligned cold filaments are then able to form by local thermal instability. Viscous dissipation during the formation of a cold filament produces accompanying hot filaments, which can be searched for in deep {\it Chandra} observations of nearby cool-core clusters. In the case of the MTI, anisotropic viscosity maintains the coherence of magnetic-field lines over larger distances than in the inviscid case, thereby providing a natural lower limit for the scale on which the field can fluctuate freely. In the nonlinear state, the magnetic field exhibits a folded structure in which the field-line curvature and field strength are anti-correlated. These results demonstrate that, if the HBI and MTI are relevant for shaping the properties of the ICM, one must self-consistently include anisotropic viscosity in order to obtain even qualitatively correct results.
\end{abstract}


\section{Introduction}

Clusters of galaxies are filled with hot and tenuous plasma, the intracluster medium (ICM), the detailed properties of which governs such important physics as heat and momentum transport, magnetogenesis, and thermodynamic stability. These properties are complicated by the fact that the ICM is only weakly collisional: while the particle mean free path $\mfp$ is $\sim$$10$--$10^3$ times smaller than the thermal-pressure scale height $H$, it is nevertheless $\sim$$10^{11}$--$10^{13}$ times larger than the ion gyroradius $r_{\rm g,i}$ \citep[e.g.][and references therein]{sc06}. As such, the material properties of the ICM are strongly anisotropic with respect to the magnetic-field direction, despite the fact that the strength of the intracluster magnetic field is relatively weak ($\sim$$0.1$--$10~\mu{\rm G}$, which constitutes only $\sim$$0.01$--$1\%$ of the thermal energy; for a review, see \citealt{ct02}).

This anisotropy fundamentally changes the convective stability properties of the ICM \citep{balbus00}. Temperature gradients, rather than entropy gradients, become the discriminating quantities that determine stability, regardless of whether temperature increases \citep{balbus00,balbus01} or decreases \citep{quataert08} in the direction of gravity. As non-isothermal clusters generally exhibit both regions of positive and negative temperature gradients, the entire ICM ought to be linearly unstable to convective motions.

The temperature in the cores of non-isothermal clusters decreases in the direction of gravity due to efficient radiative cooling at high densities \citep[e.g.][]{fabian94,pjkt05,vmmjfs05}, and any alignment of conducting magnetic-field lines and gravity there can lead to a heat-flux buoyancy-driven instability \citep[HBI;][]{quataert08}. Nonlinear numerical simulations of the HBI have revealed that the instability acts in such a way as to quiescently shut itself off, gradually reorienting magnetic field to be perpendicular to the temperature gradient and thus stifling the heat flux that gave rise to the instability in the first place \citep{pq08,mpsq11}. In the presence of radiative cooling, this field-line reorientation ultimately insulates the core, exacerbating the cooling-flow problem \citep{pqs09,brbp09,mty11} unless field lines are re-opened by sufficient turbulent stirring \citep{pqs10,ro10,mpsq11}.

Beyond the cooling radius, the temperature increases in the direction of gravity due to virialized gravitational infall. In this case, any {\em mis}alignment of magnetic-field lines and gravity can lead to a magnetothermal instability \citep[MTI;][]{balbus00,balbus01}. In the presence of a sustained temperature gradient, the MTI leads to vigorous subsonic turbulence and a radially biased magnetic field \citep{ps05,ps07,mpsq11}. The former may provide up to $5$--$30\%$ of the pressure support beyond $r_{500}$, reducing the observed Sunyaev-Zel'dovich signal and biasing X-ray mass estimates of clusters \citep{pmqs11}. The latter can lead to efficient radial heat transport, resulting in large-scale temperature profiles flatter than those expected from structure formation calculations \citep{psl08}.

These studies of the HBI and MTI did not include an important feature of weakly collisional plasmas: changes in magnetic-field strength and/or density that occur on timescales much greater than the inverse of the cyclotron frequency result in pressure anisotropies (i.e. the gas pressure perpendicular and parallel to the local magnetic field become unequal). These pressure anisotropies manifest themselves as anisotropic viscous stresses, which target precisely those motions originally responsible for the anisotropies themselves. By means of a linear stability analysis, which self-consistently accounted for the dynamical effects of both anisotropic conduction and viscosity, \citet[][hereafter K11]{kunz11} found that the HBI and MTI, when subject to the pressure anisotropies they induce, are qualitatively and quantitatively changed from what earlier studies had suggested.

In brief, instabilities that depend upon the convergence/divergence of magnetic-field lines to generate unstable buoyant motions (the HBI) are suppressed over much of the wavenumber space, whereas those that are otherwise impeded by field-line convergence/divergence (the MTI) are strengthened (K11). This not only reduces HBI growth rates but also increases the wavelengths of the fastest-growing modes to $\lambda_{\rm HBI} / H \sim 0.2$--$1$ (increasing outwards) for typical cool-core parameters. Taking into consideration the non-local nature of these modes, \citet{lk12} conjectured that the field-line insulation thought to be a nonlinear consequence of the HBI would be attenuated in all but the innermost $\sim$$20\%$ of cluster cores. Perhaps not coincidentally, these regions tend to be dominated by strong radio-mode feedback from powerful central dominant galaxies \citep[e.g.][]{burns90,mhrc09,sun09,bcsrm10}. The fastest-growing linear MTI modes, on the other hand, escape the effects of pressure anisotropy by orientating their velocities perpendicular to the magnetic field. However, anisotropic viscosity couples Alfv\'{e}n and magnetosonic waves in such a way that damped slow-mode perturbations excite a buoyantly unstable Alfv\'{e}nic response when the temperature increases in the direction of gravity. Consequently, many wavenumbers previously considered MTI-stable or slow-growing are in fact maximally unstable (see fig. 2 of K11 for an example).

These changes raise a number of questions. Is the field-line insulation thought to be a nonlinear consequence of the HBI attenuated by anisotropic viscosity? If so, can anisotropic viscosity help conduction stave off a cooling catastrophe over astrophysically relevant timescales? What role does anisotropic conduction and viscosity play in the generation of the cold filaments commonly observed in cluster cores \citep[e.g.][]{lynds70,fabian08}? Does anisotropic viscosity significantly affect the ability of the HBI and MTI to amplify magnetic fields and drive turbulence? Are the resulting field strengths, magnetic-field topologies, and turbulent velocities compatible with those observationally inferred \citep[e.g.][]{sfmbb04,ve05,sfs11}? In this paper, we employ numerical simulations to understand these instabilities and the implications they have for the observable structure and evolution of a weakly collisional ICM.

An outline of the paper is as follows. In Section \ref{sec:equations} we present the basic equations describing a weakly collisional ICM, identifying the important dimensionless free parameters that govern the plasma physics. Section \ref{sec:numerics} describes our numerical approach and details our treatment of the microscale plasma instabilities that inevitably develop in our simulations. We then present our results concerning the non-radiative HBI (Section \ref{sec:hbi}), the radiative HBI (Section \ref{sec:radhbi}), and the MTI (Section \ref{sec:mti}). Finally, in Section \ref{sec:summary} we provide a brief summary of these results, a discussion of their astrophysical implications, a comparison with related work, and an outlook of what is required improve our understanding of the ICM.

\section{Basic equations}\label{sec:equations}

The fundamental equations of motion may be written in conservative form as
\begin{equation}\label{eqn:continuity}
\D{t}{\rho} + \grad \cdot ( \rho \bb{v} ) = 0 ,
\end{equation}
\begin{equation}\label{eqn:momentum}
\D{t}{ ( \rho \bb{v} ) } + \grad \cdot \left( \rho \bb{v} \bb{v} + \msb{P}^\ast \right)  = \rho \bb{g} ,
\end{equation}
\begin{equation}\label{eqn:energy}
\D{t}{E} + \grad \cdot \left( E \bb{v} + \msb{P}^\ast \cdot \bb{v} + \bb{q} \right) = \rho \bb{g} \cdot \bb{v} - \rho \mc{L} ,
\end{equation}
\begin{equation}\label{eqn:induction}
\D{t}{\bb{B}} + \grad \cdot \left( \bb{v} \bb{B} - \bb{B} \bb{v} \right) = 0 , 
\end{equation}
where $\rho$ is the mass density, $\bb{v}$ is the velocity, $\bb{g}$ is the gravitational acceleration, $\bb{B}$ is the magnetic field,
\begin{equation}
E = \frac{1}{2} \rho v^2 + \frac{p}{\gamma - 1} + \frac{B^2}{8\pi}
\end{equation}
is the total (kinetic + internal + magnetic) energy density, and $p$ is the gas pressure. The ratio of specific heats $\gamma = 5/3$. We consider a hydrogenic plasma with equal ion and electron temperatures, $T_{\rm i} = T_{\rm e} = T$, and number densities, $n_{\rm i} = n_{\rm e}$, so that the mean mass per particle is $m_{\rm p} / 2$. The thermal speed of the ions is then $v_{\rm th} \equiv ( p / \rho )^{1/2} = (2 \kb  T / m_{\rm p} )^{1/2}$.

In Equations \ref{eqn:momentum} and \ref{eqn:energy}, we have introduced the total (gas + magnetic) pressure tensor
\begin{equation}\label{eqn:ptensor}
\msb{P}^\ast =  \left( p_\perp + \frac{B^2}{8\pi} \right) \msb{I} - \left( p_\perp - p_{\|} + \frac{B^2}{4\pi} \right) \eb \eb ,
\end{equation}
where $p_\perp$ ($p_\|$) is the gas pressure perpendicular (parallel) to the magnetic field and $\eb = \bb{B} / B$ is the unit vector in the direction of the magnetic field. The total gas pressure satisfies
\begin{equation}
p = \frac{2}{3} p_\perp + \frac{1}{3} p_{\|} .
\end{equation}
Differences between the perpendicular and parallel gas pressure arise from the conservation of the first and second adiabatic invariants for each particle on timescales much greater than the inverse of the gyrofrequency, $\Omega^{-1}_{\rm g}$ \citep{cgl56}. When the ion--ion collision frequency $\nu_{\rm ii}$ is much larger than the rates of change of all fields, an equation for the pressure anisotropy can be obtained by balancing its production by adiabatic invariance with its relaxation via collisions:
\begin{equation}\label{eqn:anisotropy}
p_\perp - p_{\|} = 0.960 \times \frac{p_{\rm i}}{\nu_{\rm ii}} \deriv{t}{} \ln \frac{B^3}{\rho^{2}} ,
\end{equation}
where $p_{\rm i}$ is the ion gas pressure \citep[e.g.][]{cs04}. Defining the (ion) parallel viscous diffusivity,
\begin{eqnarray}\label{eqn:nu}
\nu_{\|} & \equiv & 0.960 \times \frac{1}{2} \frac{\vthsq }{\nu_{\rm ii}} \\*
& \simeq & 0.031 \left( \frac{n_{\rm i}}{0.01~{\rm cm}^{-3}} \right)^{-1}  \left( \frac{\kb  T}{2~{\rm keV}} \right)^{5/2} ~ {\rm kpc}^2 ~ {\rm Myr}^{-1} , \nonumber
\end{eqnarray}
and using Equations (\ref{eqn:continuity}) and (\ref{eqn:induction}) to replace the time derivatives of density and magnetic-field strength with velocity gradients, the pressure anisotropy (Equation \ref{eqn:anisotropy}) may be written
\begin{equation}\label{eqn:braginskii}
p_\perp - p_{\|} = 3 \rho \nu_{\|} \left( \eb\eb - \frac{1}{3} \msb{I} \right) \bb{:} \grad \bb{v} .
\end{equation}
This pressure anisotropy is the physical effect behind what is known as \citet{braginskii65} viscosity -- the restriction of the viscous damping (to dominant order in the Larmor radius expansion) to the motions and gradients parallel to the magnetic field. In an incompressible fluid, small-amplitude parallel-velocity fluctuations with parallel wavenumber $k_{\|}$ are damped at a rate 
\begin{eqnarray}\label{eqn:omegavisc}
\omega_{\rm visc} & \equiv & 3 k^2_{\|} \nu_{\|} \\* 
& \simeq & 0.61 \left( \frac{ k_{\|} H }{10} \right)^2  \left( \frac{n_{\rm i}}{0.01~{\rm cm}^{-3}} \right)^{-1} \left( \frac{\kb  T}{2~{\rm keV}} \right)^{5/2} ~{\rm Gyr}^{-1} . \nonumber
\end{eqnarray}
Motions that do not affect the magnetic-field strength to linear order (e.g. Alfv\'{e}n waves) are allowed at subviscous scales. In the weak-field regime, these motions take the form of plasma instabilities (see Section \ref{sec:microscale}).

The vast disparity between the gyro- and collision frequencies also implies that the heat flux $\bb{q}$ is anisotropic with respect to the magnetic field \citep{braginskii65}:
\begin{equation}\label{eqn:heatflux}
\bb{q} = - \rho \kappa_{\|} \, \eb \eb \cdot \grad \vthsq  ,
\end{equation}
where
\begin{eqnarray}\label{eqn:kappa}
\kappa_{\|} & \equiv & 1.581 \times \frac{1}{2} \frac{v^2_{\rm th,e}}{\nu_{\rm ee}} \\*
& \simeq & 1.67 \left( \frac{n_{\rm i}}{0.01~{\rm cm}^{-3}} \right)^{-1} \left( \frac{\kb  T}{2~{\rm keV}} \right)^{5/2}  ~ {\rm kpc}^2 ~ {\rm Myr}^{-1} \nonumber 
\end{eqnarray}
is the (electron) thermal parallel diffusivity, $v^2_{\rm th,e} = 2 \kb  T / m_{\rm e}$ is the square of the electron thermal velocity, and $\nu_{\rm ee}$ is the electron--electron collision frequency \citep[e.g.][]{cs04}. Equation (\ref{eqn:heatflux}) states that heat is transported along magnetic-field lines when there is a component of the temperature gradient aligned with the magnetic field. Field-aligned temperature fluctuations with parallel wavenumber $k_{\|}$ are diffused away at a rate
\begin{eqnarray}\label{eqn:omegacond}
\cond & \equiv & \frac{2}{5} k^2_{\|} \kappa_{\|} \\*
& \simeq & 4.1 \left( \frac{ k_{\|} H }{10} \right)^2 \left( \frac{n_{\rm i}}{0.01~{\rm cm}^{-3}} \right)^{-1}  \left( \frac{\kb  T}{2~{\rm keV}} \right)^{5/2} ~{\rm Gyr}^{-1}. \nonumber
\end{eqnarray}
For future reference, we also define the (electron) parallel thermal conductivity $\chi_{\|} \equiv p \kappa_{\|} / T \propto T^{5/2}$.

The ratio of the viscous and thermal diffusivities is known as the Prandtl number ${\rm Pr}$, which is roughly constant:
\begin{equation}\label{eqn:prandtl}
{\rm Pr} \equiv \frac{\nu_{\|}}{\kappa_{\|}} = 0.607  ~ \frac{\Lambda_{\rm e}}{\Lambda_{\rm i}} \left( \frac{2 m_{\rm e}}{m_{\rm i}} \right)^{1/2} \simeq 0.02 ,
\end{equation}
where $\Lambda_{\rm e}$ ($\Lambda_{\rm i}$) is the electron (ion) Coulomb logarithm. This implies that viscous forces operate on a timescale that is a fixed number greater than the timescale on which conduction operates. In addition, there are two more important dimensionless parameters: the plasma beta,
\begin{eqnarray}
\beta & \equiv & \frac{8 \pi p}{B^2} \\*
& \simeq & 1610  \left( \frac{B}{1~\mu{\rm G}} \right)^{-2} \left( \frac{n_{\rm i}}{0.01~{\rm cm}^{-3}} \right) \left( \frac{\kb  T}{2~{\rm keV}} \right) ; \nonumber
\end{eqnarray}
and the inverse of the Knudsen number,
\begin{eqnarray}\label{eqn:knudsen}
{\rm Kn}^{-1} & \equiv & \frac{H}{\lambda_{{\rm mfp}}} \\*
& \simeq& 1207 \left( \frac{g}{10^{-8}~{\rm cm~s}^{-2}} \right)^{-1} \left( \frac{n_{\rm i}}{0.01~{\rm cm}^{-3}} \right)  \left( \frac{\kb  T}{2~{\rm keV}} \right)^{-1} ,\nonumber
\end{eqnarray}
where $H = \vthsq  / g$ is the thermal-pressure scale height.\footnote{In determining the numerical value of ${\rm Kn}$, we have assumed force balance between thermal pressure and gravity -- an assumption that will hold as an initial condition in all of our simulations.} The Knudsen number is a dimensionless measure of collisionality, and determines whether a fluid (rather than kinetic) description may be used. Introducing the dynamical frequency 
\begin{eqnarray}
\dyn & \equiv & \left( \frac{g}{H} \right)^{1/2}  \\*
& \simeq & 5.1 \left( \frac{g}{10^{-8}~{\rm cm~s}^{-2}} \right) \left( \frac{\kb  T}{2~{\rm keV}} \right)^{-1/2} ~{\rm Gyr}^{-1} , \nonumber
\end{eqnarray}
the Knudsen number may equivalently be expressed as a ratio of frequencies: ${\rm Kn} = \dyn / \nu_{\rm ii}$. In terms of Kn, the viscous and thermal parallel diffusivities are
\begin{equation}
\nu_{\|} = 0.48 \, {\rm Kn} \, v_{\rm th} H ,
\end{equation}
\begin{equation}
\kappa_{\|} \simeq 24 \, {\rm Kn} \, v_{\rm th} H ,
\end{equation}
respectively.

Finally, the last term in Equation \ref{eqn:energy} represents radiative losses. The cooling in the ICM is dominated by thermal Bremsstrahlung above temperatures $\sim$$1~{\rm keV}$, for which the radiative cooling rate (per unit volume) is
\begin{equation}\label{eqn:bremsstrahlung}
\rho \mc{L} \simeq 10^{-27} \left( \frac{n_{\rm i}}{0.01~{\rm cm}^{-3}} \right)^2 \left( \frac{\kb  T}{2~{\rm keV}} \right)^{1/2} ~ {\rm ergs~cm}^{-3}~{\rm s}^{-1}
\end{equation}
\citep{rl79}. For an isobaric perturbation, Equations (\ref{eqn:energy}) and (\ref{eqn:bremsstrahlung}) imply a cooling frequency 
\begin{eqnarray}
\cool & \equiv & \left. \frac{2}{5} \D{\ln T}{ ( \rho \mc{L} / p ) } \right|_{p} = - \frac{3}{5} \frac{\rho \mc{L}}{p} \\*
& \simeq & - 0.3 \left( \frac{n_{\rm i}}{0.01~{\rm cm}^{-3}} \right) \left( \frac{\kb  T}{2~{\rm keV}} \right)^{-1/2} ~ {\rm Gyr}^{-1} . \nonumber
\end{eqnarray}
The fact that the cooling frequency is negative indicates that isobaric thermal instability is possible \citep{field65}. Conduction suppresses this instability (at least along field lines) for parallel wavelengths smaller than the Field length,
\begin{eqnarray}
\lambda_{\rm F} & \equiv & 2 \pi \left( \frac{2}{5} \frac{\kappa_{\|}}{\cool} \right)^{1/2} \\*
\mbox{} & \simeq & 300  \left( \frac{n_{\rm i}}{0.01~{\rm cm}^{-3}} \right)^{-1} \left( \frac{\kb  T}{2~{\rm keV}} \right)^{3/2}~ {\rm kpc} . \nonumber
\end{eqnarray}
We employ radiative cooling in two of our HBI simulations.

\section{Numerical approach}\label{sec:numerics}

\subsection{Integration Scheme}

We integrate equations (\ref{eqn:continuity})--(\ref{eqn:induction}) using the conservative MHD code Athena \citep{sgths08}. Details concerning the MHD algorithms may be found in \citet{gs05,gs08}. The directionally unsplit corner transport upwind (CTU) integration method and the Roe Riemann solver are used in all of our simulations. Following \citet{sh07} and \citet{ds09}, respectively, anisotropic conduction and Braginskii viscosity are implemented via operator splitting using slope limiters on the transverse heat and viscous fluxes to ensure stability. The conduction algorithm is sub-cycled with respect to the main integrator with a time step $\Delta t_{\rm cond} = C ( \Delta x )^2 / [ 2d ( \gamma -1 ) \kappa_{\|} ]$, where $C \approx 0.5$ is the Courant number and $d$ is the number of spatial dimensions being solved. In order to prevent impulsive driving due to abrupt changes in pressure, we restrict the sub-cycling routine to take no more than 10 sub-cycles per global timestep. 

When radiative cooling is included in our study of the HBI (see Section \ref{sec:radhbi}), we employ the exact integration scheme detailed in \citet{townsend09}. The cooling source term is added to the reconstruction and interface-state correction steps in the CTU integrator, so that the cooling is fully second-order accurate. In order to prevent the formation of an unresolved cold phase, we follow \citet{spq10} and \citet{msqp12} in adopting a temperature floor $T_{\rm floor} = T_0 / 20$, where $T_0$ is the initial minimum temperature in our model atmosphere. This temperature floor is based on the reasonable assumption that, once a thermally unstable fluid element cools below $T_{\rm floor}$, it is unlikely to enter back into the hot phase. Because our focus is on the evolution of the HBI, subject to radiative cooling and Braginskii viscosity, and not on the detailed nature of multiphase gas in a thermally unstable ICM, this simplification should not significantly affect our results.

\subsection{Microscale Instabilities}\label{sec:microscale}

When the pressure anisotropy violates the inequalities
\begin{equation}\label{eqn:pabounds}
- \frac{B^2}{4\pi} \lesssim p_\perp - p_{\|} \lesssim \frac{B^2}{8\pi} ,
\end{equation}
rapidly growing microscale instabilities (firehose and mirror, respectively) are triggered and the Braginskii-MHD equations become ill-posed \citep[see][and references therein]{sckhs05}. Without finite Larmor radius effects taken into account, the fastest growing microscale modes formally occur at infinitely small scales, which in practice translates to scales near the grid where the microscale instabilities may be unresolved. Exactly what to do in this situation is not obvious and is currently under investigation (A. Schekochihin, private communication). In the mean time, because the pressure anisotropy controls the rate of viscous dissipation -- which in turn affects the large-scale dynamics -- some measures must be taken in order to capture the microscale influence on the pressure anisotropy, particularly in the weak-field regime. In this paper we choose two approaches, both of which are supported by strong evidence in the solar wind and magnetosheath (plasmas in many ways similar to the ICM) that microinstabilities isotropize the plasma to marginally-stable levels \citep[e.g.][]{sk07,bkhqss09}.

Our first approach is based upon on the theory that, once triggered, microscale fluctuations grow in such a way as to compensate on average the ``excess'' pressure anisotropy generated by the large-scale motions, thus maintaining marginal stability \citep{sckrh08,rsrc11}. We simply allow the microscale instabilities to self-consistently develop over the course of our simulations and to naturally regulate the pressure anisotropy, with the expectation that numerical viscosity will prevent the relatively small structures from getting out of hand. Our simulations are deliberately chosen with high enough resolution to not only resolve the HBI and MTI, but also to ensure healthy time- and lengthscale separations between the HBI/MTI and the firehose/mirror instabilities. As a result, the firehose fluctuations we resolve in our simulations grow fast enough to rapidly enforce marginal stability and self-consistently provide a hard-wall limiter on negative pressure anisotropies. Unfortunately, the same cannot be said for the mirror instability. The Braginskii version of the mirror instability grows substantially slower than the kinetic mirror instability, having a growth rate smaller than the parallel rate-of-strain of the viscous scale eddies. While negative pressure anisotropies are efficiently regulated, positive pressure anisotropies may not be.

It is important to note that, because we are not able to simultaneously resolve the ion Larmor radius and thermal-pressure scale height, which requires $\gtrsim$$13$ orders of magnitude in scale separation, the microscale instabilities triggered throughout the course of our simulations do not grow as fast as they would otherwise grow in nature. In our Braginskii-MHD simulations the maximum growth rate of the firehose instability, $k_{||{\rm ,max}} v_{\rm th} | \Delta + 2 / \beta |^{1/2} \sim N \dyn | \Delta + 2 / \beta |^{1/2}$, occurs at $k_{\|} H \sim N$, where $\Delta \equiv (p_\perp - p_{\|})/p$ is the fractional pressure anisotropy and $N$ is the number of grid zones per thermal-pressure scale height. By contrast, a kinetic calculation including FLR effects reveals that the parallel firehose actually has a maximum growth rate $\sim$$\Omega_{\rm g,i} | \Delta + 2/\beta |$ occurring at $k_{\|} r_{\rm g,i} \sim | \Delta + 2/\beta|^{1/2}$, spreading to larger scales as the pressure anisotropy approaches marginality. Even for our highest-resolution simulation ($N = 512$), we are underestimating the maximum growth rate of the firehose instability by a factor $\sim$$10^{11} | \Delta + 2 / \beta |^{1/2}$. This is one example of the fact that the nonlinear saturation of microscale fluctuations and the consequent regulation of pressure anisotropy occurs in our simulations on a timescale much longer than it would in nature, where microscale fluctuations grow to $\delta B / B \sim 1$ on a timescale comparable to the turnover time of the turbulent motions. A potentially serious consequence of not resolving the microscale instabilities at their natural scales is that we may be overestimating the conductivity and viscosity of the plasma by a factor $\sim$$\mfp / r_{\rm g,i} \sim 10^{10}$--$10^{11}$ \citep[see final paragraph of][]{sckrh08}. 

Our second approach is motivated by the work of \citet{shqs06}, who numerically investigated the nonlinear evolution of the collisionless magnetorotational instability and accounted for the effects of microscale instabilities by artificially limiting the pressure anisotropy to lie within the bounds given by Equation (\ref{eqn:pabounds}). The computational advantage of this approach is that microscale fluctuations are never triggered during the simulation. This closure rests on the following plasma-physical rationale \citep[e.g.][]{sc06}. Once these thresholds are crossed, microscale instabilities will produce a fluctuation ``foam'' off of which particles may pitch-angle scatter, break adiabatic invariance on the extremely short cyclotron timescale, and thereby isotropize the pressure (provided such fluctuations can penetrate down to the ion gyroscale). \citet{shqs06} modeled this process by a large effective collisionality, which was activated in regions where and when the microscale stability boundaries were sufficiently exceeded, its magnitude being proportional to the product of a large frequency $\nu_{\rm p} \gg 1/ \Delta t$ and the pressure anisotropy excess. This effectively raises the Reynolds number of the plasma and makes it more collisional.

Note that when and where this occurs the pressure anisotropy is no longer connected to the large-scale turbulent stretching of the magnetic field that gave rise to the pressure anisotropy in the first place, nor does this approach capture the effects of the microscale contribution to the total rate-of-strain of the plasma. Moreover, the heating associated with relaxation of the pressure anisotropy is not correctly captured, as the assumed rapid pitch-angle scattering and consequent pressure isotropization has no associated heating term in our energy equation. Indeed, deciding exactly what to do with this energy is not trivial \citep[see][]{sqhs07}.

In summary, our two approaches amount to two different interpretations of Equation (\ref{eqn:anisotropy}) in the presence of microscale instabilities, with antithetical implications for the viscous dissipation of macroscale motions. In the first approach, the collision frequency of the plasma remains constant while the microscale instabilities modify on the average the (parallel) rate-of-strain so as to offset the pressure anisotropy caused by the changing macroscale fields (i.e. $\overline{{\rm d} \ln B / {\rm d} t} \propto \nu_{\rm ii} / \beta$). At the macroscales, the plasma behaves as though it were more viscous. In the second approach, the microscale instabilities break adiabatic invariance, effectively increasing the collision frequency ($\nu_{\rm ii} \rightarrow \nu_{\rm p} \sim \omega_{\rm dyn} \beta$) and returning the pressure anisotropy to marginally stable values. At the macroscales, the plasma behaves as though it were less viscous. The important question of which of these interpretations is correct boils down to the (unanswered) question of whether or not such microscale fluctuations can reach the ion Larmor radius in a driven, initially Maxwellian system.

\subsection{Choice of Dimensionless Plasma Parameters}\label{sec:parameters}

Our choice of dimensionless parameters is motivated by considerations of both the physical conditions in actual galaxy clusters and the numerical constraints related to the above pressure-anisotropy concerns. Since short-wavelength modes with $k_{\|} H \gtrsim \beta^{1/2}$ are stabilized by magnetic tension (unless $\Delta + 2/\beta < 0$), one would ideally like to construct simulations with relatively large $\beta$ so that a healthy spectrum of HBI and MTI modes may grow unabated (at least in their linear phase). However, such $\beta$ not only are much larger than the observationally estimated ICM $\beta \sim 10^2$--$10^4$ \citep[for a review, see][]{ct02}, but also place steep constraints on how long the HBI and MTI can be evolved without our simulation results being plagued by the aforementioned microphysical uncertainties. One can estimate from Equations (\ref{eqn:anisotropy}) and (\ref{eqn:pabounds}) how large $\delta B_{\|} / B$ can grow in the linear phase before microscale instabilities are triggered:
\begin{equation}\label{eqn:bprl}
\frac{\delta B_{\|}}{B} \lesssim \frac{\nu_{\rm ii}}{\beta \sigma_{\rm HBI}} \sim \frac{1}{\beta \, {\rm Kn}} .
\end{equation}
In cluster cores ${\rm Kn} \sim 10^{-3}$--$10^{-2}$, increasing outwards, and a choice of very large $\beta$ implies that one cannot go far beyond the linear regime without running the risk that the microphysical closures we have employed significantly influence the subsequent large-scale dynamics. We therefore choose $\beta  \sim 10^4$--$10^5$ in our simulations, which puts them in a regime in which Braginskii viscosity is more important than magnetic tension and in which the instability can develop $\delta B_{\|} / B \sim$ a few percent before firehose and mirror instabilities are triggered. In the outer regions of galaxy clusters, ${\rm Kn} \sim 10^{-2}$--$10^{-1}$ and it is almost trivial to violate Equation (\ref{eqn:bprl}).

All this being said, we believe that our simulation results represent a step forward in our understanding of convective instability and thermal conduction in the ICM. In lieu of a full kinetic simulation that can resolve $\gtrsim$$13$ orders of magnitude in spatial and temporal scale or a sub-grid model that can correctly capture the complex interplay between the micro- and macroscales, this seems to be the best we can do at this stage.

\section{Non-Radiative HBI}\label{sec:hbi}

\begin{table*}
\centering \caption{\label{table:runs} Parameters Used in HBI and MTI Simulations}
\begin{tabular}{c c c c c c c c}
\hline
\hline
Run & $d$ & Box Size & Resolution & Field Configuration & ${\rm Kn}^{-1}_0$ & Notes  & Section \\
\hline
H2dBrag & 2 & $H_0$$\times$$2 H_0$ & 512$\times$1024 & vertical; $\beta_0 = 10^5$ & 1500 & Braginskii viscosity & \ref{sec:hbi2d} \\
H2dIsoP &  2 & $H_0$$\times$$2 H_0$ & 512$\times$1024 & vertical; $\beta_0 = 10^5$ & 1500 & isotropic pressure & \ref{sec:hbi2d} \\
H2dBLim &  2 & $H_0$$\times$$2 H_0$ & 512$\times$1024 & vertical; $\beta_0 = 10^5$ & 1500 & artificially limited Braginskii viscosity & \ref{sec:hbi2d} \\
H3dBrag &  3 & $H_0$$ \times$$H_0$$\times$$2 H_0$ & 128$\times$128$\times$256 & vertical; $\beta_0 = 10^5$ & 1500 & Braginskii viscosity & \ref{sec:hbi3d} \\
H2dBRad &  2 & $H_0$$\times$$2 H_0$ & 512$\times$1024 & vertical; $\beta_0 = 10^5$ & 1500 & Braginskii viscosity and thermal Bremsstrahlung & \ref{sec:radhbi} \\
H2dIRad &  2 & $H_0$$\times$$2 H_0$ & 512$\times$1024 & vertical; $\beta_0 = 10^5$ & 1500 & isotropic pressure and thermal Bremsstrahlung & \ref{sec:radhbi} \\
\hline
M2dBrag &  2 & $H_0$$\times$$2 H_0$ & 512$\times$1024 & horizontal; $\beta_0 = 10^5$ & 200 & Braginskii viscosity & \ref{sec:mti2d} \\
M2dIsoP &  2 & $H_0$$\times$$2 H_0$ & 512$\times$1024 & horizontal; $\beta_0 = 10^5$ & 200 & isotropic pressure & \ref{sec:mti2d} \\
M2dBLim &  2 & $H_0$$\times$$2 H_0$ & 512$\times$1024 & horizontal; $\beta_0 = 10^5$ & 200 & artificially limited Braginskii viscosity & \ref{sec:mti2d} \\
M3dBrag &  3 & $H_0$$ \times$$H_0$$\times$$2 H_0$ & 128$\times$128$\times$256 & horizontal; $\beta_0 = 10^5$ & 200 & Braginskii viscosity & \ref{sec:mti3d} \\
M3dIsoP &  3 & $H_0$$ \times$$H_0$$\times$$2 H_0$ & 128$\times$128$\times$256 & horizontal; $\beta_0 = 10^5$ & 200 & isotropic pressure & \ref{sec:mti3d} \\
\hline \\
\end{tabular}
\end{table*}

\subsection{Background Equilibrium and Initial Perturbations}

We consider a non-radiative, plane-parallel plasma stratified in both density and temperature in the presence of a uniform gravitational acceleration in the vertical direction, $\bb{g} = -g \ez$. The plasma is threaded by a uniform background magnetic field oriented along $\ez$ and is assumed initially Maxwellian so that $p_\perp = p_{\|} = p$ in the background state. Force balance then implies
\begin{equation}\label{eqn:dynamicalequilibrium}
\deriv{z}{p} = - g \rho .
\end{equation}
There is a heat flux in the background state given by
\begin{equation}\label{eqn:backgroundQ}
\bb{q} = - \chi_{\|} \deriv{z}{T} \ez .
\end{equation}
In order to preserve thermal equilibrium, the background heat flux must be divergence-free:
\begin{equation}\label{eqn:noheating}
\deriv{z}{} \left( \chi_{\|} \deriv{z}{T} \right) = 0 .
\end{equation}
Enforcing $T = T_0$ at $z=0$ and $T = T_Z$ at $z = Z$, Equation (\ref{eqn:noheating}) may be integrated to yield the temperature profile
\begin{equation}\label{eqn:hbitemperature}
T(z) = T_0 \left( 1 + \zeta \frac{z}{Z} \right)^{2/7} ,
\end{equation}
where $\zeta \equiv ( T_Z / T_0 )^{7/2} - 1$ measures the magnitude of the steady heat flux through the atmosphere ($q \propto \zeta$). Combining this result with Equation (\ref{eqn:dynamicalequilibrium}) determines the pressure profile
\begin{equation}\label{eqn:hbipressure}
p(z) = p_0 \exp \left\{ - \frac{7}{5} \frac{Z}{\zeta H_0} \left[ \left( 1 + \zeta \frac{z}{Z} \right)^{5/7} - 1 \right] \right\} ,
\end{equation}
where $H_0 = v^2_{\rm th,0} / g$ is the thermal-pressure scale height and $p_0$ is the thermal pressure, both evaluated at $z = 0$. Note that ${\rm d}\ln T/ {\rm d}z$ is largest at $z=0$, and so HBI modes will naturally grow fastest at small $z$.

This equilibrium is characterized by two dimensionless free parameters: $\zeta$, which is $\sim$$10$--$50$ in cool-core clusters, and 
\begin{eqnarray}
\mc{G} & \equiv &\frac{Z}{H_0} \\*
&\simeq & 2.0 \left( \frac{Z}{250~{\rm kpc}} \right) \left( \frac{g}{10^{-8} ~{\rm cm~s}^{-2}} \right) \left( \frac{\kb  T_0}{2~{\rm keV}} \right)^{-1} , \nonumber
\end{eqnarray}
which is a measure of the height of the atmosphere. We choose $\mc{G} = 2.0$ and $T_Z / T_0 = 2.5$ (i.e. $\zeta \simeq 23.7$), which implies $\rho_Z / \rho_0 \simeq 0.14$ by Equation (\ref{eqn:hbipressure}). These numbers are characteristic of the cool-core cluster A1795 with $\kb  T_0 \approx 2.5~{\rm keV}$, $\kb  T_Z \approx 6.3~{\rm keV}$, and $Z \approx 250~{\rm kpc}$ \citep{efaj02}.

We apply Gaussian-random velocity perturbations to our background equilibrium, having a flat spatial power spectrum and a standard deviation of $10^{-4}v_{\rm th,0}$. Such perturbations are sufficiently subsonic to ensure linear evolution from the outset. These initial conditions are not  representative of those conditions found in actual clusters, in which galaxy motions, major and/or minor mergers, and feedback from active galactic nuclei (AGN) stir the plasma.

\subsection{Numerical Setup and Boundary Conditions}\label{sec:hbisetup}

The equations are put in dimensionless form by choosing units natural to the problem. The units of velocity $[v]$, density $[\rho]$, and magnetic-field strength $[B]$ are, respectively, the initial values of the thermal speed $v_{\rm th,0}$, density $\rho_0$, and magnetic field $B_0$ at the bottom of the box ($z=0$). The unit of length $[\ell]$ is $H_0$, so that the implied unit of time $[t]$ is $H_0 / v_{\rm th,0}$, the initial sound-crossing time across a thermal-pressure scale height. Since pressure balance applies in the equilibrium state, $g=1$ in these units. The units of length and time have the scalings
\begin{equation}\label{eqn:lengthunit}
[ \ell ] \simeq 124 \left( \frac{g}{10^{-8}~{\rm cm}~{\rm s}^{-2}} \right)^{-1} \left( \frac{\kb  T_0}{2~{\rm keV}} \right) ~{\rm kpc}
\end{equation}
and
\begin{equation}\label{eqn:timeunit}
[t] \simeq 196 \left( \frac{g}{10^{-8}~{\rm cm}~{\rm s}^{-2}} \right)^{-1} \left( \frac{\kb  T_0 }{2~{\rm keV}} \right)^{1/2} ~{\rm Myr} ,
\end{equation}
respectively. Aside from the free parameters $\zeta$ and $\mc{G}$ associated with the background equilibrium, the linear evolution of the HBI depends only upon the plasma beta and the Knudsen number. We choose $\beta_0 = 10^5$ and ${\rm Kn}^{-1}_0 = 1500$, so that $B_0 \simeq 0.016$, $\nu_{\|} \simeq 3.2 \times 10^{-4} ~ T^{5/2} \rho^{-1}$, and $\kappa_{\|} \simeq 1.6 \times 10^{-2} ~ T^{5/2} \rho^{-1}$ in dimensionless units. These imply initial values of $\nu_{\|} \simeq 0.023$ and $\kappa_{\|} \simeq 1.1$ at $z=Z$; the latter causes our simulations to be rather expensive.

Linear analysis of the HBI with Braginskii viscosity has shown that the only modes to evade strong suppression are confined to a thin band in wavenumber space in which conduction is fast but viscous damping is small: $\cond \gtrsim \dyn \gtrsim \visc$, or, using the definitions (\ref{eqn:omegavisc}) and (\ref{eqn:omegacond}), $k_{||} H \lesssim \sqrt{{\rm Kn}} \lesssim 3 k_{||} H$ (K11). Taking the initial temperature and pressure profiles, Equations (\ref{eqn:hbitemperature}) and (\ref{eqn:hbipressure}) respectively, and using equations (49) and (50) from K11, we expect the fastest growth at small $z$ on parallel wavelengths $\lambda_{||} \approx 0.2H_0$. However, the lower collisionality at larger $z$ shifts the fastest-growing modes to wavelengths comparable to the thermal-pressure scale height \citep{lk12}. In order to capture this global behavior, we choose box sizes $H_0$$\times$$2 H_0$ (in 2d) and $H_0$$\times$$H_0$$\times$$2 H_0$ (in 3d), so that $L_z = Z = 2 H_0$. We have also run local simulations with the size of the simulation domain much smaller than the thermal-pressure scale height similar to those presented in \S4.1 of  \citet{mpsq11} and found that only extremely slow-growing HBI modes fit into the box, in agreement with linear theory. Our 2d runs have a resolution of $512$$\times$$1024$. Our 3d run necessarily has lower resolution ($128$$\times$$128$$\times$$256$) due to the stiff numerical constraints imposed by heat and momentum diffusion in three dimensions (a single 3d run at this resolution requires $\sim$40,000~CPU-hrs).

The boundary conditions are the same as in \citet{mpsq11}. The temperatures at the upper and lower boundaries of our computational domain are fixed for all times, $T(z=0,t) = T_0$ and $T(z=Z,t) = T_Z$, while the pressure is extrapolated into the upper and lower ghost zones in such a way as to ensure hydrostatic equilibrium at those boundaries. These choices are motivated by the observation that many galaxy clusters in the local universe are observed to have non-negligible temperature gradients \citep[e.g.][]{pjkt05,vmmjfs05}. The magnetic field is constrained to cross the upper and lower boundaries normally, although its strength there is allowed to adjust according to the local dynamics. Periodic boundary conditions are imposed in the horizontal direction(s).

Here we present results from four non-radiative HBI simulations. H2dIsoP is a 2d simulation with isotropic pressure and serves as a reference run, enabling us to draw conclusions about the effects of Braginskii viscosity. H2dBrag is a 2d simulation with Braginskii viscosity and H2dBLim is a 2d simulation in which the pressure anisotropy is artificially limited using the Sharma et al.~closure described in Section \ref{sec:microscale}. H3dBrag is a 3d simulation with Braginskii viscosity. The parameters in these simulations are summarized in Table \ref{table:runs}.

\subsection{2d Simulations}\label{sec:hbi2d}

\begin{figure}
\centering
\includegraphics[width=3in]{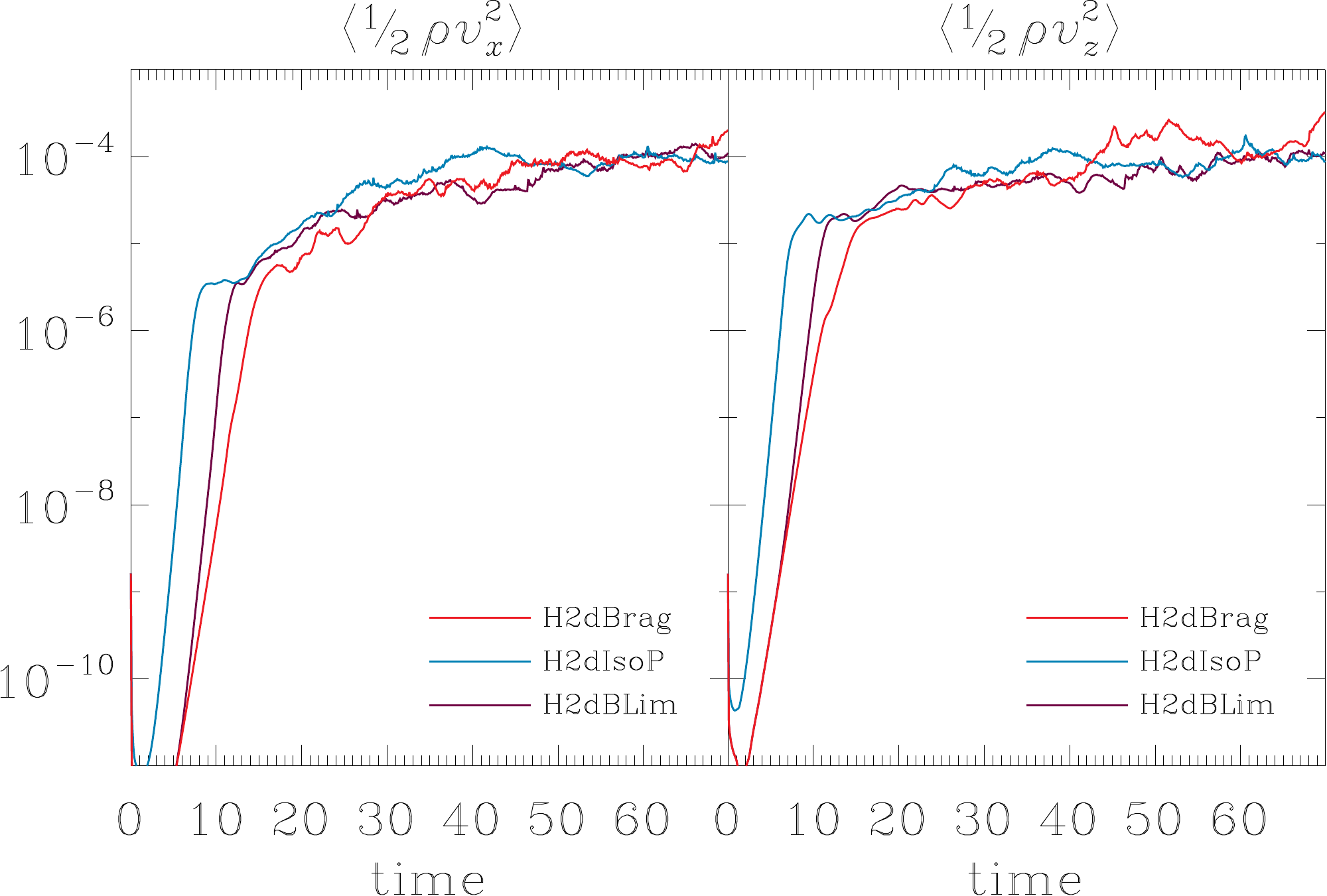}
\newline\newline 
\includegraphics[width=3in]{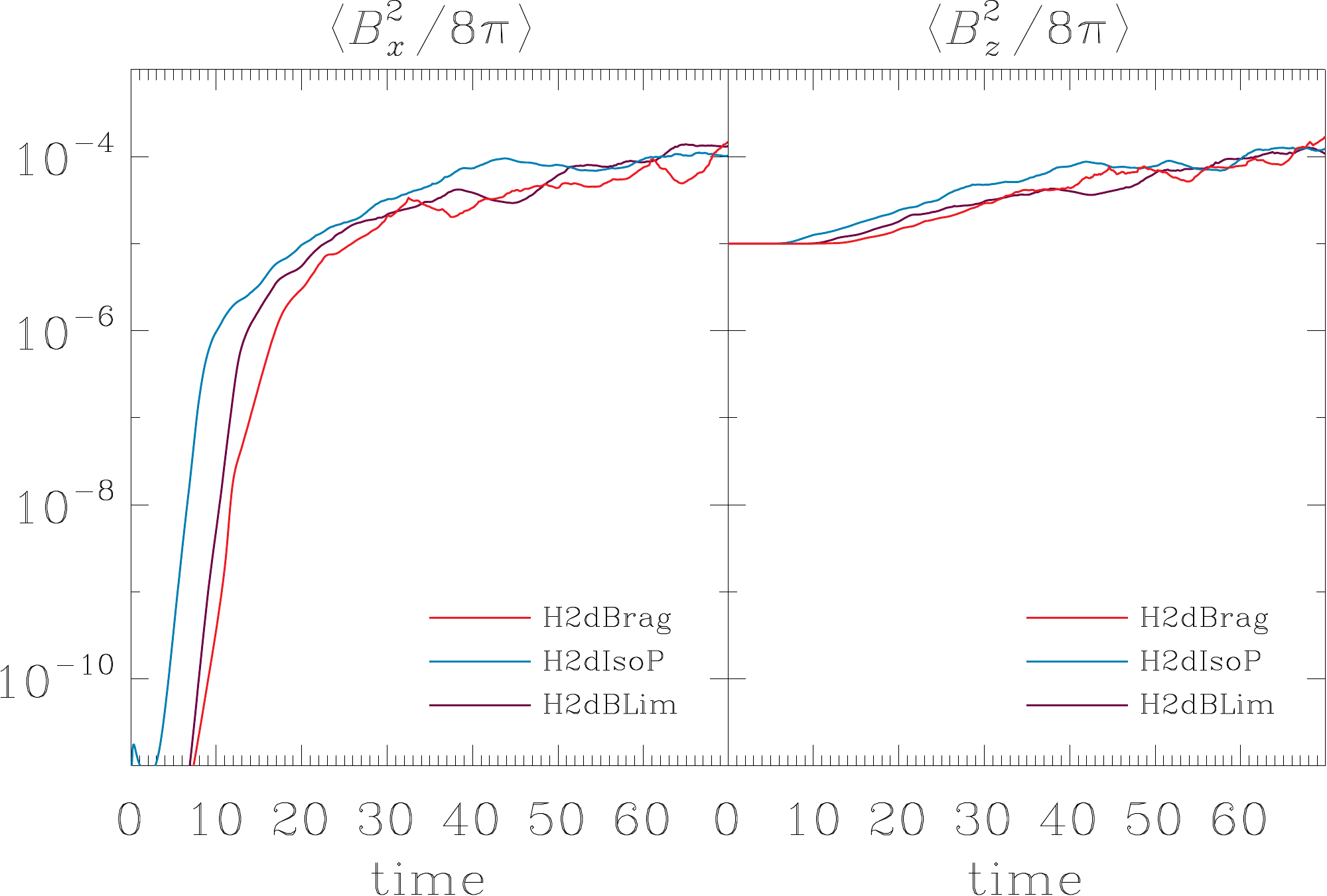}
\newline
\caption{Temporal evolution of the box-averaged horizontal ($x$) and vertical ($z$) kinetic and magnetic energy densities in runs H2dBrag (red lines), H2dIsoP (blue lines), and H2dBLim (purple lines). The units of energy density and time are, respectively, $\rho_0 v^2_{\rm th,0}$ and $H_0 / v_{\rm th,0}$ (see Equation \ref{eqn:timeunit}).}
\label{fig:2dHBI:energy}
\end{figure}
\begin{figure*}
\centering
\includegraphics[height=6.5in,angle=90]{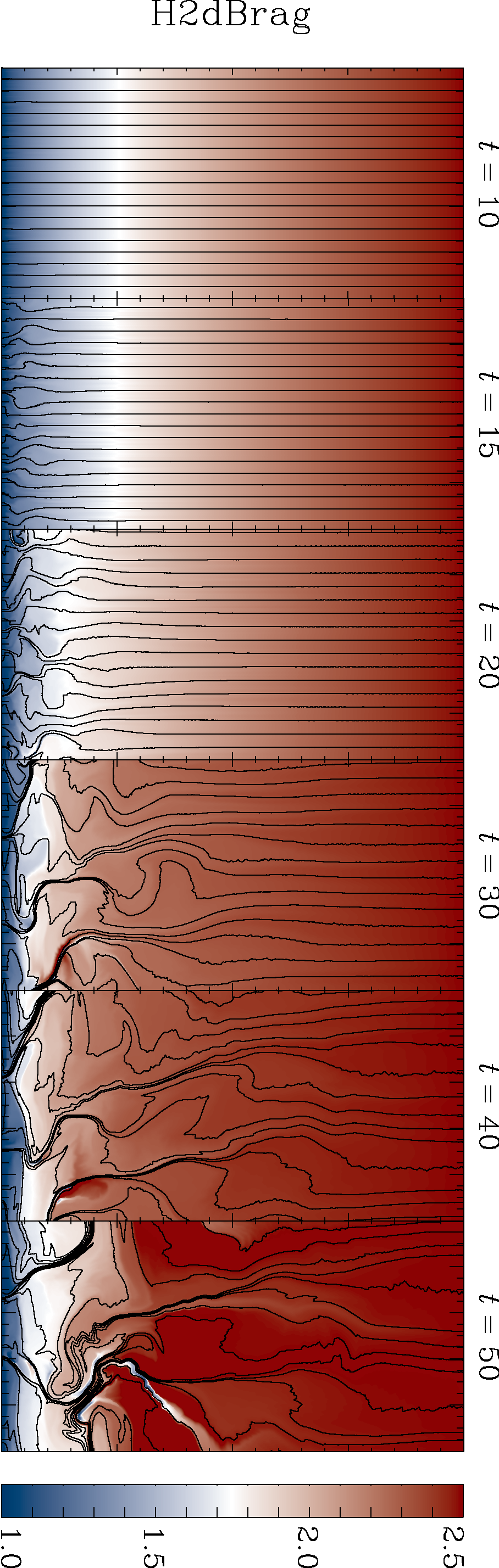}
\newline\newline 
\includegraphics[height=6.5in,angle=90]{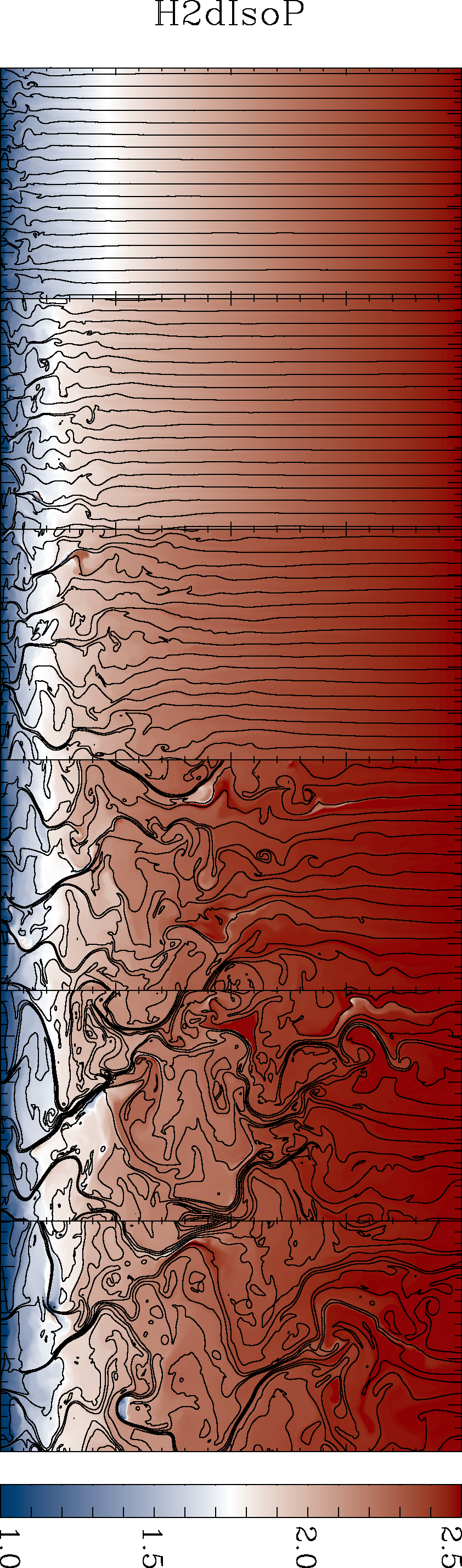}
\newline\newline 
\includegraphics[height=6.5in,angle=90]{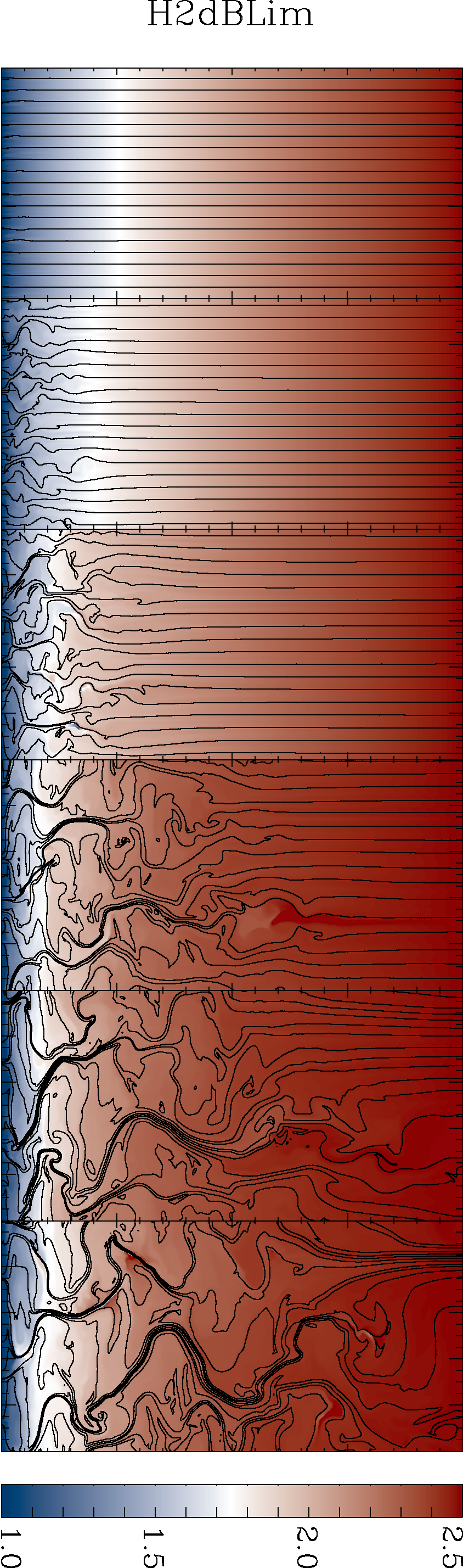}
\newline 
\caption{Spatial and temporal evolution of the HBI with Braginskii viscosity (top row), without Braginskii viscosity (middle row), and with limited Braginskii viscosity (bottom row) in two dimensions. The temperature (color) and magnetic-field lines (black lines) are shown at times $t = 10$, $15$, $20$, $30$, $40$, and $50$ (in units of $H_0 / v_{\rm th,0}$; see Equation \ref{eqn:timeunit}). The computational domain has dimensions $L_x$$\times$$L_z = 1$$\times$$2$ (in units of $H_0$; see Equation \ref{eqn:lengthunit}). We have suppressed the imaging of temperatures beyond the fixed color-bar limits.}
\label{fig:2dHBI:global}
\end{figure*}

In Figure \ref{fig:2dHBI:energy} we present the evolution of the box-averaged kinetic and magnetic energy densities. Runs in which pressure anisotropies are allowed to develop (red and purple lines) initially exhibit a growth rate $\approx$$0.6$ times smaller than that of the run without Braginskii viscosity (blue line), in agreement with predictions from linear theory (K11). A comparison between the growth rates of individual Fourier modes in run H2dBrag and those predicted by a quasi-global linear stability analysis of our model atmosphere shows excellent agreement all the way to $k_x H \sim 3000$ \citep[see fig. 5 of][]{lk12}. The growth rate in run H2dBLim (purple line) departs from that of runH2dBrag (red line) once the hard-wall pressure-anisotropy limiters become active and regulate the pressure anisotropy. The kinetic energy thereafter grows similarly to the run without Braginskii viscosity. All three runs reach an approximately saturated kinetic energy density corresponding to a box-averaged Mach number of a few percent (although local velocities can range up to $\approx$$50\%$ of the local thermal speed). The total magnetic energy increases by a factor of $\sim$$10$ over the course of the runs, similar to the increase in the total kinetic energy during the non-exponential phase of evolution. Horizontal and vertical energies reach approximate equipartition by the end of the simulation.

While all three runs appear similar, box-averaged quantities can be deceiving. The overall spatial and temporal evolution of the atmosphere during each of our 2d runs is shown in Figure \ref{fig:2dHBI:global}. The temperature (color) and magnetic-field lines (black lines) are displayed in each of the six frames, which show the atmosphere at the different times $t = 10$, $15$, $20$, $30$, $40$, and $50$ (in units of $H_0 / v_{\rm th,0}$; see Equation \ref{eqn:timeunit}).

In each of the runs, the HBI develops first at low altitude where the temperature gradient is largest, eventually progressing to higher altitudes where the temperature gradient is shallower. Runs with Braginskii viscosity (top and bottom rows) show a significant delay in the development of the HBI, particularly at higher altitudes. There are three reasons for this difference. First, early viscous damping of the seed velocity perturbations causes the instability to grow from smaller amplitudes than in the run with isotropic pressure. Second, viscous damping of motions that compress and rarefy the magnetic field lines results in a reduced HBI growth rate. Third, the increasing importance of viscous damping with height (recall $\nu_{\|} \propto T^{5/2} \rho^{-1}$) shifts the HBI to successfully larger parallel wavelengths, which eventually become comparable to the local scale height and lead to secular (rather than purely exponential) growth due to non-local effects. As a result, there is substantial difference in the saturated-state temperature profile and the topology of the magnetic-field lines for $z\gtrsim 0.1 L_z$.

In run H2dBrag, regions of negative pressure anisotropy (i.e. decreasing magnetic-field strength) produce firehose instabilities near the grid scale when the local value of $(p_\perp - p_{\|} + B^2 / 4 \pi ) < 0$. The firehose fluctuations grow exponentially until they compensate for the excess pressure anisotropy, after which they grow secularly. Once the local pressure anisotropy is regulated to be $\simeq$$-B^2/4\pi$, the firehose turbulence moves to longer wavelengths. This can be seen clearly by comparing the structure of the firehose modes near the top of the $t=50$ panel with those of the $t=20$ panel. In run H2dBLim, such firehose fluctuations do not exist by construction: the Braginskii pressure anisotropy is limited by hand to lie within the microscale stability boundaries. Rather than produce firehose instabilities, regions of large negative pressure anisotropy effectively eliminate the magnetic tension. As a result, sharp folds in magnetic fields are allowed to develop in regions of decreasing magnetic-field strength. Note that there is much less numerical reconnection in runs H2dBrag and H2dBLim than in run H2dIsoP, since small-scale motions that change the magnetic-field strength are viscously damped.

\begin{figure}
\centering
\includegraphics[height=3in,angle=90]{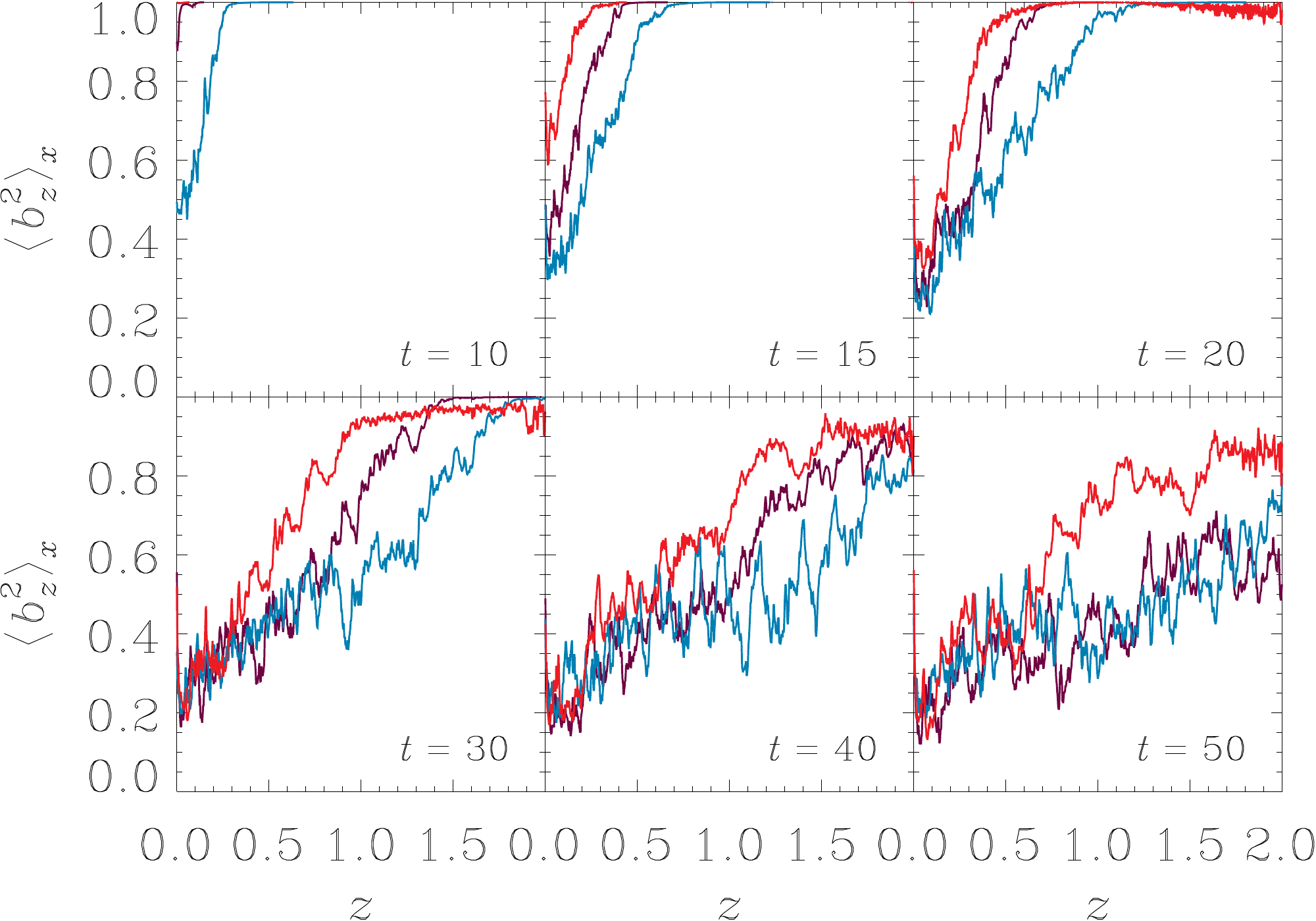}
\caption{Spatial and temporal evolution of the horizontally averaged magnetic-field angle at each height in runs H2dBrag (red lines), H2dIsoP (blue lines), and H2dBLim (purple lines). The units of length and time are, respectively, $H_0$ (see Equation \ref{eqn:lengthunit}) and $H_0 / v_{\rm th,0}$ (see Equation \ref{eqn:timeunit}).}
\label{fig:2dHBI:angle}
\end{figure}

The horizontally averaged magnetic-field angle as a function of height, calculated as $\langle b^2_z \rangle_x$, is shown in Figure \ref{fig:2dHBI:angle} at the same times as in Figure \ref{fig:2dHBI:global}. In all three cases (H2dBrag: red line; H2dIsoP: blue line; H2dBLim: purple line), the HBI grows by reorienting the magnetic field to be more and more horizontal and $\langle b^2_z \rangle_x$ generally decreases in time at all heights. However, there are significant differences between each of the runs. As was evidenced in Figure \ref{fig:2dHBI:global}, Braginskii viscosity impedes appreciable field-line reorientation for $z\gtrsim 0.1 L_z$, where the viscous and dynamical frequencies become comparable. At these heights, the large (parallel) wavenumbers required to keep the HBI in action as the magnetic field becomes more and more horizontal are strongly suppressed by the pressure anisotropy they generate. For example, at $t=30$ ($\simeq$$6.6~{\rm Gyr}$ for $g = 10^{-8}~{\rm cm~s}^{-2}$ and $\kb  T_0 = 2.5~{\rm keV}$) the magnetic-field angle in the Braginskii and non-Braginskii runs are similar only in the innermost $\approx$$20\%$ of the core. Beyond this height, the relatively straight field lines in run H2dBrag allow heat conduction to remain active at a rate comparable to the field-free Spitzer value. The magnetic-field angle in run H2dBLim, in which the pressure anisotropy was artificially limited so as to prevent microscale instabilities from growing, is intermediate between that of run H2dBrag and H2dIsoP. This is because the limiters restrict how effective Braginskii viscosity can be at suppressing the HBI. 

\begin{figure}
\centering
\includegraphics[width=2.1in]{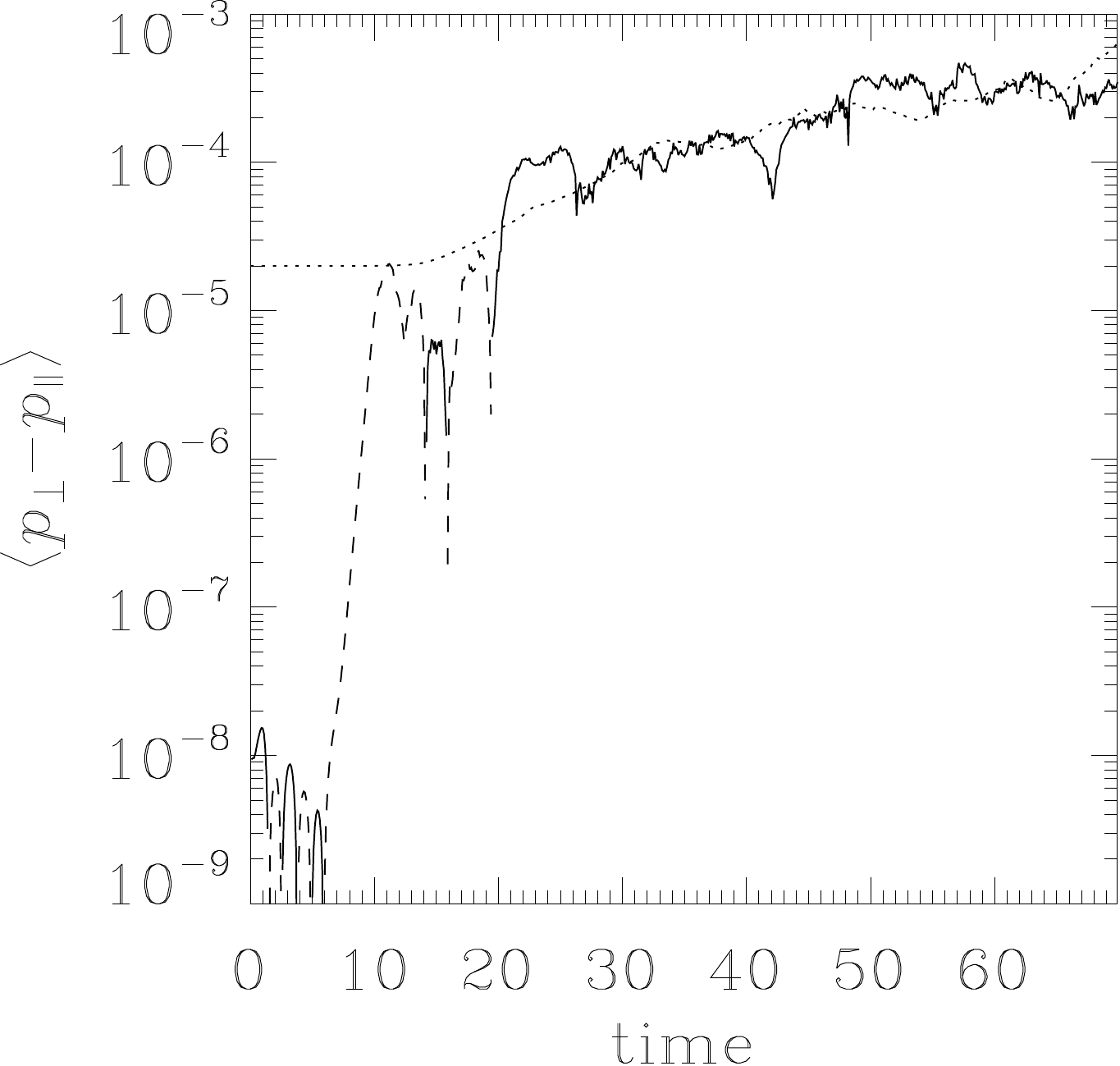}
\caption{Temporal evolution of the box-averaged pressure anisotropy in run H2dBrag. Positive (negative) pressure anisotropies are denoted by a black solid (dashed) line. The dotted line traces the box-averaged value of $B^2 / 4 \pi$, a quantitative measure of microscale stability (see Section \ref{sec:microscale}). The units of pressure and time are, respectively, $\rho_0 v^2_{\rm th,0}$ and $H_0 / v_{\rm th,0}$ (see Equation \ref{eqn:timeunit}).}
\label{fig:2dHBI:panis}
\end{figure}

In Figure \ref{fig:2dHBI:panis} we plot the box-averaged pressure anisotropy as a function of time in run H2dBrag. The solid (dashed) black line denotes an average positive (negative) pressure anisotropy, while the dotted line traces the box-averaged value of $B^2 / 4\pi$ (a quantitative measure of microscale stability; see Section \ref{sec:microscale}). Initially, the pressure anisotropy grows exponentially because $p_\perp - p_{\|} \propto \delta B_{\|} / B_0$. During this phase, there are more regions of decreasing field strength ($\delta B_{\|} < 0$) than increasing field strength ($\delta B_{\|} > 0$) and so the box-averaged pressure anisotropy is negative. This is because regions with $\delta B_{\|} < 0$ correspond to downward displacements, which predominate by taking advantage of the steeper temperature profile at smaller $z$ (note that ${\rm d}^2 T / {\rm d} z^2 < 0$). Once these regions of negative pressure anisotropy satisfy $p_\perp - p_{\|} < -B^2_0 / 4\pi$, rapidly growing firehose fluctuations efficiently reduce the pressure anisotropy to marginal stability. This is why the dashed line in Figure \ref{fig:2dHBI:panis} never appreciably crosses the dotted line. Once the HBI settles into its nonlinear phase, the box-averaged pressure anisotropy is positive since, in general, $p_\perp - p_{\|} \propto {\rm d}\ln B/{\rm d}t > 0$. Because the mirror instability is not accurately captured in our Braginskii-MHD simulations, it is not as efficient at regulating the pressure anisotropy as it would be in nature. Nevertheless, the box-averaged pressure anisotropy always stays within a factor of a few of $B^2/4\pi$, due to efficient firehose and not-so-efficient mirror regularization.

We note here that the evolution of the kinetic energy shown in our Figure \ref{fig:2dHBI:energy} is qualitatively different than that presented in figure 3 of \citet{mpsq11}. Those authors found that the energy in the vertical motion was in the form of stable oscillations that decay nonlinearly, whereas the horizontal kinetic energy persisted once the magnetic field became predominantly horizontal. While it is not surprising that our Braginskii-HBI simulations do not show this tendency, it is rather intriguing that our non-Braginskii-HBI simulations do not either. We attribute this to their choice of $\beta = 10^{12}$. At such small magnetic-field strengths, the magnetic field exerts essentially no dynamical feedback upon the gas motions, even as the magnetic field acquires a sharp folded structure. A more careful and dedicated study of this difference will be presented in Avara et al. (2012, in preparation).

\subsection{3d Simulation}\label{sec:hbi3d}

\begin{figure}[t]
\centering
\includegraphics[width=3in]{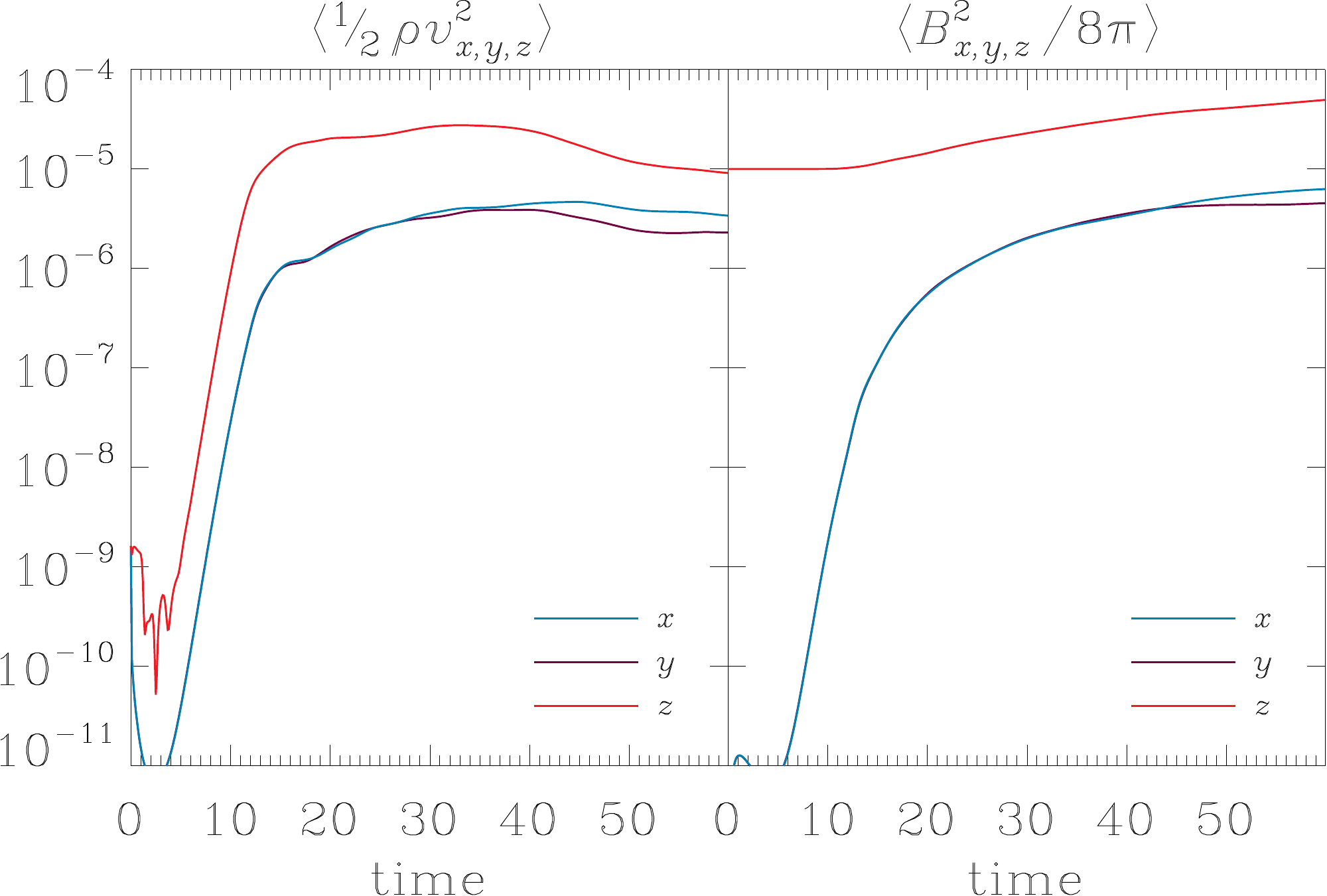}
\caption{Temporal evolution of the box-averaged kinetic and magnetic energies in run H3dBrag. The units of energy density and time are, respectively, $\rho_0 v^2_{\rm th,0}$ and $H_0 / v_{\rm th,0}$ (see Equation \ref{eqn:timeunit}).}
\label{fig:3dHBI:energy}
\end{figure}

Figure \ref{fig:3dHBI:energy} presents the evolution of the box-averaged kinetic and magnetic energy densities from run H3dBrag. There are a few differences between the behavior shown in this figure and that shown in its 2d analog (Figure \ref{fig:2dHBI:energy}). Some of these are genuine differences due to the increased dimensionality of the simulation, while others are attributable to the factor of 4 difference in resolution. To determine which differences are genuine and which are not, we have run a $128$$ \times$$256$ version of run H2dBrag that is similar to run H3dBrag in every way except dimensionality. Comparing these two runs with the original $512$$\times$$1024$ run H2dBrag, we are able to conclude that run H3dBrag is not fully converged (the energies in both the 2d and 3d runs are generally too small by a factor of a few). While this is a lesson in itself -- that properly simulating the HBI under actual cluster conditions requires more than $128$ zones per thermal-pressure scale height -- we believe there is much to be gained from nevertheless presenting our 3d results.

Of those differences that are clearly attributable to the increased dimensionality, one is a marked deficit of energy in the horizontal components of the velocity and magnetic field. This is because small-wavelength perturbations whose wavevectors have a component perpendicular to both gravity and the magnetic field behave like modified Alfv\'{e}n waves that are only slowly growing or decaying (depending on their exact wavevector orientation; see \S4.1.2 of K11). In other words, as the magnetic field becomes on the average more horizontal, the extra degree of freedom allows Braginskii viscosity to reorient these perturbations so as to minimize field-line compressions and rarefactions.

\begin{figure}
\centering
\includegraphics[height=3in,angle=90]{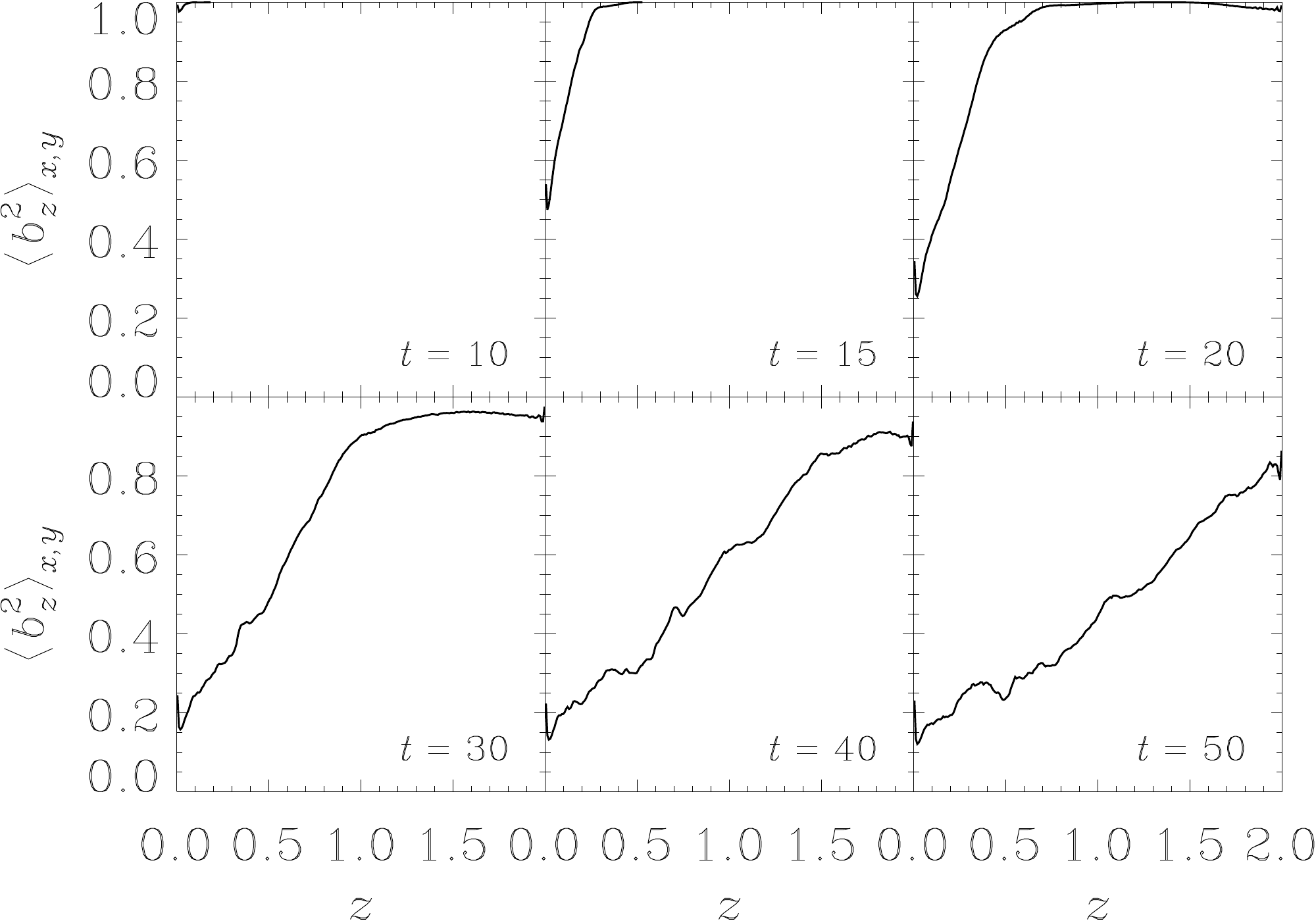}
\caption{Spatial and temporal evolution of the horizontally averaged magnetic-field angle at each height in run H3dBrag. The units of length and time are, respectively, $H_0$ (see Equation \ref{eqn:lengthunit}) and $H_0 / v_{\rm th,0}$ (see Equation \ref{eqn:timeunit}).}
\label{fig:3dHBI:angle}
\end{figure}

Another difference is in the horizontally averaged magnetic-field angle as a function of height and time, shown in Figure \ref{fig:3dHBI:angle}. While the curves for $t \le 40$ are very similar to those shown as red lines in Figure \ref{fig:2dHBI:angle} (run H2dBrag), there is one noticeable difference: the magnetic-field angle, $\langle b^2_z \rangle_{x,y}$, is smaller for $z \lesssim 0.1$ in the 3d run. We attribute this to interchange motions, which allow horizontally inclined magnetic field lines to slip past one another (a similar effect occurs in 3d simulations of the Rayleigh-Taylor instability; \citealt{sg07}). It is in this region where the field lines have been considerably reoriented, since the effect of Braginskii viscosity is greatly reduced there (recall $\nu_{\|} \propto T^{5/2} \rho^{-1}$). Elsewhere in the atmosphere, these interchange motions do not occur as readily, since the magnetic field does not become predominantly horizontal due to effective parallel-viscous suppression of short-wavelength HBI modes. While the $t=50$ panel in Figure \ref{fig:3dHBI:angle} appears different than the red line in the $t=50$ panel of Figure \ref{fig:2dHBI:angle}, with a relative decrease in $\langle b^2_z \rangle_{x,y}$ at all heights, we can safely attribute this to insufficient resolution: our $128$$\times$$256$ 2d test simulation shows a similar decrease.

\begin{figure}
\centering
\includegraphics[height=3.5in]{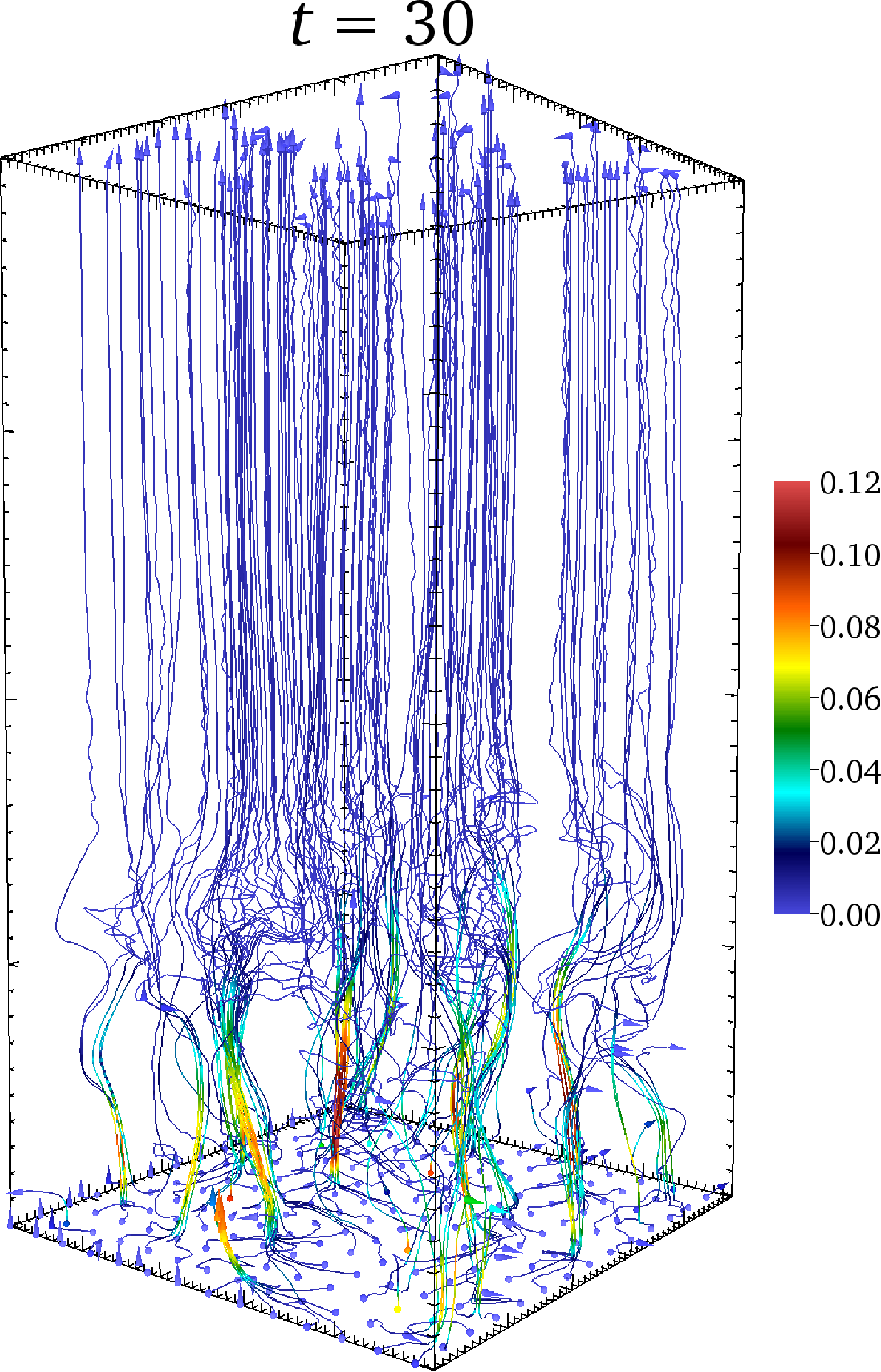}
\newline\newline
\includegraphics[height=3.5in]{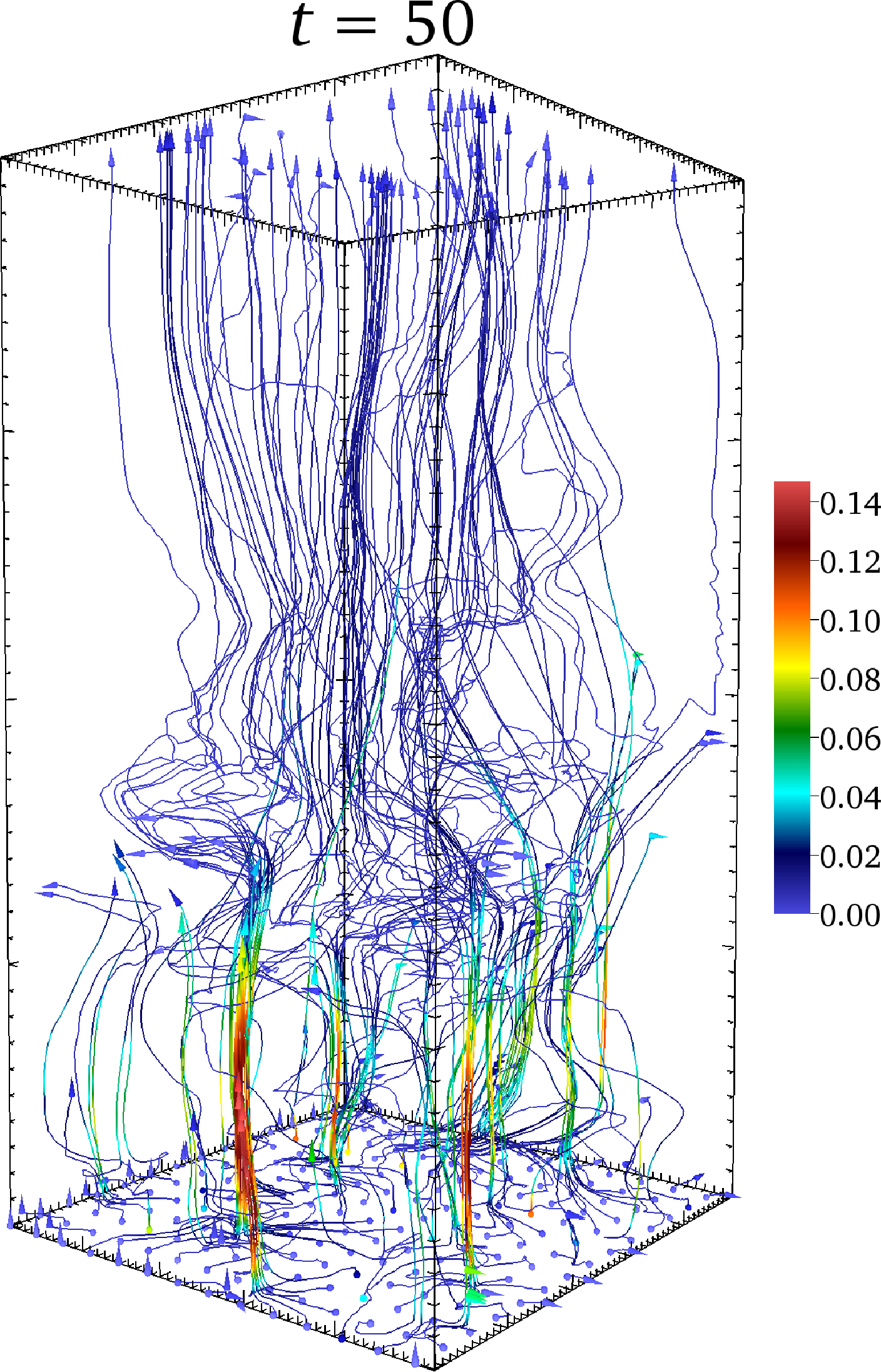}
\newline
\caption{Magnetic-field lines (color-coded according to the local field strength) from run H3dBrag at times $t = 30$ and $50$ (in units of $H_0 / v_{\rm th,0}$; see Equation \ref{eqn:timeunit}). The computational domain has dimensions $L_x$$\times$$L_y \times L_z = 1$$\times$$1$$\times 2$ (in units of $H_0$; see Equation \ref{eqn:lengthunit}). The magnetic-field strength is in units of $(4\pi p_0)^{1/2}$.}
\label{fig:3dHBI:global}
\end{figure}

These properties can be also be seen in Figure \ref{fig:3dHBI:global}, which shows the magnetic-field lines (color-coded according to the local field strength) at times $t=30$ and $50$. At $t=30$ the HBI has yet to significantly affect the magnetic field for $z \gtrsim H_0$. For $0.1 H_0 \lesssim z \lesssim 0.6 H_0$, the field lines have been strongly compressed in some regions and strongly rarefied in others. The flow of heat across these heights occurs along magnetic sheaths (or filaments), inside of which $\beta \sim 10^2$. The resulting tension in these strong-field regions is responsible for straightening these field lines out. For $z \lesssim 0.1 H_0$, the HBI has successfully reoriented the field lines to be predominantly horizontal. By $t=50$, the HBI is active to various extents throughout the entire atmosphere. The magnetic filaments have become longer and more prominent, while the field lines below $z \approx 0.1 H_0$ remain horizontally inclined. These magnetic bundles were also seen in our 2d simulations, but are more pronounced here in 3d. The reason is that it is more efficient to gather field lines together in 3d than in 2d.

\section{Radiative HBI}\label{sec:radhbi}

We have also run two 2d HBI simulations including radiative cooling, one with Braginskii viscosity (H2dBRad) and one with isotropic pressure (H2dIRad). The parameters used are summarized in Table \ref{table:runs}. 

Our numerical setup is the same as that detailed in Section \ref{sec:hbi}, aside from two important differences. First, when radiative cooling is considered, thermodynamic equilibrium requires
\begin{equation}\label{eqn:nocooling}
\deriv{z}{} \left( \chi_{\|} \deriv{z}{T} \right) = \rho \mc{L} \propto \rho^2 T^{1/2}.
\end{equation}
In contrast with Equation (\ref{eqn:noheating}), this equation cannot be integrated analytically. Instead, we employ a shooting method to simultaneously solve Equations (\ref{eqn:dynamicalequilibrium}), (\ref{eqn:backgroundQ}), and (\ref{eqn:nocooling}) subject to three boundary conditions: $T=T_0$ and $\rho = \rho_0$ at $z=0$, and $T = T_Z = T_0 ( 1 + \zeta )^{2/7}$ at $z=Z$. 

\begin{figure}
\centering
\includegraphics[width=3.2in]{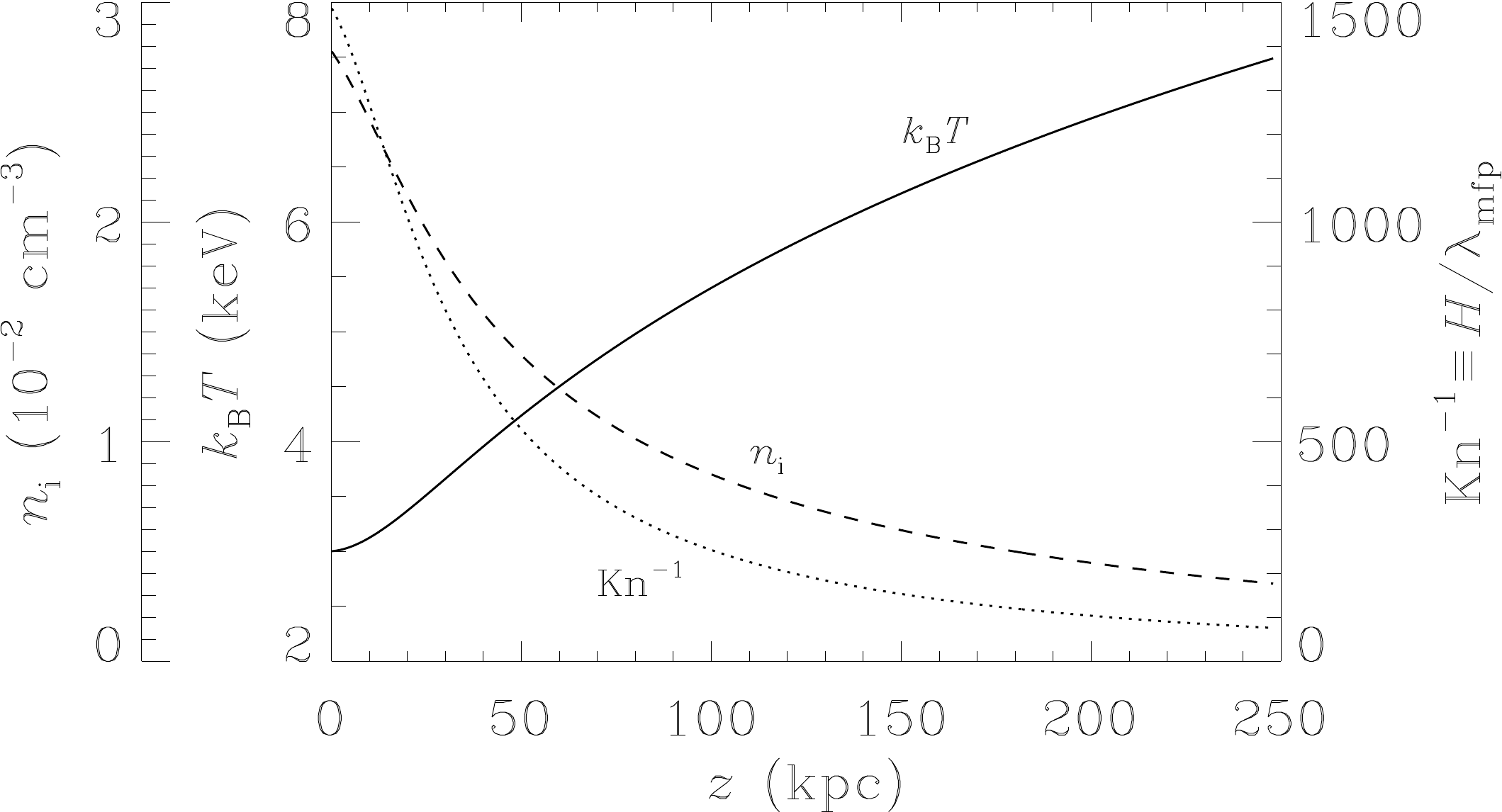}
\caption{Equilibrium atmosphere used as the initial condition in runs H2dBRad and H2dIRad. Profiles of the temperature (solid line), the ion density (dashed line), and the inverse of the Knudsen number (dotted line; see Equation \ref{eqn:knudsen}) are given in the dimensional units on the accompanying scales. These profiles are very similar to those observed in the cool-core cluster A85, especially for $r \gtrsim 50~{\rm kpc}$ \citep{cdvs09}.}
\label{fig:2dRHBI:profile}
\end{figure}

As in our non-radiative HBI simulations, we choose $\mc{G} = 2.0$, $T_Z / T_0 = 2.5$, $\beta_0 = 10^5$, and ${\rm Kn}^{-1}_0 = 1500$. Radiative cooling introduces another dimensionless free parameter,
\begin{equation}
\mc{C} \equiv \frac{ \rho \mc{L} / p }{ v_{\rm th} / H} \simeq 0.1  \left( \frac{g}{10^{-8}~{\rm cm~s}^{-2}} \right)^{-1} \left( \frac{n_{\rm i}}{0.01~{\rm cm}^{-3}} \right) .
\end{equation}
We adopt a value of $\mc{C}_0 = 0.18$, for which $\cool \simeq -0.11 \dyn$ at $t=z=0$. Note that a simultaneous choice of $\mc{G}$, $\zeta$, ${\rm Kn}_0$, $\beta_0$, and $\mc{C}_0$ implies specific dimensional values for our model cluster core: $n_{\rm i,0} \simeq 0.028~{\rm cm}^{-3}$, $\kb  T_0 \simeq 3.0~{\rm keV}$, $\kb  T_Z \simeq 7.5~{\rm keV}$, $g \simeq 1.5\times 10^{-8}~{\rm cm~s}^{-2}$, $B_0 \simeq 0.26~\mu{\rm G}$, and $Z \simeq 248~{\rm kpc}$.
\footnote{While this ``central'' density is a factor of $\sim$$2$--$3$ smaller than those found in actual cluster cores, one cannot construct a thermodynamic equilibrium between conduction and cooling with greater central densities (for our choices of $\zeta$, $\mc{G}$, and $\mc{C}_0$). This is an indication that even unbridled conduction cannot offset radiative losses in all clusters \citep[see, e.g.,][]{zn03}.} 
The corresponding units of length and time are $[\ell] \simeq 124~{\rm kpc}$ and $[t] \simeq 160~{\rm Myr}$, respectively; the former implies a grid size $\simeq$$0.24~{\rm kpc}$. The resulting thermodynamic equilibrium representing our initial conditions is shown in Figure \ref{fig:2dRHBI:profile} in dimensional units. The temperature (solid line), ion density (dashed line), and inverse Knudsen number (dotted line) are very similar to those observed in the cool-core cluster A85, especially for $r \gtrsim 50~{\rm kpc}$ \citep{cdvs09}.

Second, we must modify the temperature boundary condition at $z=0$ in order to prevent the development of sharp temperature gradients between the ghost zones and the few first active zones where the cooling rate is greatest. We choose reflective boundary conditions, which enforces a zero-gradient condition on the temperature at the bottom of the computational domain. While this no longer implies a fixed heat flux through the computational domain (as in our non-radiative HBI runs), it is more compatible with the physical conditions in actual cluster cores.

\begin{figure}
\centering
\includegraphics[width=3in]{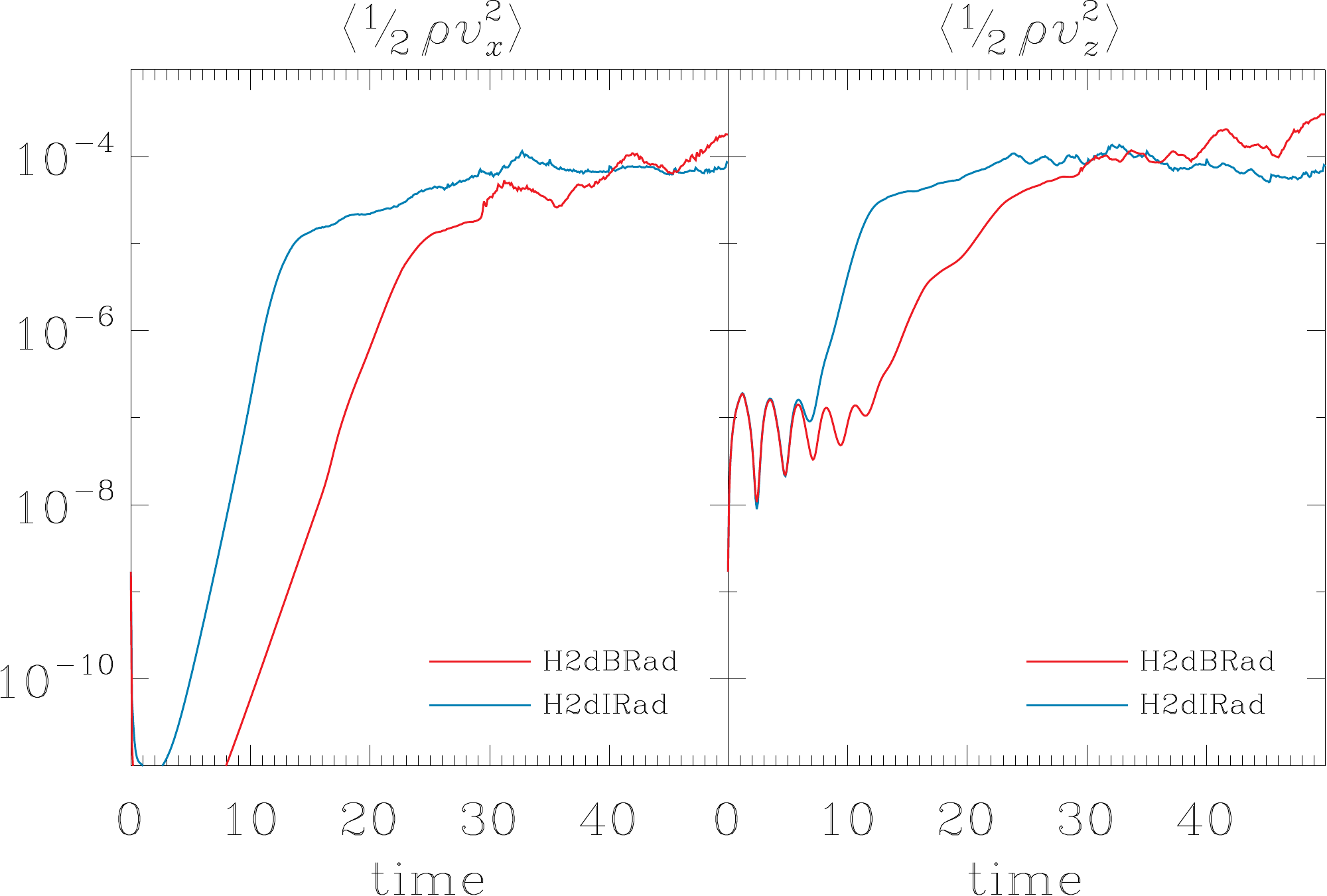}
\newline\newline 
\includegraphics[width=3in]{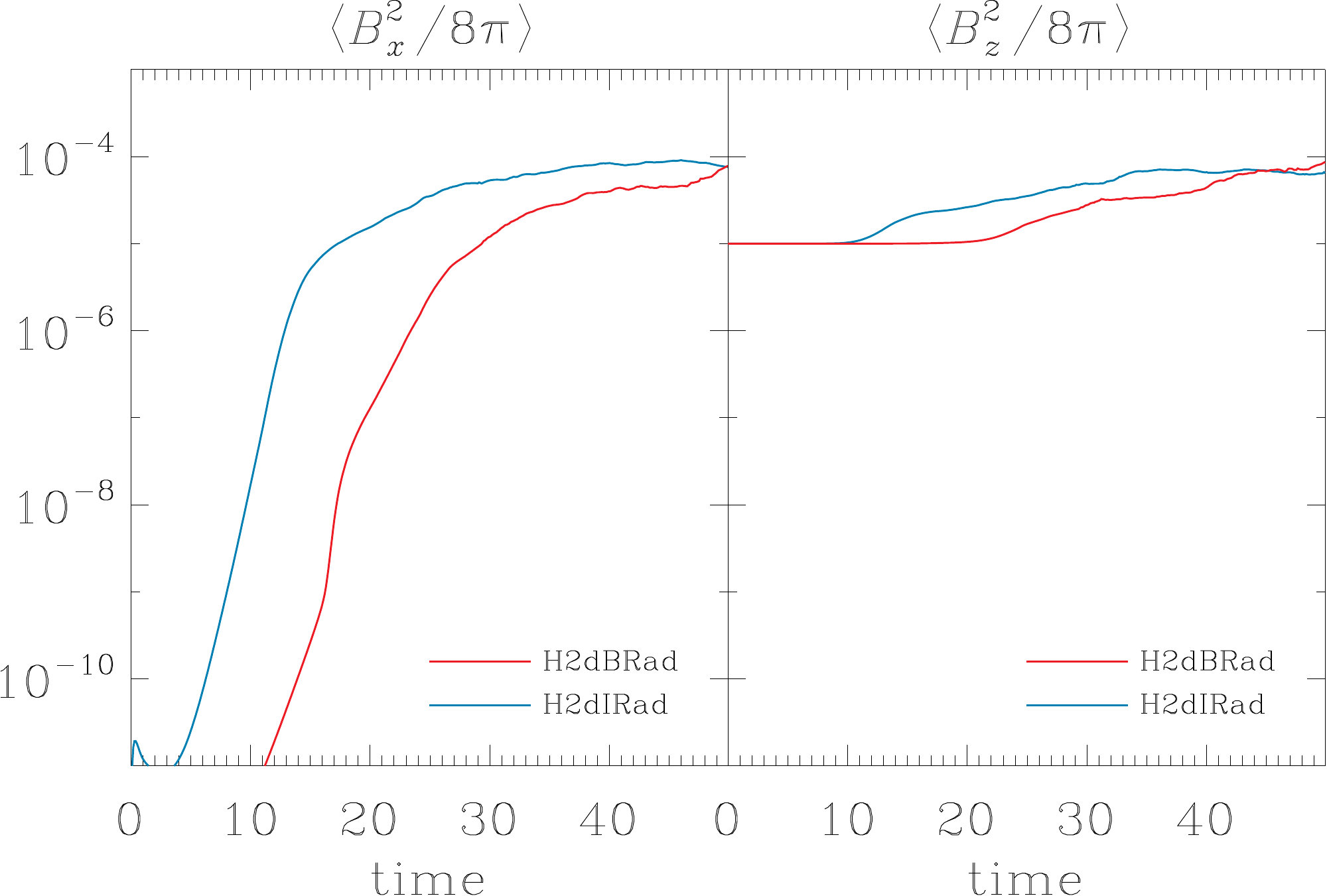}
\newline
\caption{Temporal evolution of the box-averaged horizontal ($x$) and vertical ($z$) kinetic and magnetic energy densities in runs H2dBRad (red lines) and H2dIRad (blue lines). The units of energy density and time are, respectively, $\rho_0 v^2_{\rm th,0} \simeq 2.7 \times 10^{-10}~{\rm ergs~cm}^{-3}$ and $H_0 / v_{\rm th,0} \simeq 160~{\rm Myr}$.}
\label{fig:2dRHBI:energy}
\end{figure}
\begin{figure*}
\centering
\includegraphics[height=6.5in,angle=90]{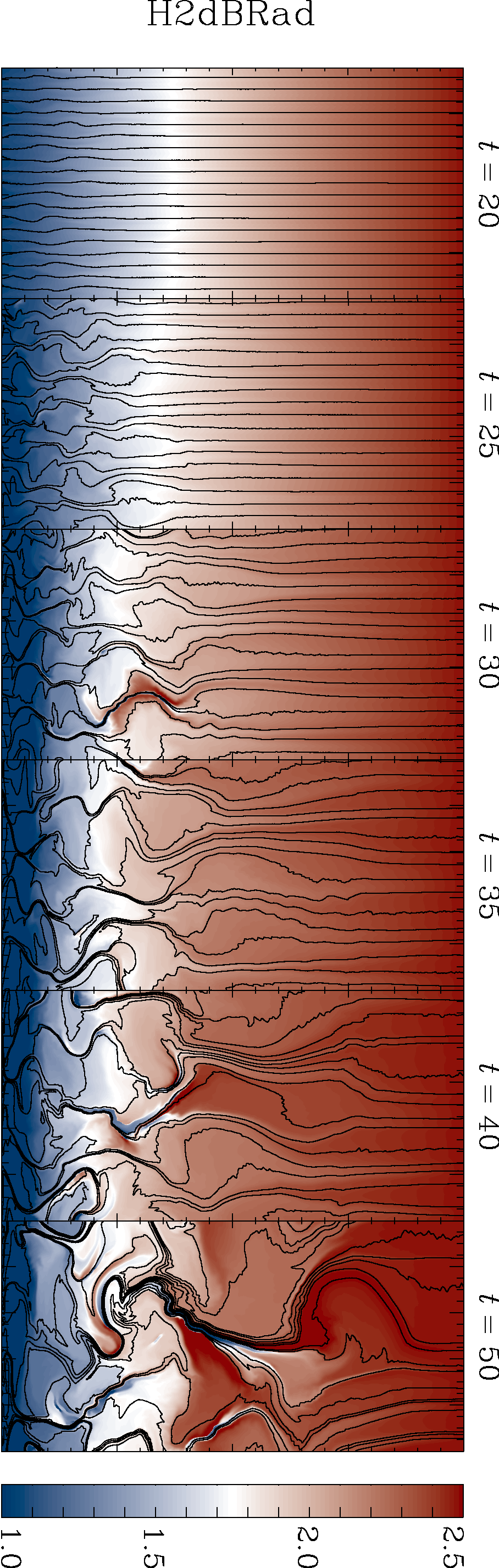}
\newline\newline 
\includegraphics[height=6.5in,angle=90]{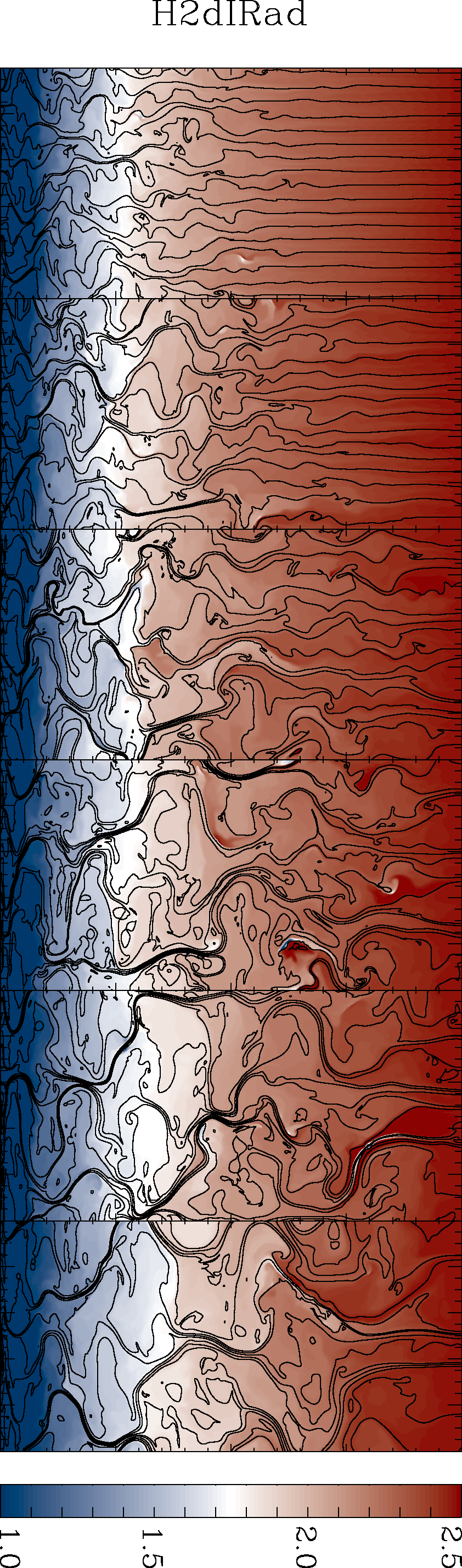}
\newline 
\caption{Spatial and temporal evolution of the radiative HBI with (top row) and without (bottom row) Braginskii viscosity. The temperature (color) and magnetic-field lines (black lines) are shown at times $t = 20$, $25$, $30$, $35$, $40$, and $50$ (in units of $H_0 / v_{\rm th,0} \simeq 160~{\rm Myr}$). The computational domain has dimensions $L_x$$ \times$$L_z = 1$$\times$$2$ (in units of $H_0 \simeq 124~{\rm kpc}$). We have suppressed the imaging of temperatures beyond the fixed color-bar limits.}
\label{fig:2dRHBI:global}
\end{figure*}

In Figure \ref{fig:2dRHBI:energy} we show the temporal evolution of the box-averaged kinetic and magnetic energy densities in runs H2dBRad (red lines) and H2dIRad (blue lines). In both runs, the instability growth rates are reduced from those in their respective non-radiative runs (see Figure \ref{fig:2dHBI:energy}). The change in growth rates is due to an interplay between two effects. First, the equilibrium atmosphere has a shallower temperature profile when cooling is taken into account. Because the HBI growth rate scales with $( {\rm d} \ln T / {\rm d} z )^{1/2}$, this naturally decreases growth rates. On the other hand, as shown in \citet{br08} by way of a local linear analysis of the radiative HBI, cooling acts to further destabilize the atmosphere (especially at longer wavelengths). Physically, this is because cooling weakens the ability of conduction to wipe away temperature fluctuations of a given wavelength along magnetic-field lines; such temperature fluctuations, a result of fluid elements' heat exchanges with the background conductive flux, are necessary for the HBI to function. The periodic oscillations seen in the early evolution of the vertical kinetic energy are due to small departures from and oscillations about thermodynamic equilibrium. As in our non-radiative HBI runs, Braginskii viscosity retards the development and growth of the HBI. The energy densities in runs H2dBRad and H2dIRad do not become comparable until $t \gtrsim  40$ ($\simeq$$6.4~{\rm Gyr}$). By the end of the simulations ($t \simeq 8~{\rm Gyr}$), the total magnetic energy in both runs has increased by a factor of $\sim$$10$.

\begin{figure}
\centering
\includegraphics[height=3in,angle=90]{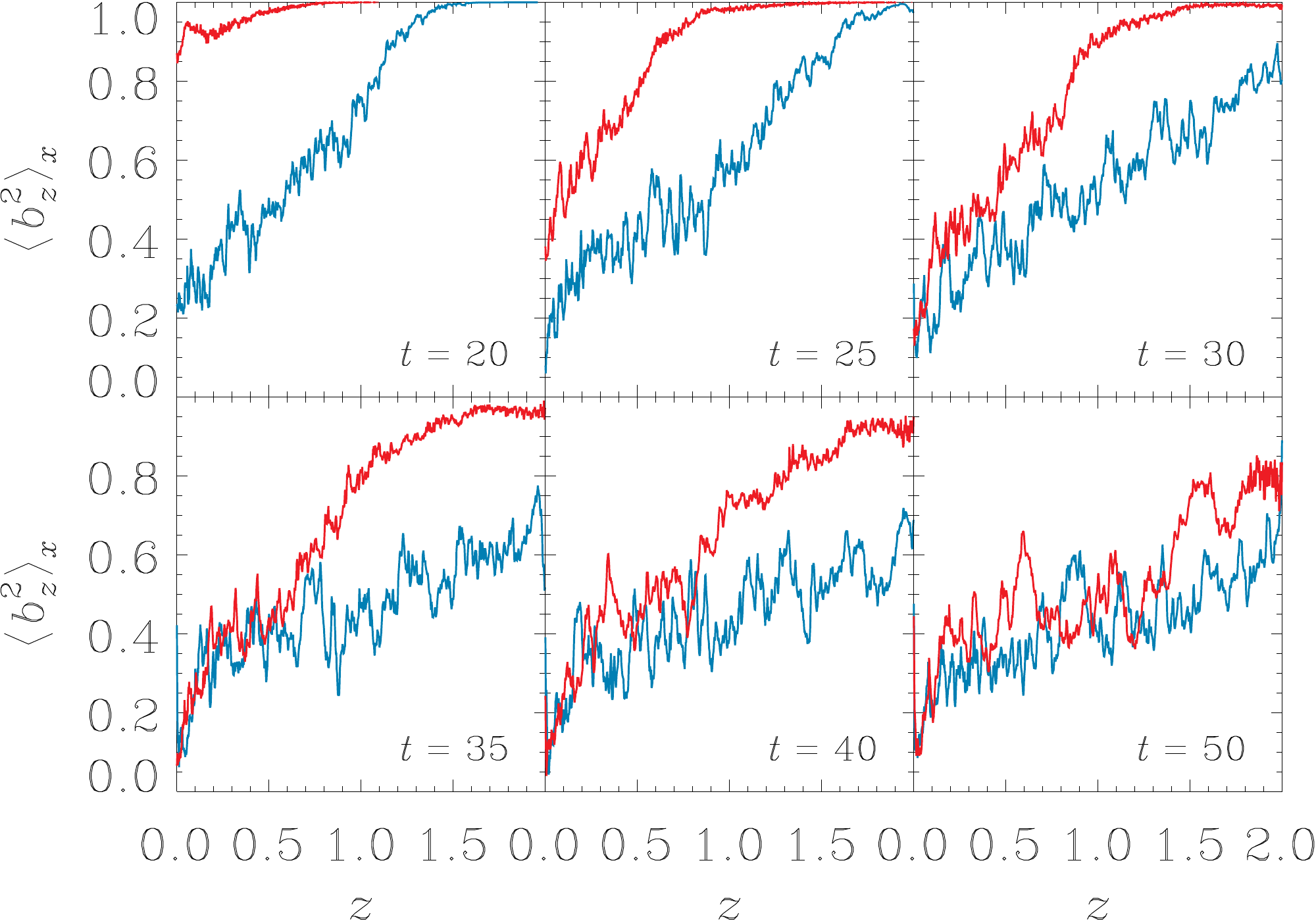}
\caption{Spatial and temporal evolution of the horizontally averaged magnetic-field angle at each height in runs H2dBRad (red lines) and H2dIRad (blue lines). The units of length and time are, respectively, $H_0 \simeq 124~{\rm kpc}$ and $H_0 / v_{\rm th,0} \simeq 160~{\rm Myr}$.}
\label{fig:2dRHBI:angle}
\end{figure}

Figure \ref{fig:2dRHBI:global} exhibits the temperature (color) and magnetic-field lines (black lines) in runs H2dBRad and H2dIRad at times $t=20$, $25$, $30$, $40$, and $50$ (in units of $H_0 / v_{\rm th,0} \simeq 160~{\rm Myr}$). As in the runs without cooling, the HBI develops first at low altitudes and subsequently spreads to higher altitudes, with Braginskii viscosity significantly affecting the structure of the magnetic field for $z \gtrsim 0.2 H_0$. This behavior is highlighted quantitatively in Figure \ref{fig:2dRHBI:angle}, which shows the horizontally averaged magnetic-field angle $\langle b^2_z \rangle_x$ as a function of height at the same times as in Figure \ref{fig:2dRHBI:global}. A comparison with Figure \ref{fig:2dHBI:angle} reveals that radiative cooling further promotes horizontal alignment of the field lines. This is achieved not only by locally increasing the temperature gradient at the bottom of the box, which increases the local growth rate of the HBI, but also because the ensuing cooling flow helps to squeeze any horizontal field lines. Radiative cooling also introduces two other notable differences.

\begin{figure}
\centering
\includegraphics[width=3in]{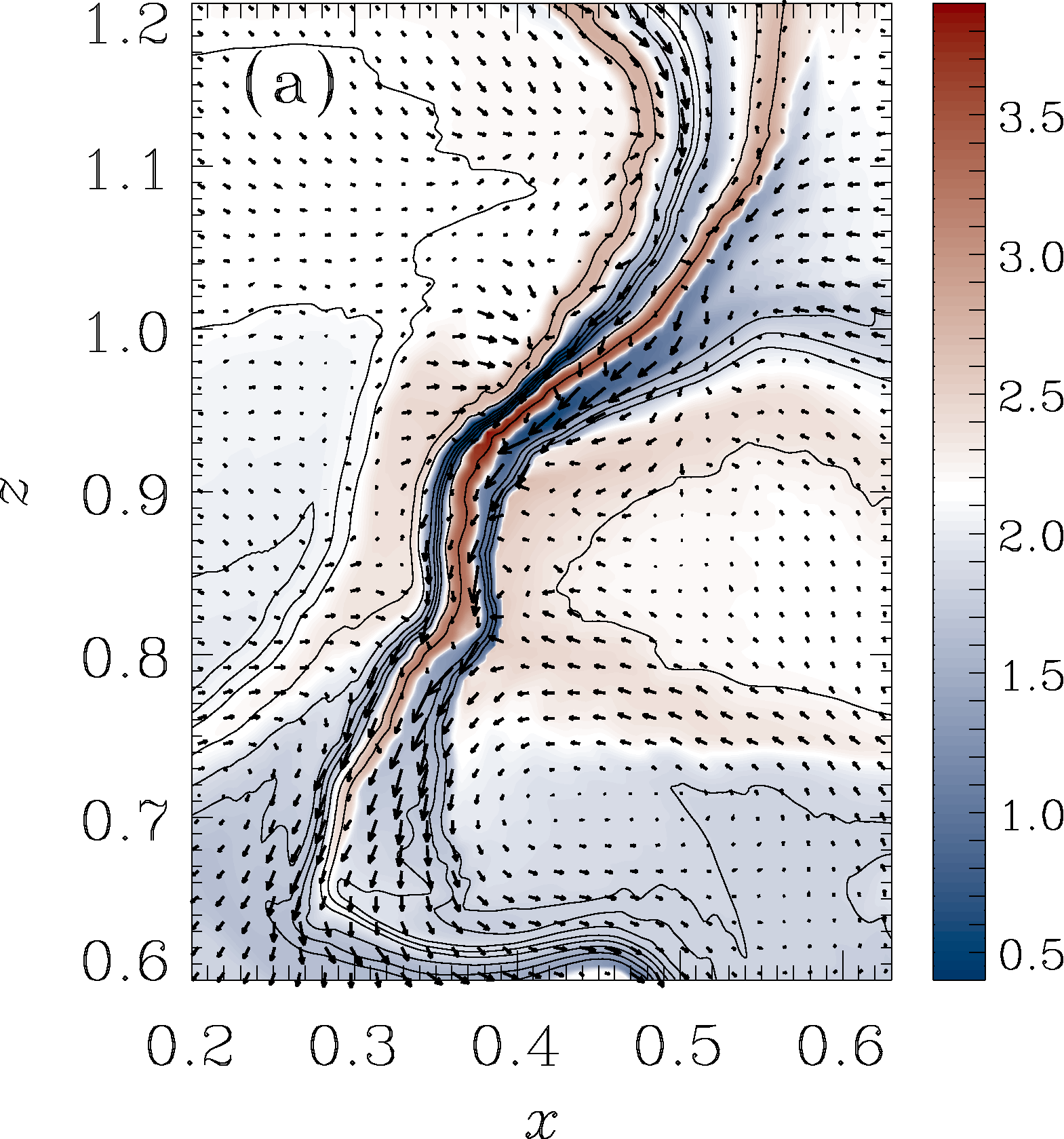}
\newline\newline
\includegraphics[width=3in]{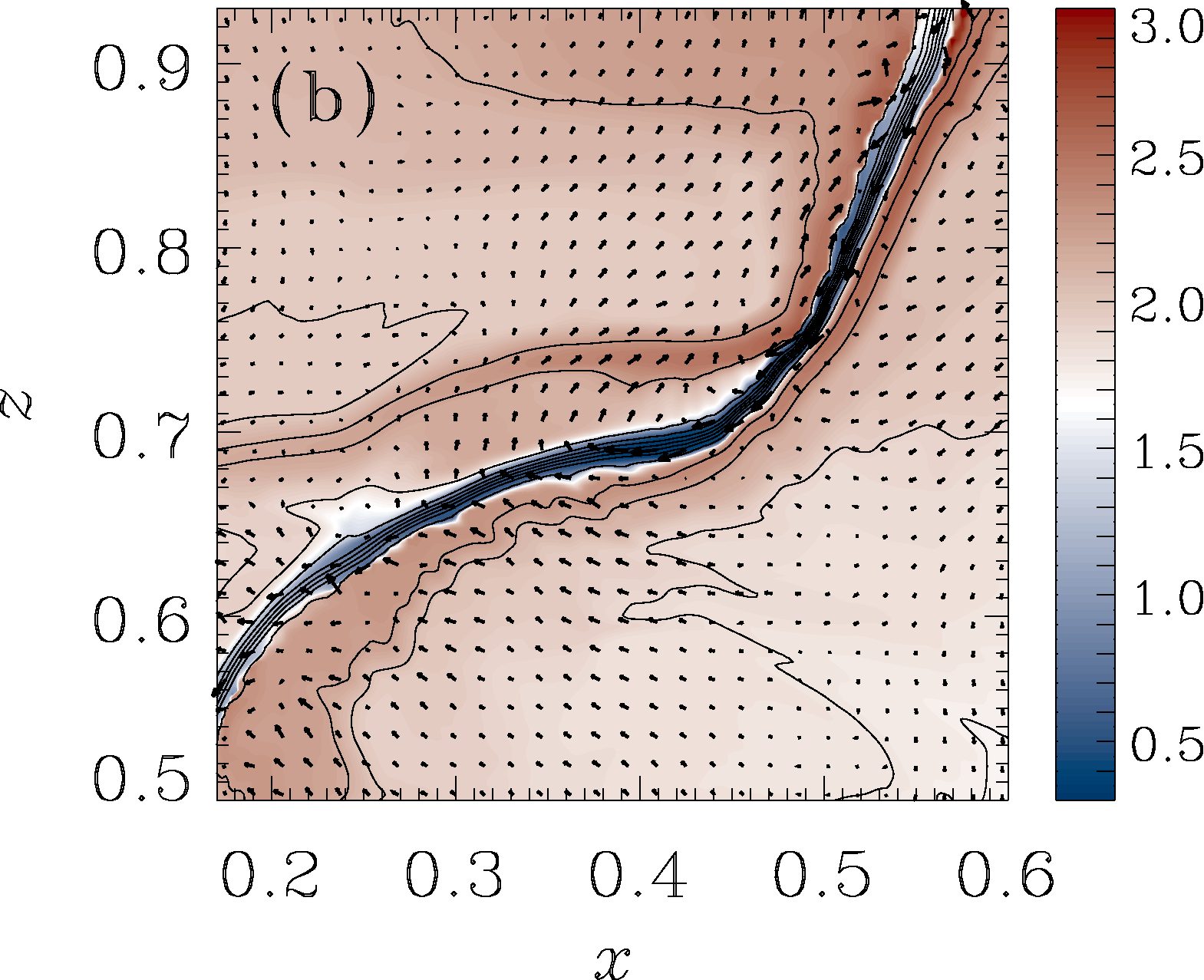}
\newline
\caption{Representative cool filaments from run H2dBRad. (a) A $\approx$$53~{\rm kpc}$$\times$$74~{\rm kpc}$ region surrounding two neighboring cool filaments at time $t \approx 7.6~{\rm Gyr}$ and location $z \approx 74$--$148~{\rm kpc}$. (b) A $\approx$$53~{\rm kpc}$$\times$$53~{\rm kpc}$ region surrounding a cool filament at time $t \approx 8.7~{\rm Gyr}$ and location $z \approx 62$--$115~{\rm kpc}$. In each of the frames, the temperature (color), magnetic-field lines (solid lines), and velocity vectors (black arrows) are shown; the temperature normalized to $\kb T_0 \simeq 3.0~{\rm keV}$. The magnetic field runs locally parallel to the filaments, with a strength $\sim$$5$--$20~\mu{\rm G}$. The maximum magnetic-field strength and speed in each frame are (a) $\simeq$$19.6~\mu{\rm G}$ and $\simeq$$296~{\rm km~s}^{-1}$, and (b) $\simeq$$13.8~\mu{\rm G}$ and $\simeq$$250~{\rm km~s}^{-1}$, respectively.}
\label{fig:2dRHBI:filament}
\end{figure}

First, thin ($\lesssim$$1~{\rm kpc}$) filaments of cool ($\lesssim$$1~{\rm keV}$) gas transiently appear throughout the course of run H2dBRad, typically lasting anywhere between $\sim$$100~{\rm Myr}$ and $\sim$$1~{\rm Gyr}$ and extending over distances $\sim$$30$--$80~{\rm kpc}$. Sometimes these filaments interact and merge with others nearby, forming relatively long chains of cool gas that wind throughout the cluster core. Sometimes filaments appear in pairs that run alongside one another over distances of a few tens of kpc.

In Figure \ref{fig:2dRHBI:filament}, we focus in on three representative cool filaments. Figure \ref{fig:2dRHBI:filament}a presents a $\approx$$53~{\rm kpc}$$\times$$74~{\rm kpc}$ region surrounding two neighboring cool filaments at time $t \approx 7.6~{\rm Gyr}$ and location $z \approx 74$--$148~{\rm kpc}$. Figure \ref{fig:2dRHBI:filament}b presents a $\approx$$53~{\rm kpc}$$\times$$53~{\rm kpc}$ region surrounding a cool filament at time $t \approx 8.7~{\rm Gyr}$ and location $z \approx 62$--$115~{\rm kpc}$. In all cases, these filaments follow the local magnetic-field lines, which serve to insulate them from the surrounding warm gas and which tend to become relatively isothermal over distances comparable to the Field length in the cool gas. This morphology is in agreement with dedicated numerical studies of local thermal instability in globally stable, anisotropically conducting plasmas by \citet{spq10}. The velocities in the filaments are also aligned with the local magnetic field and are well-ordered with speeds $\sim$$100$--$300~{\rm km~s}^{-1}$, depending upon their orientation (vertically oriented filaments tend to have larger bulk velocities due to gravitational acceleration). The magnetic-field strength is enhanced during the formation of these filaments, reaching typical values of $\sim$$5$--$20~\mu{\rm G}$. This enhancement extends over a region that is much longer than the extent of the cold gas itself, since cool gas becomes compressed along field lines and evacuates regions of plasma. One consequence of this enhancement is the local production of hot ($\approx$$7$--$10~{\rm keV}$) gas due to parallel viscous heating, which often envelopes the cool filament and produces a sharper temperature change across the filament. We discuss the astrophysical implications of these filaments in Section \ref{sec:summary}, as well as their agreement with current observational estimates.

\begin{figure}
\centering
\includegraphics[height=1.23in]{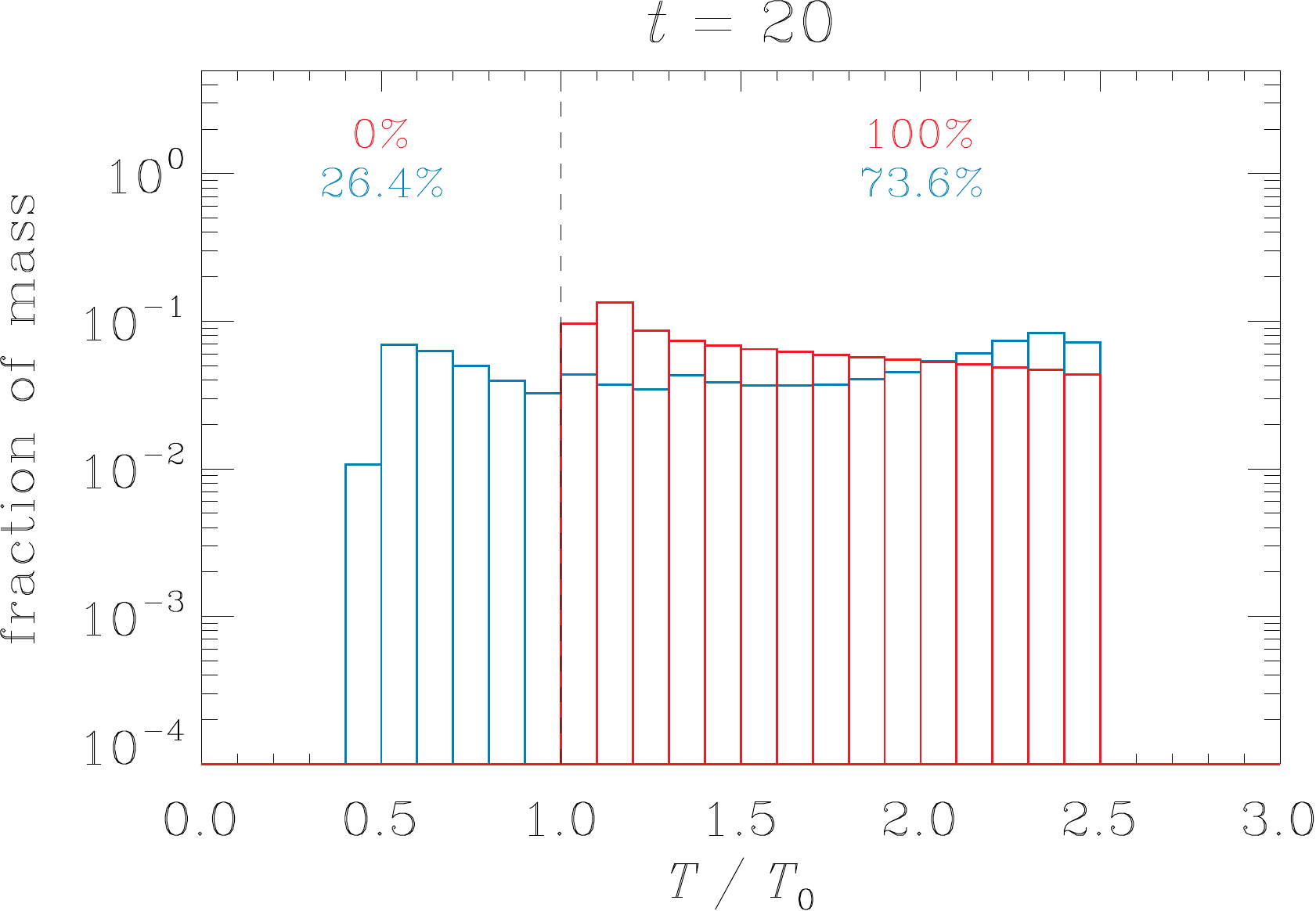}
\includegraphics[height=1.23in]{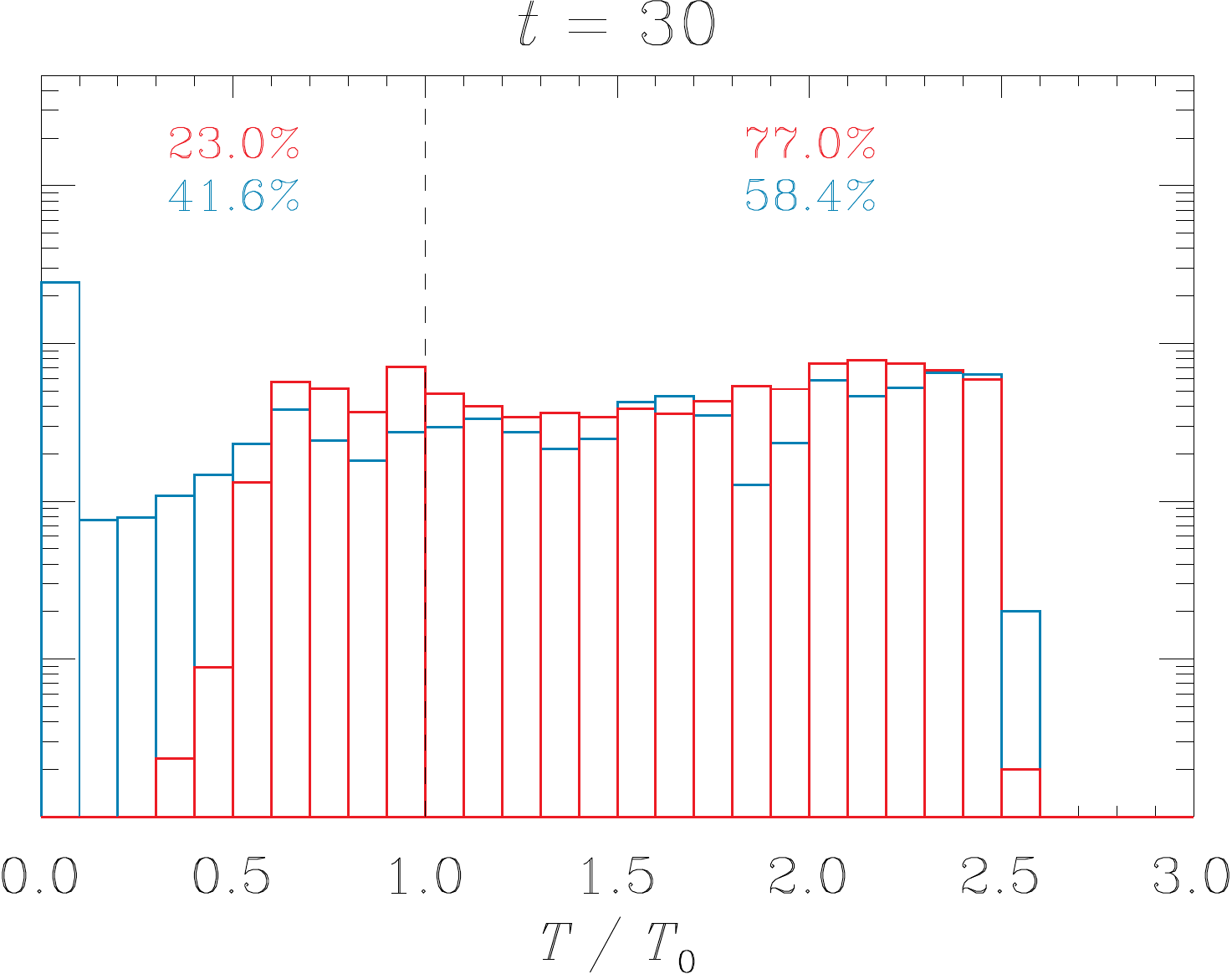}
\newline\newline 
\includegraphics[height=1.23in]{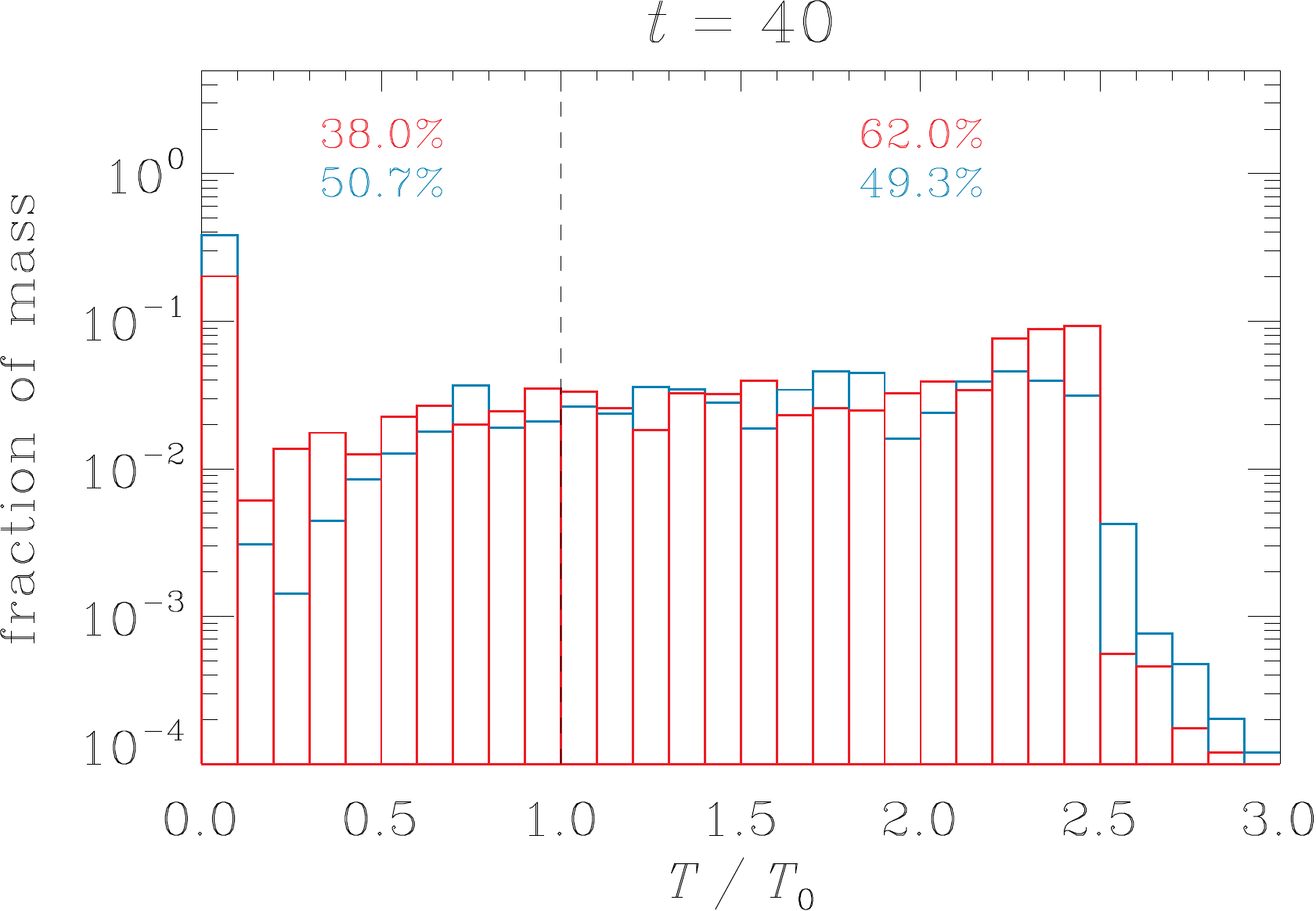}
\includegraphics[height=1.23in]{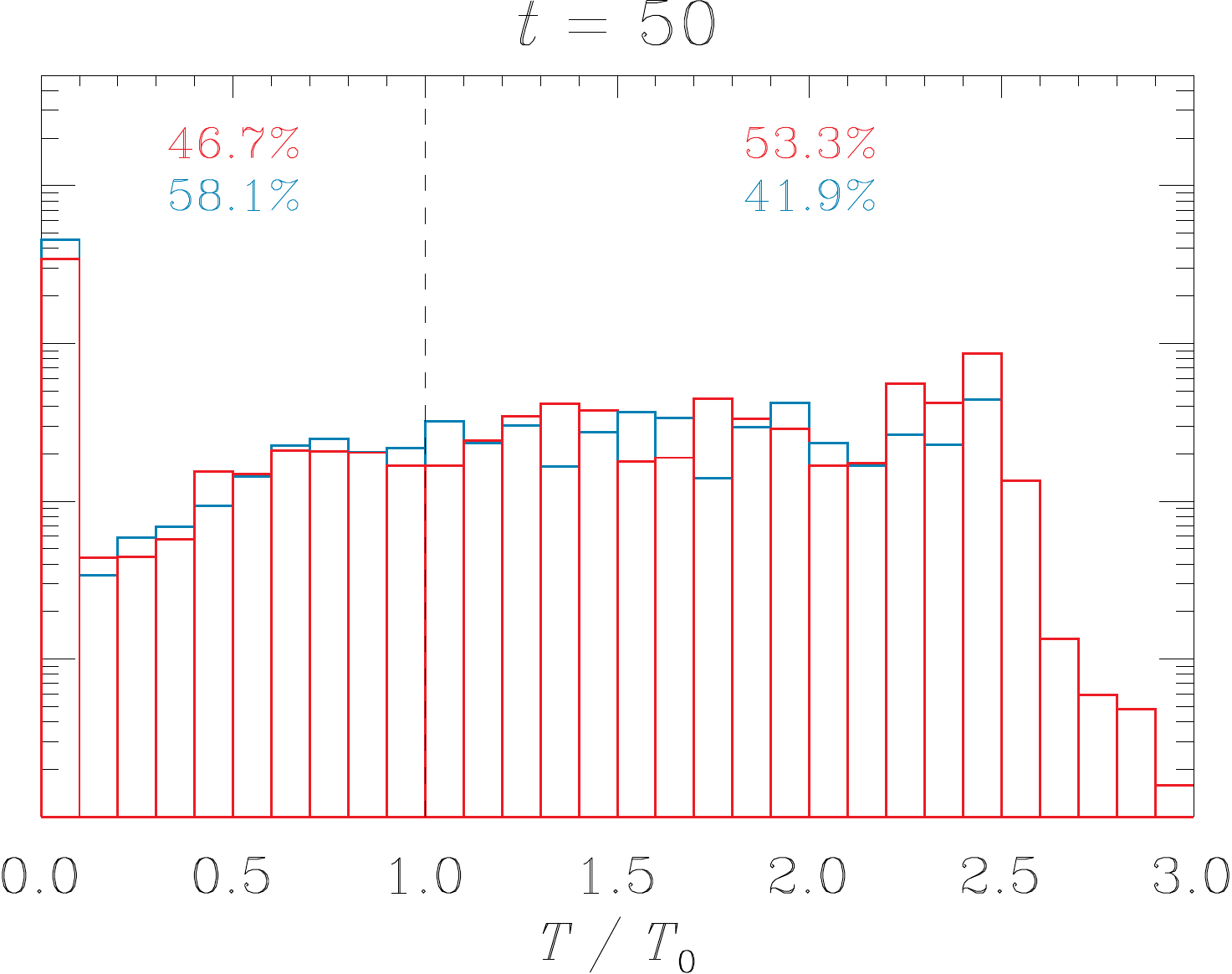}
\newline
\caption{Fraction of the total mass in each thermal phase (`cool' refers to $T < T_0 \simeq 3~{\rm keV}$; `warm' refers to $T \ge T_0$) at times $t = 20$, $30$, $40$, and $50$ (in units of $H_0 / v_{\rm th,0} \simeq 160~{\rm Myr}$) for runs H2dBRad (red) and H2dIRad (blue). The vertical dashed line denotes the temperature boundary separating the `cool' and `warm' phases. The total mass fraction in each of the phases is given as a percentage for each run.}
\label{fig:2dRHBI:tdist}
\end{figure}

Second, while a cooling catastrophe inevitably occurs in both runs, the amount of time until the cooling catastrophe occurs, as well as the amount of cold mass at any given time, are different. In Figure \ref{fig:2dRHBI:tdist}, we show histograms of the fractions of the total mass in each thermal phase (`cool' refers to $T < T_0 \simeq 3~{\rm keV}$ and `warm' refers to $T \ge T_0$) at times $t = 20$, $30$, $40$, and $50$ (in units of $H_0 / v_{\rm th,0} \simeq 160~{\rm Myr}$) for runs M2dBRad (red) and M2dIRad (blue). Mass is binned by temperature in intervals of width $\Delta (T/T_0) = 0.1$. The mass fraction in each phase is shown in each plot as a percentage. Braginskii viscosity delays a cooling catastrophe by reducing the efficacy of the HBI. As a result, the amount of mass in the cool phase is always less when Braginskii viscosity is included. While the temperature distributions become comparable by $t = 50$ ($\simeq$$8~{\rm Gyr}$), we note that in run M2dBrag ({\it i}) a significant portion of the cool gas is in the form of filaments located away from $z=0$ and ({\it ii}) there is hot gas with temperatures $\gtrsim$$7~{\rm keV}$ due to parallel viscous heating.

Why are there no cool filaments in run H2dIRad? The difference has to do with the fact that local thermal instability does not grow exponentially in dynamically evolving atmospheres \citep[e.g][]{bs89}. Without Braginskii viscosity suppressing the small-scale evolution of the HBI in the majority of the cluster core and thereby allowing conductive heating to offset a significant portion of the global radiative losses, a cooling flow readily develops and advects potentially unstable cold gas along with the bulk flow. Since a global equilibrium state is approximately preserved by potent conduction outside of the innermost $\sim$$20\%$ of the core, local thermal instability can proceed there. 

\begin{figure}
\centering
\includegraphics[width=3in]{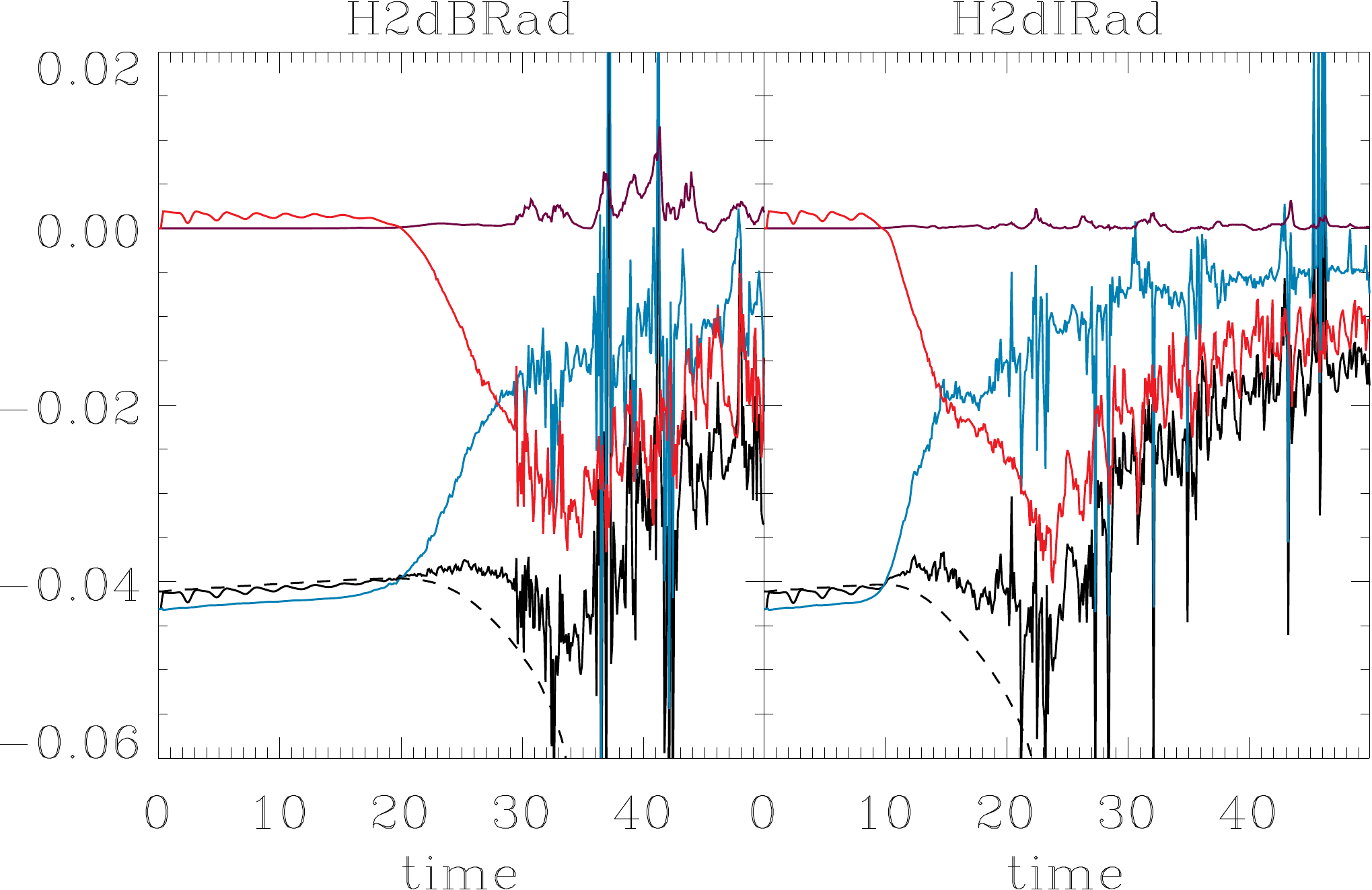}
\newline
\caption{Temporal evolution of the advective (red lines), convective (purple lines), conductive (blue lines), and total (black solid lines) fluxes through $z = 0.4$ ($\simeq$$50~{\rm kpc}$) in runs H2dBRad and H2dIRad. The black dashed lines denote the radiative cooling rate integrated over the region $0 \leq z \leq 0.4$, which must be balanced by the energy fluxes for the core to be in equilibrium. Negative (positive) fluxes correspond to downward (upward) energy transport. The units of time and energy flux are, respectively, $H_0 / v_{\rm th,0} \simeq 160~{\rm Myr}$ and $\rho_0 v^3_{\rm th,0} \simeq 0.02~{\rm ergs~s}^{-1}~{\rm cm}^{-2}$.}
\label{fig:2dRHBI:fluxes}
\end{figure}

Finally, in Figure \ref{fig:2dRHBI:fluxes} we present the advective (red lines), convective (purple lines), conductive (blue lines), and total (black solid lines) energy fluxes through $z = 0.4$ ($\simeq$$50~{\rm kpc}$) in runs H2dBRad and H2dIRad. (See equations 20--22 of \citealt{brbp09} for definitions of these fluxes.) The black dashed lines denote the radiative cooling rate integrated over the region $0\leq z \leq 0.4$, which must be balanced by the energy fluxes for the core to be in equilibrium. Negative fluxes correspond to downward energy transport and vice versa for positive fluxes. Figure \ref{fig:2dRHBI:fluxes} shows that, in both cases, the atmosphere evolves from the initial equilibrium at the time when the amount of inward conductive flux transported begins to dwindle. The most striking difference between runs H2dIRad and H2dBRad is that the suppression of the conductive flux occurs much later in the run with Braginskii viscosity ($t \approx 20$ vs. $t \approx 10$). As a result, the conductive and total energy fluxes at the end of run H2dBRad are double those in run H2dIRad. 

After the onset of the HBI, run H2dBRad exhibits brief episodes of substantial convective flux, at times reaching $\sim$$20\%$ of the total flux. This reverse convective flux acts as a coolant in the energy equation. The convective flux maxima correspond to the strong episodes of heat conduction towards the cool core, which can be traced to times when filaments cross the referent surface ($z=0.4$) where the fluxes are evaluated. The presence of Braginskii viscosity promotes the formation of such filaments and, consequently, this mode of heat conduction and convective feedback. A possibility that a cool core mitigates abrupt changes in its thermal state via such a feedback loop has been proposed by \citet{br08} as a mechanism to regulate thermal conductivity in hot (i.e. massive) galaxy clusters.

\section{MTI}\label{sec:mti}

\subsection{Background Equilibrium and Initial Perturbations}

We consider a non-radiative, plane-parallel plasma stratified in both density and temperature in the presence of a uniform gravitational acceleration in the vertical direction, $\bb{g} = -g \ez$. We thread the plasma with a uniform background magnetic field oriented along $\ex$, so that there is no heat flux in the background state (i.e. $q = 0$). While a subcritical transition to turbulence exists for an initially vertical magnetic field subject to modest stirring \citep{mpsq11}, we focus only on the simplest background from which the linear MTI grows the fastest. Since $q = 0$, thermal equilibrium is trivially satisfied and we are free to choose a linearly decreasing temperature profile \citep{ps05,mpsq11}:
\begin{equation}\label{eqn:mtiregion2temp}
T(z) = T_0 \left( 1 - \frac{z}{3H_0} \right) .
\end{equation}
Force balance then implies
\begin{equation}\label{eqn:mtiregion2dens}
\rho(z) = \rho_0 \left( 1 - \frac{z}{3H_0} \right)^2 .
\end{equation}
\begin{equation}\label{eqn:mtiregion2pres}
p(z) = p_0 \left( 1 - \frac{z}{3H_0} \right)^3 .
\end{equation}
Because the MTI induces large vertical displacements in the plasma, we attempt to minimize the effects of our boundary conditions (\S\ref{sec:mtisetup}) by sandwiching the unstable volume of plasma between two buoyantly neutral layers, following \citet{ps07}. We divide the vertical box size $L_z$ into three regions: region I ($0 < z < L_z/4$), region II ($L_z/4 \le z \le 3L_z/4$), and region III ($3L_z/4 < z < L_z$). Region II is described by Equations (\ref{eqn:mtiregion2temp})--(\ref{eqn:mtiregion2pres}) with $z \rightarrow z - L_z/4$, so that $T = T_0$, $\rho = \rho_0$, and $p = p_0$ at the base of the MTI-unstable region. Regions I and III are isothermal atmospheres with $T_{\rm I} = T_0$ and $T_{\rm III} = T_0 ~ ( 1 - L_z / 6 H_0 )$, respectively. Requiring force balance and continuity yields the following density and pressure distributions:
\begin{equation}
\frac{\rho_{\rm I} (z)}{\rho_0} = \frac{p_{\rm I}(z)}{p_0} = \exp\left( - \frac{z - L_z / 4}{H_0} \right) ,
\end{equation}
\begin{equation}
\frac{\rho_{\rm III} (z)}{\rho_0} = \frac{p_{\rm III}(z)}{p_0} = \rho_0 \left( 1 - \frac{L_z}{6 H_0} \right)^2 \exp\left( - \frac{z - 3 L_z / 4}{H_0 - L_z / 6} \right) ,
\end{equation}
We further prescribe isotropic conduction in regions I and III in order to stabilize the MTI there. 

We apply Gaussian-random velocity perturbations to our background equilibrium, having a flat spatial power spectrum and a standard deviation of $10^{-4} v_{\rm th,0}$. While we have assumed equal ion and electron temperatures throughout this paper, we caution here that, due to the very low collisionality in cluster outskirts, the assumption of equal ion and electron temperatures may not hold.

\subsection{Numerical Setup and Boundary Conditions}\label{sec:mtisetup}

We non-dimensionalize our equations using the same units chosen in Section \ref{sec:hbisetup}. Note, however, that typical temperatures at the base of the MTI-unstable portion of the ICM, where the temperature begins to decrease outwards, are $\kb  T_0 \approx 5$--$8~{\rm keV}$, so that $[\ell] \sim 400~{\rm kpc}$ and $[t] \sim 350~{\rm Myr} $ are more representative numbers for the units of length and time, respectively. We choose the free parameters $\beta_0 = 10^5$ and ${\rm Kn}^{-1}_0 = 200$, so that $B_0 \simeq 0.016$, $\nu_{\|} \simeq 2.4 \times 10^{-3} ~ T^{5/2} \rho^{-1}$, and $\kappa_{\|} \simeq 0.12 ~ T^{5/2} \rho^{-1}$ in dimensionless units. While ${\rm Kn}_0$ under actual cluster-outskirt conditions is $\sim$$5$ larger than our chosen value, we have found that the implied $\kappa_{\|} (z = L_z, t = 0) \sim 1$ makes the computations unnecessarily expensive.

Although linear analyses without \citep{balbus01} and with (K11) Braginskii viscosity have shown that the fastest-growing MTI modes satisfy $1 \ll k_{\|} H \ll \beta^{1/2}$ and are therefore local, recent work on the nonlinear development of the MTI \citep{mpsq11} has indicated that simulations with $L / H \sim 0.1$ significantly underestimate the magnitude of the turbulent velocities in the saturated state. We therefore choose box sizes $H_0$$\times$$2 H_0$ (in 2d) and $H_0$$\times$$H_0$$\times$$2 H_0$ (in 3d); the MTI-unstable layer then has a vertical size equal to $H_0$.

Our boundary conditions are the same as in \citet{mpsq11}. The temperatures at the upper and lower boundaries of our computational domain are fixed for all times, while the pressure is extrapolated into the upper and lower ghost zones in such a way as to ensure hydrostatic equilibrium at those boundaries. Periodic boundary conditions are imposed in the horizontal direction(s). We further maintain horizontal magnetic fields at the upper and lower boundaries on the computational domain. Note that this does not insulate the ghost zones from the computational domain, as we have prescribed isotropic conduction in regions I and III. Thus a fixed temperature difference is imposed across the vertical length of our model atmosphere. In this situation the MTI cannot exhaust the source of free energy, which is being constantly replenished by the boundary conditions and thus continuously drives MTI turbulence. If the temperature difference across the computation domain were allowed to relax (i.e. Neumann boundary conditions), the atmosphere would become isothermal before the MTI could fully develop \citep{ps05,ps07,psl08,mpsq11}. 

Here we present results from five MTI simulations. M2dIsoP is a 2d simulation with isotropic pressure and serves as a reference run, enabling us to draw conclusions about the effects of Braginskii viscosity. M2dBrag is a 2d simulation with Braginskii viscosity and M2dBLim is a 2d simulation in which the pressure anisotropy is artificially limited using the Sharma et al.~closure described in Section \ref{sec:microscale}. M3dBrag is a 3d simulation with Braginskii viscosity and run M3dIsoP is a 3d simulation with isotropic pressure. The parameters in these simulations are summarized in Table \ref{table:runs}.

\subsection{2d Simulations}\label{sec:mti2d}

\begin{figure}
\centering
\includegraphics[width=3in]{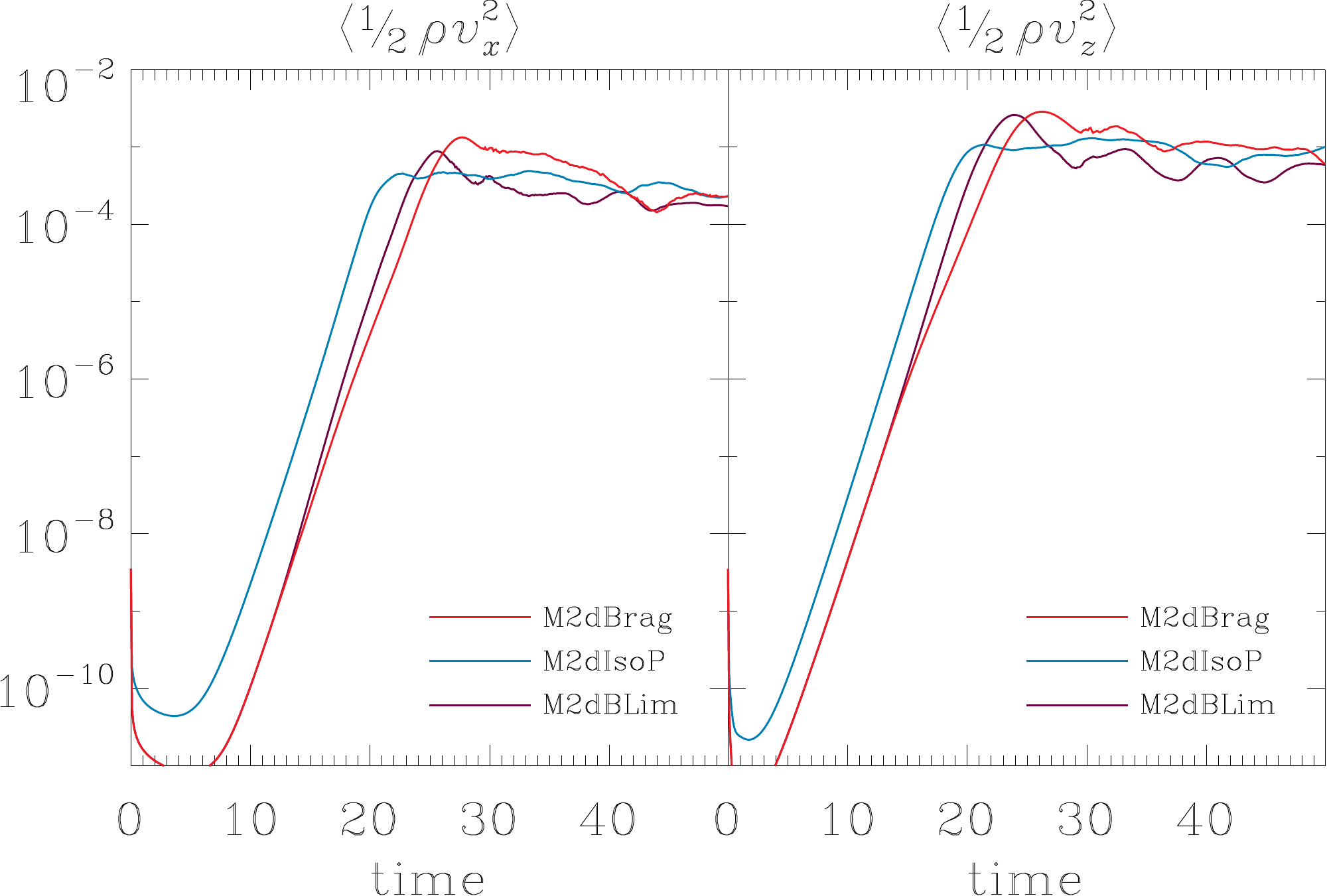}
\newline\newline 
\includegraphics[width=3in]{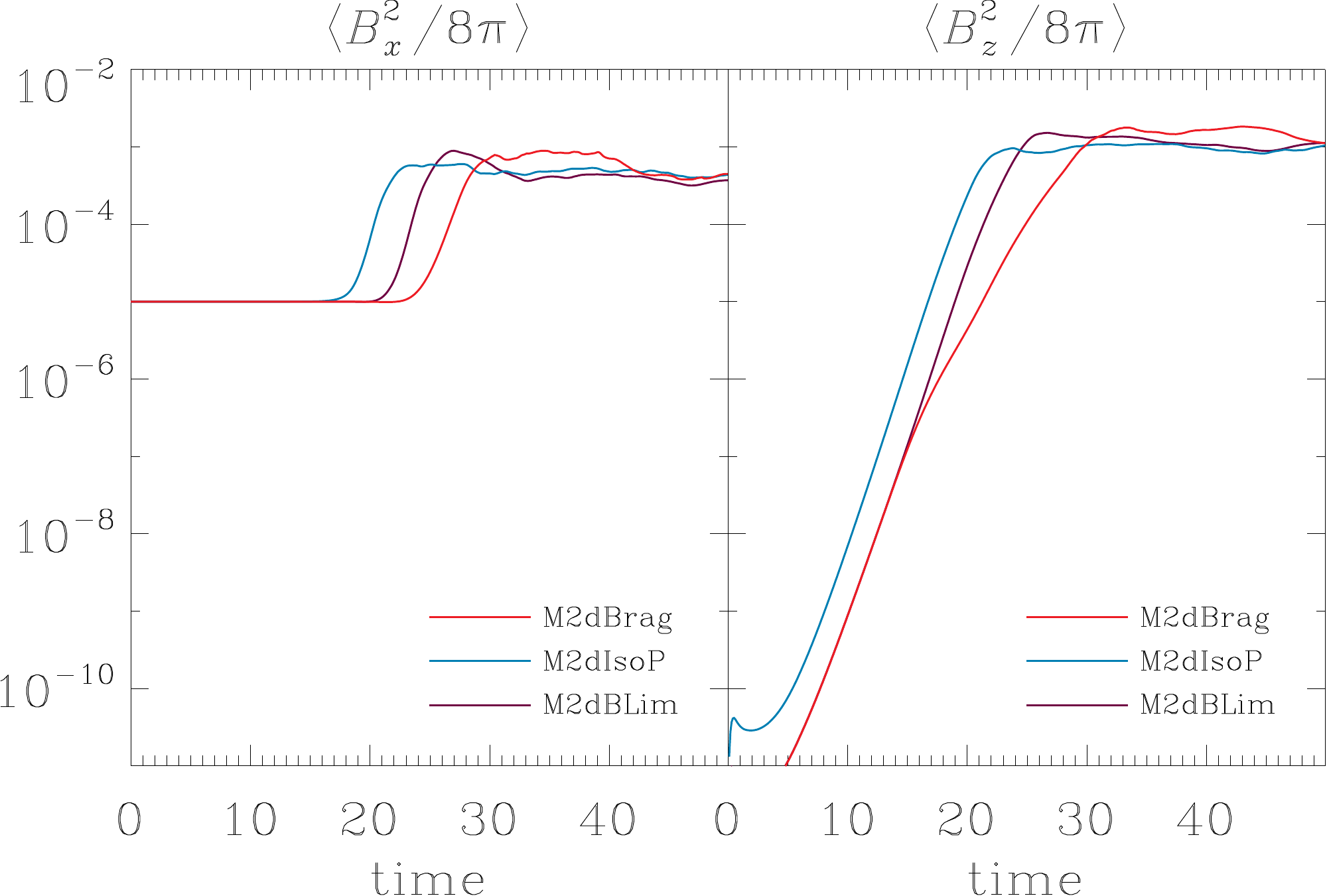}
\newline
\caption{Temporal evolution of the horizontal ($x$) and vertical ($z$) kinetic and magnetic energies, averaged over the MTI-unstable portion of the box ($0.5 \le z \le 1.5$), in runs M2dBrag (red lines), M2dIsoP (blue lines), and M2dBLim (purple lines). The units of energy density and time are, respectively, $\rho_0 v^2_{\rm th,0}$ and $H_0 / v_{\rm th,0}$ (see Equation \ref{eqn:timeunit}).}
\label{fig:2dMTI:energy}
\end{figure}
\begin{figure}
\centering
\includegraphics[width=2.1in]{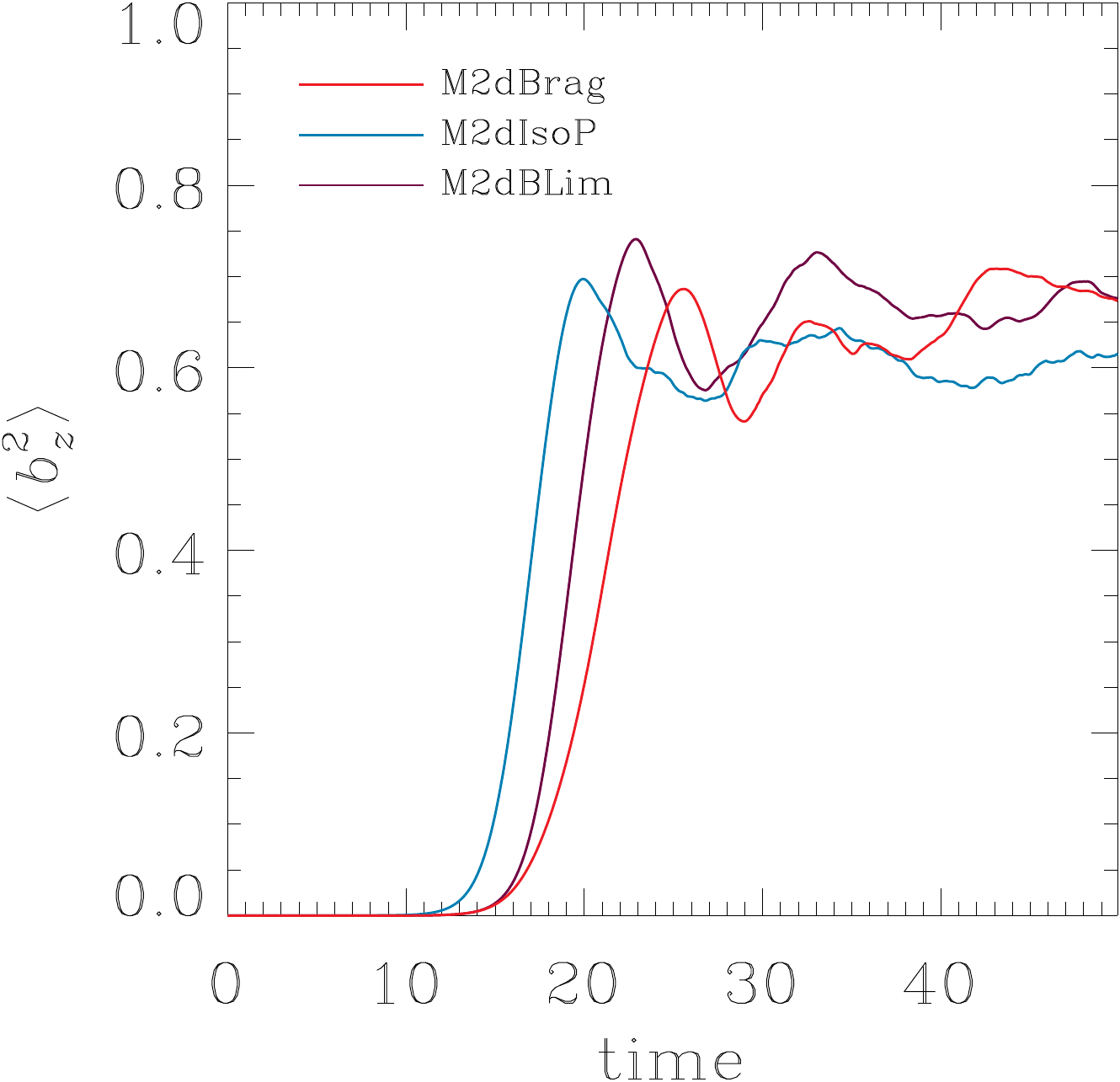}
\caption{Temporal evolution of the magnetic-field angle $b^2_z$, averaged over the MTI-unstable portion of the box ($0.5 \le z \le 1.5$), in runs M2dBrag (red line), M2dIsoP (blue line), and M2dBLim (purple line). The unit of time is $H_0 / v_{\rm th,0}$ (see Equation \ref{eqn:timeunit}).}
\label{fig:2dMTI:bz2}
\end{figure}
\begin{figure*}
\centering
\includegraphics[height=6.5in,angle=90]{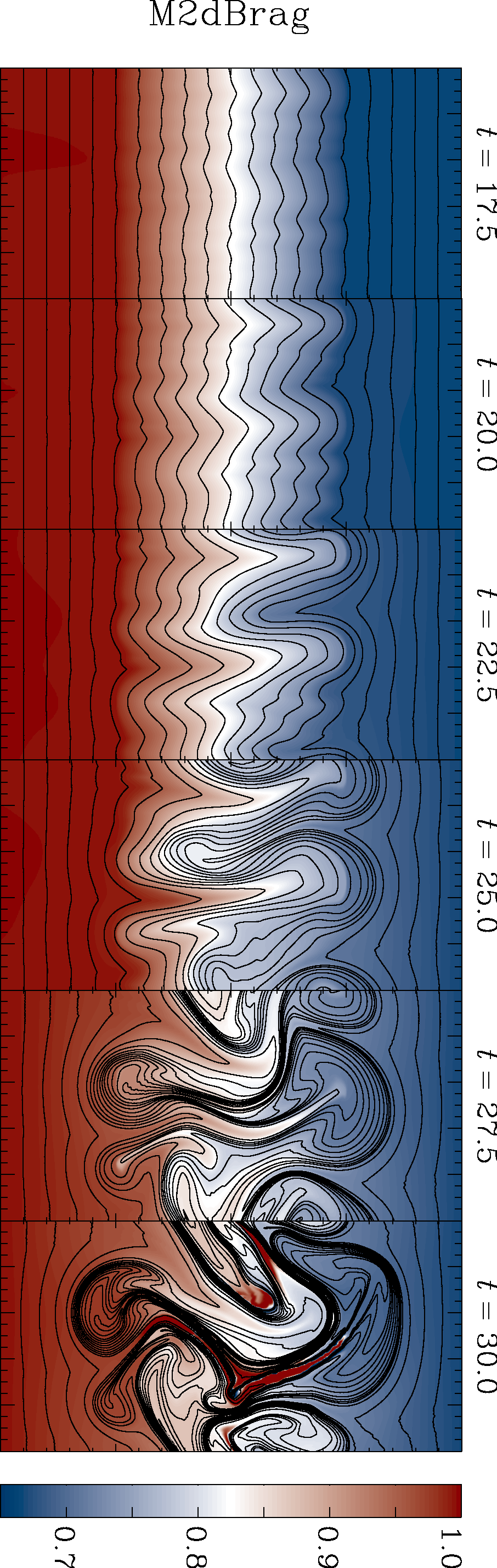}
\newline\newline 
\includegraphics[height=6.5in,angle=90]{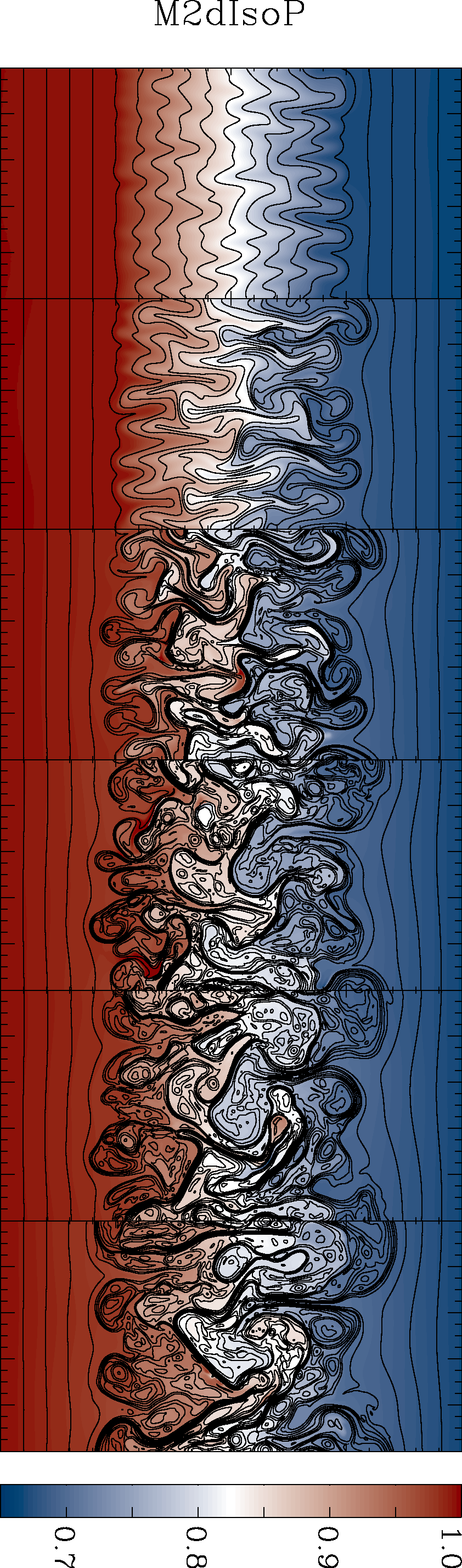}
\newline\newline 
\includegraphics[height=6.5in,angle=90]{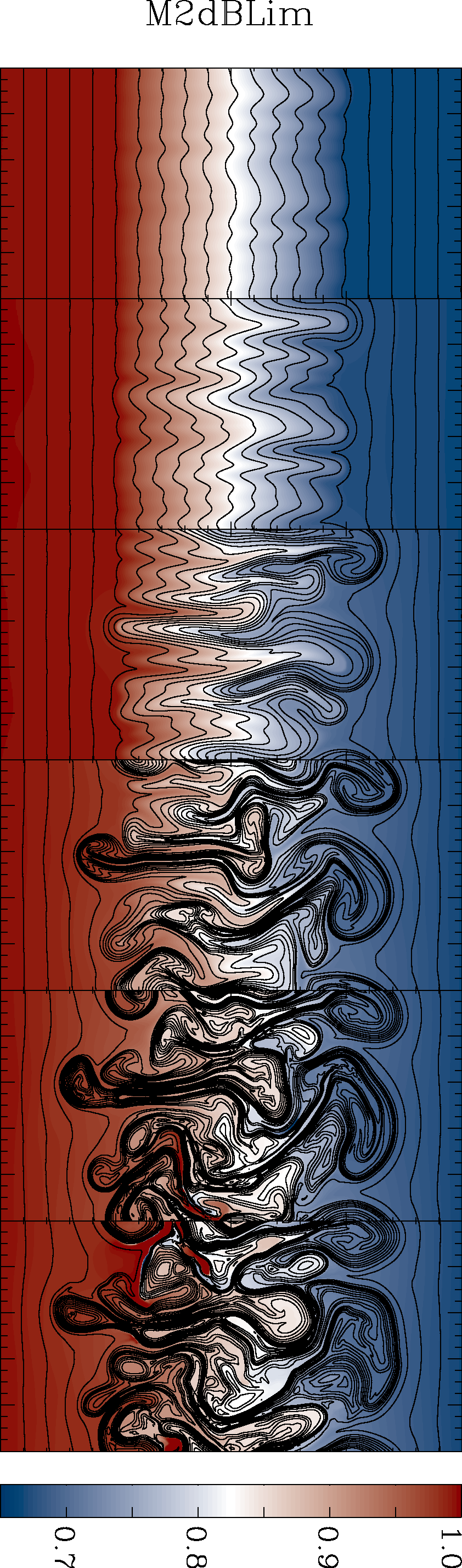}
\newline 
\caption{Spatial and temporal evolution of the MTI with Braginskii viscosity (top row), without Braginskii viscosity (middle row), and with limited Braginskii viscosity (bottom row). The temperature (color) and magnetic-field lines (black lines) are shown at times $t = 17.5$, $20.0$, $22.5$, $25.0$, $27.5$, and $30.0$; see Equation \ref{eqn:timeunit}). The computational domain has dimensions $L_x$$\times$$L_z = 1$$\times$$2$ (in units of $H_0$; see Equation \ref{eqn:lengthunit}). We have suppressed the imaging of temperatures beyond the fixed color-bar limits.}
\label{fig:2dMTI:global}
\end{figure*}

Figure \ref{fig:2dMTI:energy} presents the evolution of the kinetic and magnetic energies averaged over the MTI-unstable region ($0.5 \le z \le 1.5$). Runs in which pressure anisotropies are allowed to develop (red and purple lines) initially exhibit a growth rate equal to that of the run without Braginskii viscosity (blue line), in agreement with predictions from linear theory (K11). This was expected on the grounds that the fastest-growing linear MTI modes are Alfv\'{e}nically polarized (i.e. $\delta B_{\|} = 0$) and therefore escape viscous damping by not producing a linear pressure anisotropy. However, these modes do produce a nonlinear pressure anisotropy, which begins to significantly decrease the growth rate once $( \delta B_\perp / B )^2 \gtrsim 0.1$. The growth rate in run M2dBLim (purple line) does not show this behavior, instead agreeing with that of run M2dBrag (red line). This is because the limiters do not allow the local nonlinear pressure anisotropy to become greater than the local magnetic pressure. All three runs reach an approximately saturated kinetic energy corresponding to a Mach number of several percent. The total magnetic energy increases by a factor of $\sim$$100$ over the course of the runs, with a final $\beta \sim 10^3$ and most of the energy in the vertical component. The kinetic and magnetic energies saturate in approximate equipartition.

Curiously, runs in which the pressure may become anisotropic have a slightly prolonged phase of exponential growth and, consequently, greater maximum energy densities than those found in run M2dIsoP. We believe there are two principal reasons for this difference. First, Braginskii viscosity suppresses the formation of small perpendicular scales, which results in conditions less favorable for grid-scale magnetic reconnection. Second, the suppression of perpendicular fluctuations due to Braginskii viscosity constrains the perturbed magnetic field to grow predominantly vertically in run M2dBrag. As a result, there is less interference amongst the buoyant plumes than in run M2dIsoP, where the perturbed field lines acquire a more tangled topology. Indeed, runs M2dBrag and M2dBLim exhibit greater saturated values of $\langle b^2_z \rangle$ than those in run M2dIsoP (see Figure \ref{fig:2dMTI:bz2}).

These differences are dramatically highlighted in Figure \ref{fig:2dMTI:global}, which shows the overall spatial and temporal evolution of the atmosphere during each of our 2d runs. The temperature (color) and magnetic-field lines (solid lines) are displayed in each of the six frames, which show the atmosphere at the different times $t = 17.5$, $20.0$, $22.5$, $25.0$, $27.5$, and $30.0$ (in units of $H_0 / v_{\rm th,0}$). Clear differences exist in the topology and evolution of the magnetic field between each of the runs.

The magnetic-field fluctuations in runs M2dBrag and M2dBLim emerge on larger scales than in run M2dIsoP. This is because the initial growth of modes with $\delta B_{\|} \ne 0$ induces a positive pressure anisotropy that shifts all unstable modes to larger (parallel) wavelengths. (Recall from Equation \ref{eqn:bprl} that, for our chosen plasma parameters, fluctuations in magnetic-field strength as small as $\sim$$0.1\%$ can produce a pressure anisotropy comparable to the magnetic tension.) In addition, the growing modes in runs M2dBrag and M2dBLim exhibit more of a sawtooth-like structure than do those in run M2dIsoP. This is because Braginskii viscosity targets only those motions that change the field strength. As a result, magnetic-field lines tend to remain locally parallel to one another in order to minimize field-line compressions and rarefactions. This also results in a more laminar development of the instability than in run M2dIsoP, as small-scale features along field lines are viscously damped. 

Whether the pressure anisotropy is limited (M2dBLim) or not (M2dBrag), the turbulence tends to arrange the magnetic fields in long, thin flux sheets with $\ell_{\|} \gg \ell_\perp$: the field reverses its direction at the smallest scale available to it (the numerical resistive scale) but field lines curve at the scale of the flow (the viscous scale) except in the sharp bends of the folds \citep[for an analytical theory of folded structure, see][]{scmm02}. This folded structure, a general property of random forcing in plasmas with large magnetic Prandtl numbers, allows the small-scale direction-reversing magnetic field to back react on the flow in a spatially coherent way: the velocity gradients become locally anisotropic with respect to the direction of the folds (with $\eb \eb \bb{:} \grad \bb{v}$ partially suppressed) so that the dynamo saturates at marginally stable balance between reduced ``parallel'' stretching and ``perpendicular'' mixing. Thus, the field strength and the field-line curvature are anticorrelated \citep[see][]{sctmm04}.

One consequence of reduced parallel stretching is the notable paucity of firehose instabilities in run M2dBrag. In run H2dBrag (HBI with Braginskii viscosity), such a reduction in parallel stretching and concomitant production of microscale instabilities was not possible, since correlations between field strength and field-line curvature are necessary for the HBI to function in the first place. Rather than reduce the pressure anisotropy to marginally stable values (Equation \ref{eqn:pabounds}) via the secular growth of microscale fluctuations, MTI-driven turbulence appears to avoid the production of large pressure anisotropies altogether by orienting magnetic fields primarily across local velocity gradients. Consequently, the box-averaged pressure anisotropy (not shown) is sub-marginal during the nonlinear phase of evolution.

\subsection{3d Simulations}\label{sec:mti3d}

\begin{figure}
\centering
\includegraphics[width=3in]{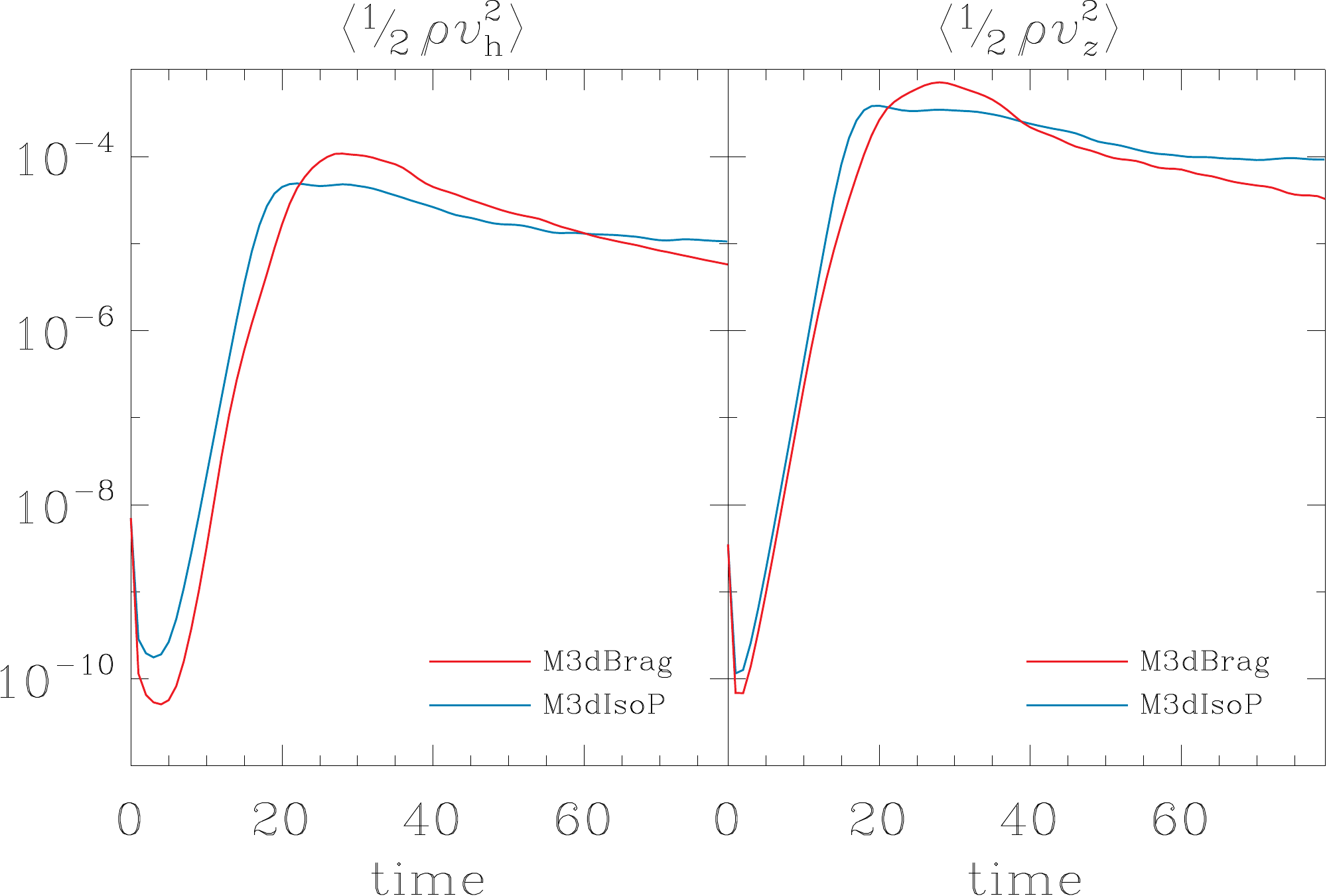}
\newline\newline 
\includegraphics[width=3in]{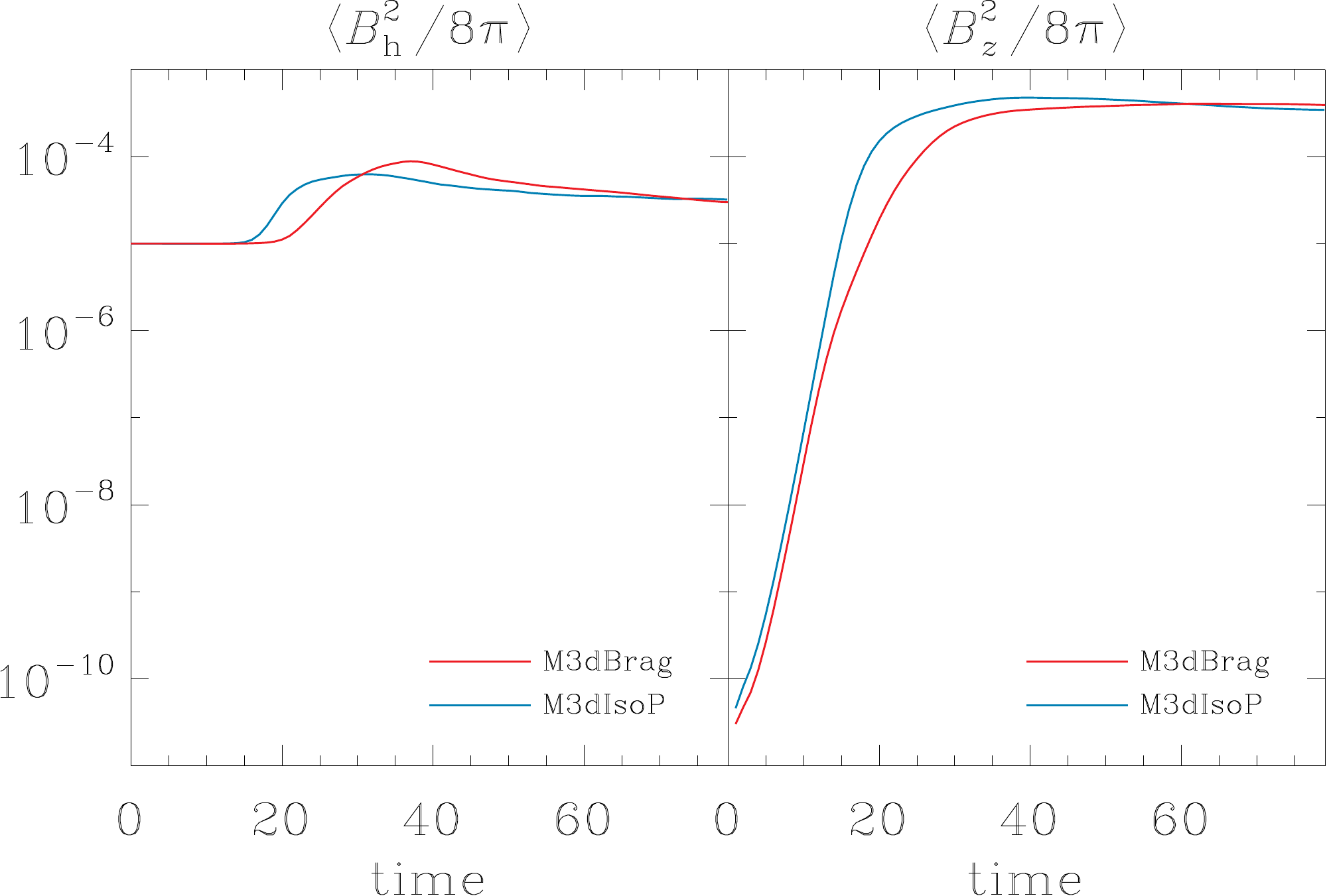}
\newline
\caption{Temporal evolution of the horizontal (``h'') and vertical (``$z$'') kinetic and magnetic energy densities, averaged over the MTI-unstable portion of the box ($0.5 \le z \le 1.5$), in runs M3dBrag (red lines) and M3dIsoP (blue lines). The units of energy density and time are, respectively, $\rho_0 v^2_{\rm th,0}$ and $H_0 / v_{\rm th,0}$ (see Equation \ref{eqn:timeunit}).}
\label{fig:3dMTI:energy}
\end{figure}

Figure \ref{fig:3dMTI:energy} shows the evolution of the kinetic and magnetic energies averaged over the MTI-unstable region ($0.5 \le z \le 1.5$) for runs M3dBrag (red lines) and M3dIsoP (blue lines). Many features are in agreement with our 2d simulations. First, the growth rates from both runs are equal until the nonlinear pressure anisotropy begins to significantly decrease the M3dBrag growth rate. Second, run M3dBrag has a slightly prolonged phase of exponential growth and greater maximum energy densities than those found in run M3dIsoP. As explained in Section \ref{sec:mti2d}, we believe this is due to Braginskii viscosity suppressing the formation of small perpendicular scales. While a certain amount of the decrease in saturated energies from 2d to 3d is likely due to the decreased resolution, it is also due to the fact that, in 3d, magnetic field lines are allowed to penetrate into the additional dimension and alleviate regions of strong magnetic pressure. Overall, the evolution of the average energies are qualitatively similar in 2d and 3d.

\begin{figure}
\centering
\includegraphics[width=2.1in]{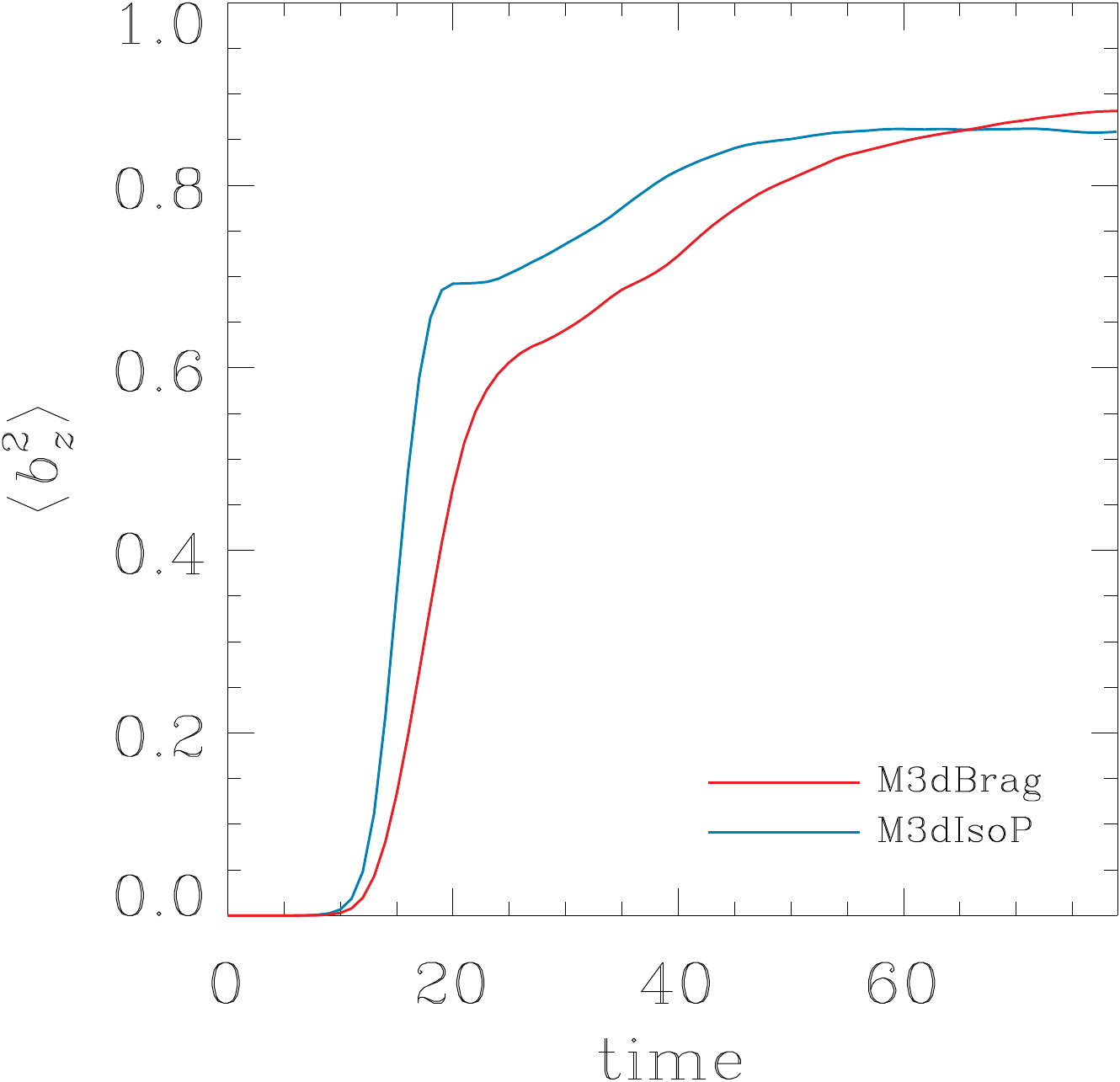}
\caption{Temporal evolution of the magnetic-field angle $b^2_z$, averaged over the MTI-unstable portion of the box ($0.5 \le z \le 1.5$), in runs M3dBrag (red line) and M3dIsoP (blue line). The unit of time is $H_0 / v_{\rm th,0}$ (see Equation \ref{eqn:timeunit}).}
\label{fig:3dMTI:bz2}
\end{figure}

The average magnetic-field angle, on the other hand, shows qualitative differences between runs M3dBrag and M3dIsoP (Figure \ref{fig:3dMTI:bz2}). In run M3dBrag the magnetic field at the end of the linear phase is oriented slightly more horizontal than in run M3dIsoP. However, $\langle b^2_z \rangle$ subsequently grows faster and ultimately becomes larger than in run M3dIsoP. We attribute this difference to the presence of the Alfv\'{e}nic MTI: as the mean field becomes more vertical, Braginskii viscosity couples $k_y \ne 0$ Alfv\'{e}n modes to damped slow modes and drives them buoyantly unstable. These modes, absent in the case of isotropic pressure, likely play a role in reorienting the magnetic field vertically at late times.

\begin{figure}
\centering
\includegraphics[height=3.5in]{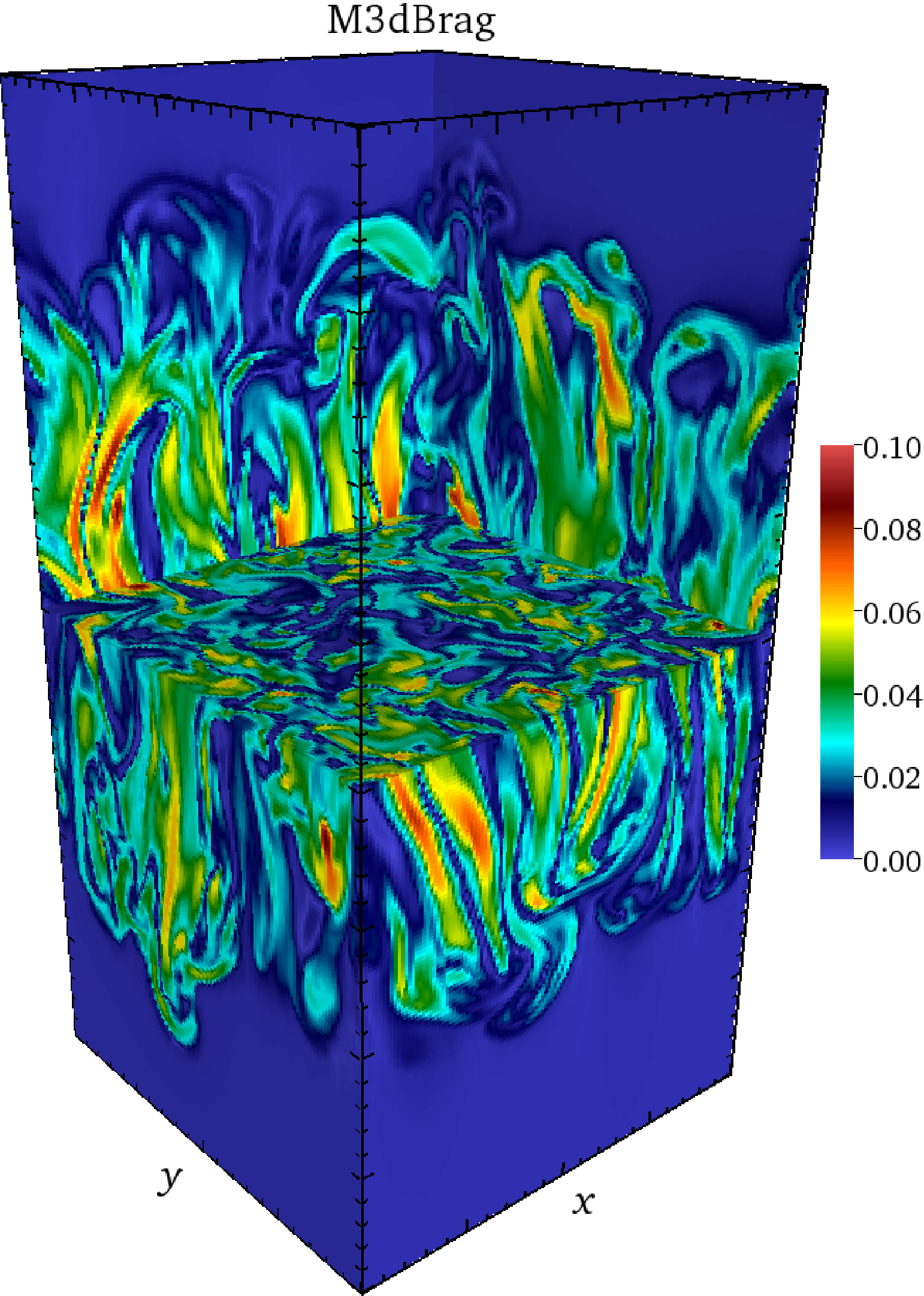}
\newline\newline
\includegraphics[height=3.5in]{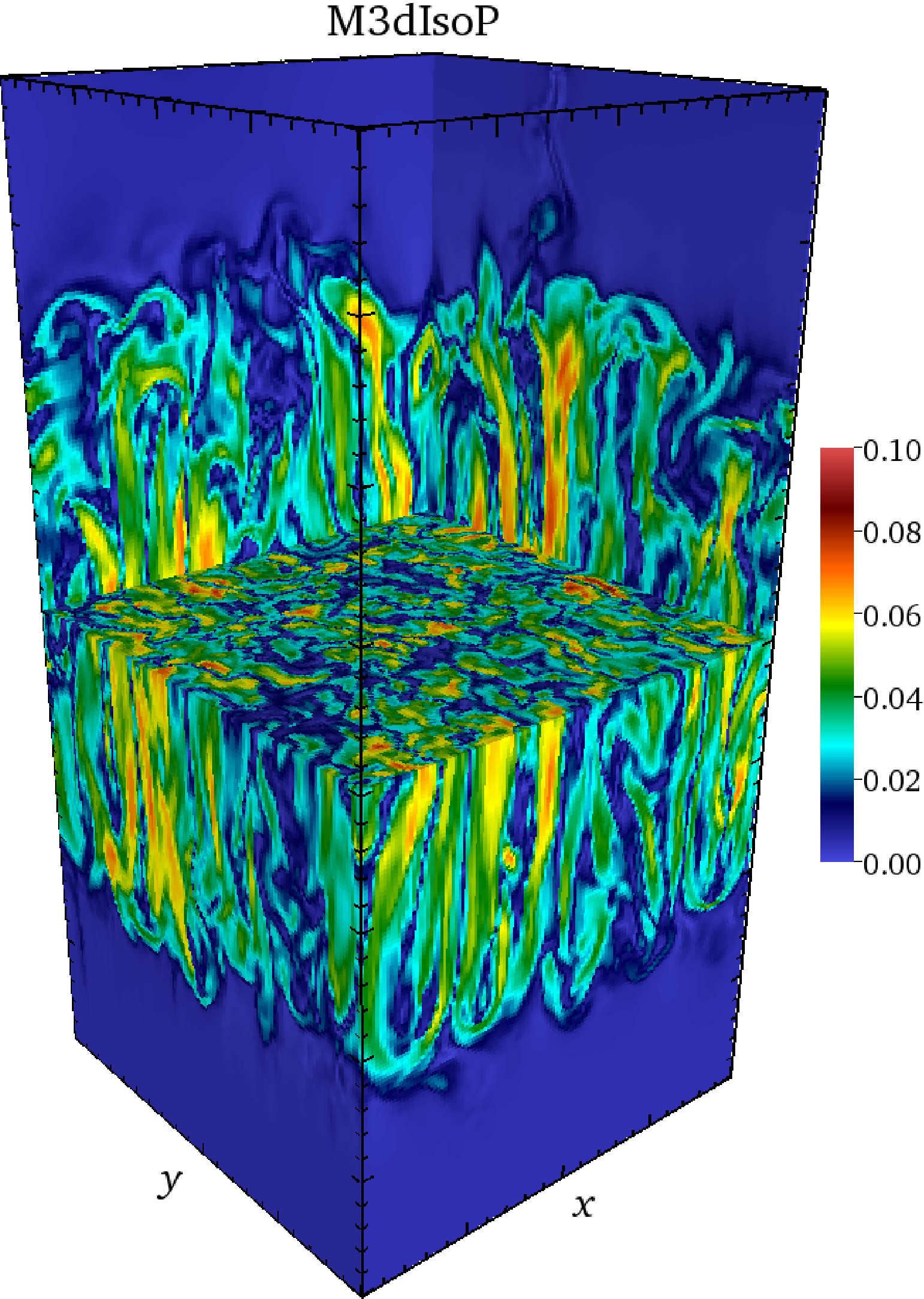}
\newline
\caption{Pseudocolor plot of magnetic-field strength (color) in runs M3dBrag ({\it top}) and M3dIsoP ({\it bottom}) at time $t = 37$ (in units of $H_0 / v_{\rm th,0}$; see Equation \ref{eqn:timeunit}). The computational domain has dimensions $L_x$$\times$$L_y$$\times$$L_z = 1$$\times$$1$$\times$$2$ (in units of $H_0$; see Equation \ref{eqn:lengthunit}). The magnetic-field strength is in units of $(4\pi p_0)^{1/2}$.}
\label{fig:3dMTI:global}
\end{figure}

Figure \ref{fig:3dMTI:global} exhibits the magnetic-field strength (color) at time $t = 37$ in runs M3dBrag and M3dIsoP. As in runs M2dBrag and M2dIsoP, the magnetic-field fluctuations in run M3dBrag emerge on larger scales than in run M3dIsoP. In addition, because small-scale features along field lines are viscously damped, there is less reconnection and the magnetic field remains coherent over longer distances. As a result, the folded structure of the field is more apparent in run M3dBrag. It is also clear from this figure that the field strength and the field-line curvature are anticorrelated.

\section{Discussion}\label{sec:summary}

\subsection{HBI}

Our simulations of the HBI that self-consistently allow for anisotropic pressure have a number of implications for the structure and evolution of cool-core clusters, some of which may be potentially observable. Perhaps the most important is our finding that field-line insulation of the entire cool core, previously found to be a nonlinear consequence of the HBI \citep[e.g.][]{pqs09,brbp09}, does not occur on astrophysically relevant timescales. The smaller degree of collisionality outside of the innermost regions of cool-core clusters ensures that the magnetic-field lines retain a strong vertical component outwards of $\sim$$50~{\rm kpc}$ from the cluster center. There, the comparatively large pressure anisotropy self-generated by the HBI forces the fastest-growing modes to have relatively long wavelengths, at which non-local effects start to play a role and curb growth \citep[see][]{lk12}. However, the relatively low temperatures and high densities in the innermost regions result in very little difference there between the inviscid and viscous cases. In the presence of radiative cooling, a cooling catastrophe inevitably occurs in these regions. This highlights the need for radio-mode feedback at these scales from a powerful central dominant galaxy.

Another important result from our simulations is the formation of cool filaments when anisotropic conduction, Braginskii viscosity, and radiative cooling are all taken into account. The pressure anisotropies generated by the HBI suppress its ability to impede the conductive flux to small radii. As a result, much of the atmosphere evolves slow enough to allow local thermal instability to set in. The physical characteristics of the magnetically aligned, cool filaments that emerge in our simulation (temperatures $\lesssim$$1~{\rm keV}$; magnetic-field strengths $\sim$$5$--$20~\mu{\rm G}$; magnetically aligned velocities $\sim$$100$--$300~{\rm km~s}^{-1}$; lifetimes $\sim$$10^2$--$10^3~{\rm Myr}$; aspect ratios $\sim$$30$--$80$) are similar to those observed or observationally inferred \citep[e.g.][]{hcjf06,mvrm10}. The magnetic field is responsible for insulating the filaments from the surrounding hot gas, with strengths capable of stabilizing the filaments and possibly delaying star formation \citep{fabian08}. In other words, our simulations are able to reproduce many of the observed properties of the cool filaments {\em without} a need for ad hoc heating prescriptions to offset radiative cooling (such as those used by \citealt{msqp12} and \citealt{smqp12}). However, we caution that our simulations may be overestimating the thicknesses of the filaments since ({\it i}) we have neglected line cooling below $T \sim 4 \times 10^7~{\rm K}$ and ({\it ii}) our spatial resolution is only $\simeq$$0.24~{\rm kpc}$. As a result, we are not able to form the very thin ($\lesssim$$100~{\rm pc}$) filaments that are sometimes observed in, for example, NGC 1275 \citep{fabian08} and Abell 1795 \citep{mvrm10}.

We have also found that the cool filaments formed in our simulations are often surrounded by gas that is hotter than average. This is because the compression of field lines during the formation of the filament induces a pressure anisotropy that leads to parallel viscous heating as it is collisionally relaxed. Temperatures in these hot filamentary ``envelopes'' are in the range $\approx$$7$--$10~{\rm keV}$. It would be interesting to examine deep Chandra observations of nearby cool-core clusters for evidence of hot filaments. Since these hot filaments would be relatively tenuous and hence have low emissivity, they could easily have been missed in existing analyses. However, the hot filaments may be revealed by constructing maps of the {\sc Fe\,xxvi}/{\sc Fe\,xxv} K-shell recombination line ratio.

Finally, the structure of the magnetic field produced by the HBI is significantly affected by Braginskii viscosity. Braginskii viscosity causes the HBI to grow efficiently only for a thin band of (parallel) wavelengths, which correlate with the local collisionality of the plasma (recall that modes with non-negligible growth rates satisfy $k_{\|} H \lesssim \sqrt{{\rm Kn}} \lesssim 3 k_{\|} H$; K11). The innermost (relatively collisional) regions of cool-core clusters are thus likely to harbor preferentially azimuthal magnetic fields due to efficient field-line reorientation by the HBI. However, AGN- and/or merger-driven turbulence may be able to randomize this field \citep{pqs10,ro10,mpsq11}. At distances $\sim$$50$--$100~{\rm kpc}$ from the cluster center, we have found that the flow of heat occurs primarily along magnetic sheaths (or filaments). Beyond $\sim$$100~{\rm kpc}$ (where the collisionality is low) our results suggest that the HBI exerts relatively little influence on the direction of the magnetic field over astrophysically relevant timescales. These findings can be further tested by well-resolved, global simulations that include anisotropic conduction, Braginskii viscosity, and turbulent stirring, as well as by cluster observations with the Expanded Very Large Array and (eventually) the Square Kilometer Array using rotation measures of background polarized radio sources \citep[e.g.][]{brm11}.

\subsection{MTI}

The consequences of including Braginskii viscosity in a treatment of the MTI are not as great as for the HBI. As a result, many of the observationally important results previously found in MTI simulations that neglected Braginskii viscosity (e.g. radially biased magnetic fields, flattened temperature profiles, vigorous subsonic turbulence that may lend an appreciable amount of pressure support; \citealt{ps05,ps07,psl08,mpsq11,pmqs11}) are for the most part unchanged. The reason is simple: the fastest-growing MTI modes are polarized such that $\delta B_{\|} = 0$ and therefore escape parallel viscous damping. There are some differences, however, that concern the emerging structure of the magnetic field.

For example, the magnetic-field fluctuations produced by the MTI with Braginskii viscosity emerge on larger scales than do those without Braginskii viscosity. This is because the generation of a positive pressure anisotropy shifts all unstable modes to larger (parallel) wavelengths. In addition, because field-line compressions and rarefactions are damped by the linear pressure anisotropy they generate, magnetic-field lines tend to remain locally parallel to one another. As a result, the magnetic field remains coherent over larger distances. Since the viscosity of the ICM is much larger than the magnetic resistivity, the turbulence tends to arrange the magnetic fields in long, thin flux sheets with $\ell_\perp \ll \ell_\| \sim \ell_{\rm visc}$, for which the field strength and the field-line curvature are anticorrelated \citep[see][]{sctmm04}. 

It follows that Braginskii viscosity implies a minimum physical scale above which the magnetic field can fluctuate freely. This is interesting in light of observations of the Faraday rotation measure, which indicate magnetic-field correlation lengths in the range $\sim$$0.1$ to a few tens of kpc \citep[e.g.][]{ve05,gmgpgrcf08,bfmggddt10,gdmfsgbvb10,ke11}. While Braginskii viscosity may not be unique in its effect on the geometry of the intracluster magnetic field, it may provide important clues about the physical processes controlling its correlation length. On the other hand, if Braginskii viscosity can be shown to be the dominant mechanism, it would provide a direct link between the macroscopic observables such as the field correlation length and the elusive, microscopic plasma processes.

\subsection{Comparison with Related Work}

Shear viscosity, whether isotropic or anisotropic, has long been suspected to play an important role in the ICM. Using {\it Chandra} X-ray observations of H$\alpha$ filaments in the Perseus cluster, \citet{fsccgw03} argued that the effective viscosity of the ICM must be large enough to explain the apparently laminar flow as rising bubbles drag up colder, inner gas. \citet{fabian03} also showed that viscosity can have an important effect in dissipating the sound energy produced by the formation of bubbles and heating the surrounding ICM. Subsequent numerical work by \citet{rmfsv05} demonstrated that viscosity may be necessary to maintain the observed integrity of AGN-blown buoyant cavities by quenching Kelvin-Helmholtz and Rayleigh-Taylor instabilities. Eliminating the need for bubbles to inflate supersonically in order to evade these instabilities, viscosity also provides a natural explanation for the observed absence of strong shocks bounding the ghost cavities. \citet{ds09} extended this work by taking into consideration Braginskii viscosity and studying the influence of different magnetic-field orientations. Finally, \citet{kscbs11} showed that Braginskii viscosity, when regulated by microscale instabilities, provides a local thermally stable heating source. Given a sufficient supply of turbulent power, this provides a physical mechanism for mitigating cooling flows and preventing cluster core collapse. Our study compliments all these efforts and lends credence to the notion that understanding the viscosity of the ICM is vital to understanding its morphology, energetics, and stability.

In parallel to the work presented here, \citet[][hereafter P12]{pmqs12} carried out an independent study of the HBI and MTI subject to Braginskii viscosity. In the areas of overlap between our work and that presented in P12, there is broad agreement. However, there are some differences worth noting, which we believe are mainly due to different choices of free parameters, initial conditions, and numerical approaches.

As discussed in Section \ref{sec:parameters}, our choice of free parameters is motivated by the physical conditions in galaxy clusters and the numerical constraints related to the impact of microscale instabilities on our results. Accordingly, we have chosen to initialize the magnetic-field strength in all of our simulations using $\beta \sim 10^4$--$10^5$. In all cases, the energy in the magnetic field saturates with $\beta \sim 10^2$--$10^3$, in approximate equipartition with the fluid motions. By comparison, in P12 the local HBI and MTI simulations have an initial $\beta \sim 10^{12}$, the fiducial global HBI simulations have an initial $\beta \sim 10^7$, and the global MTI simulations have an initial $\beta \sim 10^5$--$10^6$. None of the runs presented in P12 appear to saturate with approximate equipartition between kinetic and magnetic energies. One consequence of the different choice of $\beta$ is that the saturated magnetic field in our MTI simulations remains strongly biased in the vertical direction, whereas in the P12 simulations the magnetic field becomes nearly isotropic (a result also found by \citealt{mpsq11}). In the saturated state of our MTI runs, the horizontal kinetic energy is less than the vertical magnetic energy and so there is insufficient energy in the horizontal motions to isotropize the magnetic field.

Differences between our results and those in P12, especially concerning the evolution of the HBI, may also be due to different choices of ${\rm Pr}$. We have used ${\rm Pr} \simeq 0.02$ (see Equation \ref{eqn:prandtl}), while P12 chose ${\rm Pr} = 0.01$. Our non-radiative and radiative HBI simulations also start with atmospheres (similar to A1795 and A85) that are less collisional than the fiducial cool-core model in P12. In the notation of P12's figure 2, our cool-core atmospheres have $\cond / \omega_{\rm buoy} \sim 10$--$30$ for $r \gtrsim 20~{\rm kpc}$ (i.e. $\visc / \omega_{\rm buoy} \sim 2$--$5$). This may be the reason why we find the HBI to be appreciably suppressed beyond $\sim$$50~{\rm kpc}$, while those authors do not. However, this may also be due to the fact that all of our HBI simulations started with a magnetic field aligned with gravity, whereas P12 initialized their simulations with tangled magnetic fields on scales $30$--$50~{\rm kpc}$. Future work may resolve these differences.

Finally, in our paper we have highlighted the influence of Braginskii viscosity on the morphology of the intracluster magnetic field and on the formation of cool filaments. We have also tried to address the impact of microscale instabilities on our results by taking two different approaches to capture their macroscale effects: ({\it i}) by working at very high spatial resolution so that the microscale instabilities grow fast enough to naturally regulate the pressure anisotropy, and ({\it ii}) by employing anisotropy limiters to restrict the pressure anisotropy to stable or marginally stable values. Using these approaches, we have shown that the manner in which microscale instabilities saturate affects the properties of the intracluster magnetic field. These approaches must be considered provisional, however, as there is currently no complete microphysical theory concerning the saturation of these instabilities.

\subsection{Summary and Outlook}

In this paper, we have employed numerical simulations to investigate the linear and nonlinear dynamic and radiative stability of a weakly collisional, magnetized ICM. We have taken into consideration the effects of anisotropic heat and momentum transport, radiative cooling, magnetic tension, and microscale instabilities, and have ascertained a number of their implications for the structure of the intracluster magnetic field, the resolution of the cooling-flow problem, and the nature of convective turbulence in a dilute plasma.

Despite such progress, there are still a number of unanswered questions, some of which may be addressed by well-resolved global, three-dimensional numerical simulations of cluster cool cores and cluster outskirts. However, the efficacy of such simulations is likely to be contingent upon the implementation of a realistic sub-grid model for the microscale instabilities that captures their interplay with the computationally resolved meso- and macroscales. While formulating such a model is a rather formidable task, dedicated efforts to construct a more complete microphysical theory and to understand its bearing on heat and momentum transport, magnetogenesis, and thermodynamic stability in astrophysical systems are clearly needed.

\section*{}

Support for M.W.K. and T.B. was provided by NASA through Einstein Postdoctoral Fellowship Award Numbers PF1-120084 and PF9-00061, respectively, issued by the Chandra X-ray Observatory Center, which is operated by the Smithsonian Astrophysical Observatory for and on behalf of NASA under contract NAS8-03060. M.W.K. was supported by STFC grant ST/F002505/2 during the early phases of this work. T.B. and C.S.R. acknowledge support from NSF under grant AST-0908212. The Texas Advanced Computing Center at The University of Texas at Austin provided HPC resources under grant numbers TG-AST100030 and TG-AST030031N. This work used the Extreme Science and Engineering Discovery Environment (XSEDE), which is supported by NSF grant OCI-1053575. We thank Tobias Heinemann for assistance with plotting magnetic-field lines; Steve Balbus, Henrik Latter, and Alex Schekochihin for incisive comments on an early version of the manuscript that led to a much improved presentation; and, Mark Avara, Ian Parrish, and Eliot Quataert for useful conversations. 



\label{lastpage}

\end{document}